\documentclass[12pt,a4paper,final,headline,chapterbib,oneside,nocenter,noupper,bold]{thesis}
\newcommand{\be}{\begin{equation}}
\newcommand{\ee}{\end{equation}}
\newcommand{\ba}{\begin{eqnarray}}
\newcommand{\ea}{\end{eqnarray}}

\usepackage{graphicx}% Include   files
\usepackage{dcolumn}% Align table columns on decimal point
\usepackage{bm}% bold mat\

\begin{document}

%\preprint{APS/123-QED}

%\pacs{81.05.Rm, 82.70.-y, 83.80.Fg}

\title{On the rigidity of amorphous solids}

%\affiliation{Service de Physique de l'Etat Condens\'e (CNRS URA 2464), DSM/DRECAM, CEA Saclay,
%91191 Gif sur Yvette, France}
%\email{matthieu.wyart@m4x.org}
\maketitle

\tableofcontents

\begin{abstract}

We poorly understand the properties of amorphous systems at small
length scales, where a continuous elastic description breaks down. This
is apparent when one considers their vibrational and transport
properties, or the way forces propagate in these solids. Little is known
about the microscopic cause of their rigidity. Recently it has been
observed numerically that an assembly of elastic particles has a
critical behavior near the jamming threshold where the pressure
vanishes. At the transition such a system does not behave as a continuous
medium at any length scales. When this system is compressed, scaling is
observed for the elastic moduli,  the coordination number, but also for
the density of vibrational modes.  In the present work we derive
theoretically these results, and show that they apply to various
systems such as granular matter and silica, but also to colloidal
glasses. In particular we show that: (i) these systems present a large
excess of vibrational modes at low frequency in comparison with normal
solids,  called the ``boson peak" in the glass literature. The
corresponding modes are very different from plane waves, and their
frequency is related to the system coordination; (ii) rigidity is a
non-local property of the packing geometry, characterized by a length
scale which can be large. For elastic particles this length diverges
near the jamming transition; (iii) for repulsive systems the shear
modulus can be much smaller than the bulk modulus. We compute the
corresponding scaling laws near the jamming threshold. Finally, we
discuss the implications of these results  for the glass transition,
the transport, and the geometry of the random close packing.

\end{abstract}

\chapter{Introduction}
\label{c1}
\section{ Anomalous properties of amorphous solids}

In the last century, the development of statistical physics revolutionized our understanding of matter. It furnished a microscopic explanation of heat, and  gave a description of  different states of matter, such as the liquid or the solid state.  Later, it explained that sudden transitions between these states can occur when a parameter is slowly  tuned, despite the microscopic interactions staying the same. At the heart of these discoveries lie the concepts of equilibrium and entropy.   At equilibrium, all the possible states  with identical energy have an equal probability:  this allows to define an entropy, and a temperature. Nevertheless, many systems around us are not at equilibrium. These can be open systems crossed by fluxes of matter and heat, such as biological systems. Another case is glassy systems, such as structural glasses or spin glasses, where the characteristic times  become so slow that there are never equilibrated on  experimental time scales. Finally there are also systems where particles are too large to be sensitive to temperature, such as granular matter. These systems are still poorly understood, and one of the current goal of nowadays statistical physics  is to explain their original properties, and hopefully to find generic methods to describe them.

We do not have a satisfying description of amorphous systems such as structural glasses, colloids, emulsions or granular matter. This is particularly apparent when one considers the low temperature properties of glasses \cite{W.Phillips}. Their low-temperature specific heat has a nearly-linear temperature dependence rather than varying as $T^3$ as would be found in a crystal  \cite{W.Phillips}. The prevailing explanation for this linear specific heat is in terms of tunneling in localized two-level systems \cite{thermo}: atoms or group of atoms switch between two possible configurations by tunneling. This phenomenological model has also explain the $T^2$ dependence of the thermal conductivity at very low temperature. However, several empirical facts are still challenging the theory \cite{AndyAnderson,ludwig}. Furthermore, after 30 years of research there is yet no accepted picture of what these two-levels systems are.  At higher temperature, around typically 10 K which corresponds to the Thz frequency range for phonons, other universal properties  of glasses are not fully understood. In particular the thermal conductivity displays a plateau, which suggests that at these frequencies phonons are strongly scattered.  This effect is significant: for example in silica glass, the thermal conductivity  is several orders of magnitude smaller than in the crystal of the same composition  \cite{bergman}.

\begin{figure}
\centering
\includegraphics[angle=0,width=9cm]{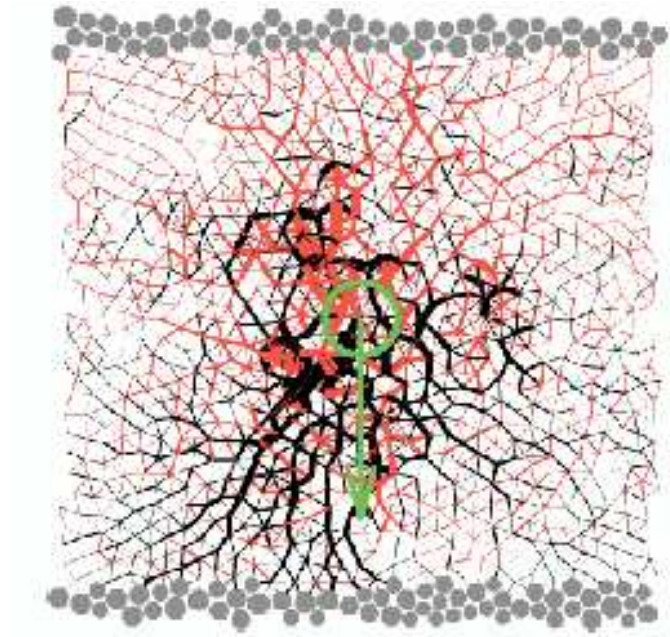}
\caption {Response to a monopole of force in a Lennard-Jones system of 1000 particles. Black (grey) lines correspond to compressive (tensile) stesses. Leonforte et al. \cite{tanguy3}  }
\label{figforcetanguy}
\end{figure} 

Athermal amorphous systems, such as granular matter, also display fascinating properties, both in their static behavior and in their rheology. The following puzzle underlines the subtlety  of force propagation in granular matter \cite{ball}: the supporting force under a conical heap of poured sand is a minimum, rather than a maximum, at the center of the pile where it is deepest. As we shall discuss in the next Chapter, it has been proposed  that in granular medium the force propagates differently than in a continuous elastic body \cite{cates1,cates2}.  It turns out experimentally  \cite{behringer,reydellet} and numerically \cite{bouchaud, tanguy3}  that an elastic-like  behavior is recovered at large distances. Curiously enough, the cross-over length can be  large in comparison with the particle size.  Fig.(\ref{figforcetanguy})  shows  the response to a point force  in a Lennard-Jones simulations \cite{tanguy3} at zero temperature. The average response is similar to the one of a continuous elastic medium, but near the source the fluctuations are of the order of the average. They decay exponentially with distance,  with a  characteristic length of roughly 30 particles sizes. One may ask what determines such a distance, below which an amorphous solid behaves as a continuous medium. More generally, what length scales characterize these systems?

The length scales we are discussing might  also affect  the rheology of granular matter. An interesting question is how grains flows, or how they compact \cite{compaction}. For example if a layer of sand is inclined, an avalanche is triggered.  Interestingly the angle $\theta$ of avalanche appears to be controlled by the width $h$ of the granular layer. $\theta$ decreases when $h$ grows  when $h$ is smaller than of the order of ten particle sizes. Similar length scales also appear in the spatial correlations of the velocities of grains in dense flows \cite{pouliquen}.

A particularity of the amorphous state is that it is not at equilibrium.  Consequently the properties and the microscopic structure of these systems depend much on their history. For example if a granular pile is made by a uniform deposition, rather than by pouring sand from the top, the supporting force does not display the minimum discussed above at the center of the pile, but rather a flat maximum.   Often amorphous solids are obtained from a fluid phase by varying some parameters such as temperature, density or applied shear stress until the system stops flowing: this is the jamming transition \cite{liu}.  As the dynamics greatly slows down once this transition is passed,  the structure of  amorphous solids  does not to differ too much from the marginally stable state at the transition.  Thus a better understanding of the microscopic features of amorphous solids requires a better knowledge of the jamming mechanisms. It is a hard and much studied problem. When a glass is cooled rapidly enough to avoid crystallization, the relaxation times rapidly grow. In some cases the relaxation times  follow an Arrhenius law with temperature; such  glasses are called ``strong''. If the relaxation times grow faster, the glass is ``fragile'' \cite{ang}. There is no available theory to compute quantitatively the  temperature dependence  of the relaxation times, and to decide a priori which glasses are strong or fragile. Recently it was observed numerically and experimentally that the relaxation in the super-cooled is very heterogeneous and involves rearrangements of particles clusters \cite{Ediger,Weeks}.  Althought several models of the glass transition predict such heterogeneities, see e.g. \cite{twbbb} and references therein,  their cause and  nature  is still a much debated question.

Although they can lead to collective dynamics, most of the spatial models of the relaxation  near the jamming threshold have purely {\it local}  rules. This is the case for example for kinetically constrained models \cite{ka} where particles are allowed to move individually if their direct neighborhood satisfies some specific conditions.  The starting point of the present work is the following remark: the stability against individual particle displacement is much less demanding than the stability toward collective motions of particles. In $d$ dimensions, $d+1$ neighbors are sufficient to pin one particle. As we shall discuss in details later, Maxwell showed that $2d$ contacts per particle {\it on average} are necessary to guarantee the stability of a solid \cite{max}.  The fact that the criterion of stability is non-local suggests that it is so for the minimal motions responsible for the relaxation. In any case, this underlines the importance of understanding what guarantees the stability of an assembly of interacting particles. In what follows we aim to furnish a microscopic description  of the rigidity characterizing amorphous solids. 

\begin{figure}
\centering
\includegraphics[angle=0,width=13cm]{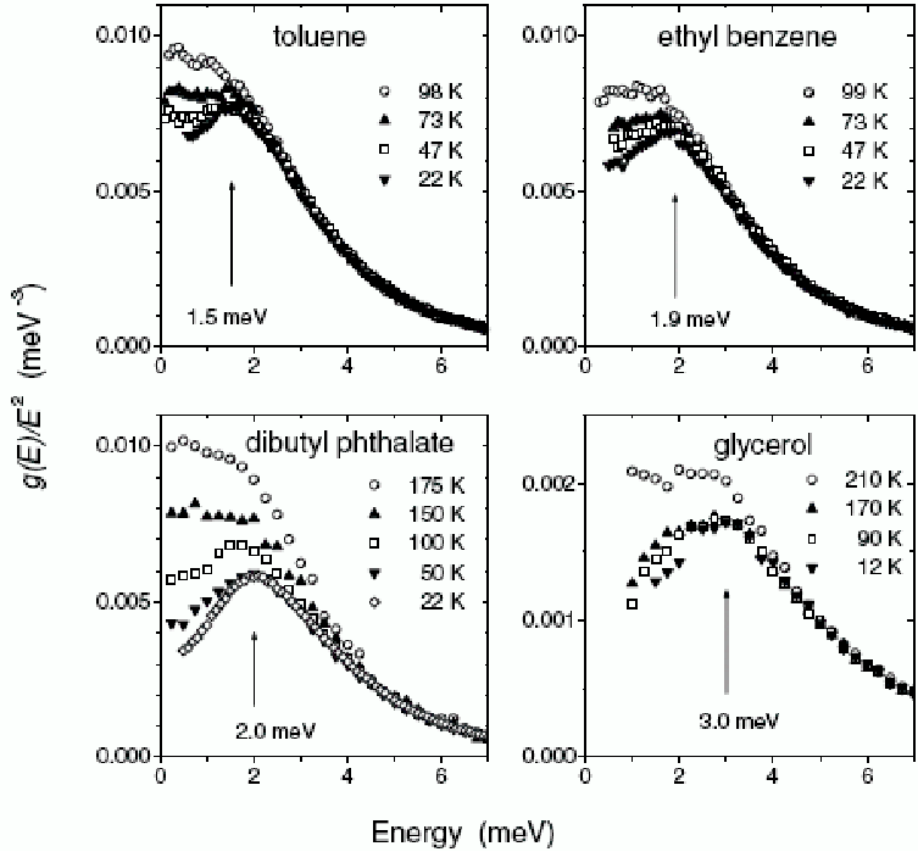}
\caption {Reduced density of states ($g(E)\approx D(\omega)$ following our notation) of collective motions in toluene, ethylbenzene, dibutylphthalate, and glycerol glasses. Arrows indicate the energy of the boson peak estimated from the data at lowest temperature. Chumakov et al., \cite{chuma}.}
\label{g}
\end{figure}

The informations about the rigidity of  a solid against collective particle motions are contained in the density of the vibrational modes $D(\omega)$:  a system is stable if there are no unstable modes. In a continuous isotropic elastic medium, the invariance by translation implies that the vibrational modes are plane waves. As a consequence in  three-dimensions the density of vibrational modes $D(\omega)$ follows  the Debye law  $D(\omega) \propto \omega^2$. By contrast, at low frequency all glasses present an excess of vibrational modes in comparison with the Debye behavior. This excess of vibrational modes is the so-called  ``boson peak''  \footnote[1]{The term  ``boson peak" was introduced because the amplitude of the scattering peak varies according to the Bose-Einstein factor at low temperature.} which appears as a maximum in $D(\omega)/\omega^2$. It is observed in particular in scattering experiments, see Fig.(\ref{g}). The frequency of the peak lies in the terahertz range,  that is typically between $\omega_D/10$ and $\omega_D/100$, where $\omega_D$ is the  Debye frequency.  

Several empirical facts suggest that the presence of these excess modes is related to many of the original properties of amorphous solids. In most glasses \cite{so,tao,eng,chuma}, with some exceptions as silica \cite{wisch},  this boson peak shifts toward zero frequency when  the glass is heated, as shown in Fig(\ref{g}). Eventually the peak reaches zero frequency, as it has been observed numerically \cite{kob} and empirically \cite{tao}. This suggests that in some glasses the corresponding modes take part in the relaxation of the system \cite{parisi,gri}. The presence of the boson peak  also affects the low-temperature properties of glasses. The plateau in the thermal conductivity appears at temperatures that correspond to the boson peak frequency \cite{W.Phillips}, which suggests that the excess-modes do not contribute well to transport. Furthermore, since these modes are soft, one expects that their non-linearities are important. Hence these modes may form two-levels systems \cite{buchenau, shlomon}.  Finally, as the linear response to any force or deformation can be expressed in terms of the vibrational modes, it is reasonable to think that the boson peak affects force propagation. The recent simulations from which Fig.(\ref{figforcetanguy}) is taken show that the length scale that appears in the response to a point force also appears in the normal mode analysis: only for larger system sizes the lowest frequency modes are the one expected from a continuous elastic description \cite{tanguy1,tanguy2}.

This excess of vibrational modes has been  studied with  diverse approaches. There are  phenomenological models also dealing with two-levels systems such as the ``soft potential theories" \cite{buchenau, karpov, gurevich}, which assumes the presence of strongly anharmonic localized soft potential with randomly distributed parameters. A second approach consists in studying  the vibrations of elastic network with disorder.  Simulations of a harmonic lattice with a random distribution of force constants \cite{schi,tu} exhibit a density of states qualitatively similar to what is observed with glasses in scattering experiments. From the theoretical point of view, models that assume spatially fluctuating elastic constants  \cite{schirmacher} show an excess of modes whose frequency decreases with the amplitude of the disorder.   Recently further developments  were proposed using the euclidean random matrix theory \cite{parisi2, parisi3, mezard}, where  an assembly of particles at infinite temperature is considered. The density of states corresponds to the spectrum of a disordered matrix, the dynamical matrix. When the density $\rho$ is infinite, the system behaves as a continuous medium. To approximate the density of states at finite $\rho$ one uses perturbation theory in the inverse density. This leads to an excess of modes whose frequency  goes to zero as the density decreases toward a finite threshold, and furnishes several exponents that describes the density of states at this transition.   A third approach uses the mode coupling theory (MCT).  MCT models the dynamics of supercooled liquids, and predicts a glass transition at finite temperature where the relaxation time diverges. In the glass region, the structure does not fully relax for any waiting time. Nevertheless if the mode coupling equations are used into this glass region, the dynamic that appears is non-trivial. It can be interpreted in terms of harmonic vibrations \cite{gotze} around a frozen amorphous structure. The corresponding spectrum displays an excess of modes that converges toward zero frequency  at the transition. 

These different approaches describe the presence of an elastic instability, and make predictions on how the density of states behaves with some parameters when this instability is  approached.  Nevertheless they have several drawbacks. In these models, disorder is the main cause for the excess density of states. This is  inconsistent with scattering data that show that some crystals also  display   excess vibrational modes \cite{nakayama,leadbetter,caplin,bilir}. In particular silica, the glass with one of the strongest boson peak, as a density of states extremely similar to the crystals of identical composition and similar density, see Fig.(\ref{cry}), as we shall discuss in more details  in Chapter \ref{c8}.  Thus the cause, and more importantly the {\it nature} of these excess modes are still unclear.  In particular one may ask what microscopic  features determine  the vibrational properties of amorphous solids at these intermediate frequencies, and what are  the signatures of marginal stability  at a microscopic level.  In what follows we attempt to answer these questions in weakly-connected amorphous solids such as an assembly of repulsive, short-range particles. Then we argue that this description also applies to other systems, such as silica glass or  particles with friction. Finally we derive some properties of colloidal glasses and discuss the possible implications of this approach for the glass transition.

\section{Critical behavior at the jamming transition}

\begin{figure}
\centering
\includegraphics[angle=0,width=10cm]{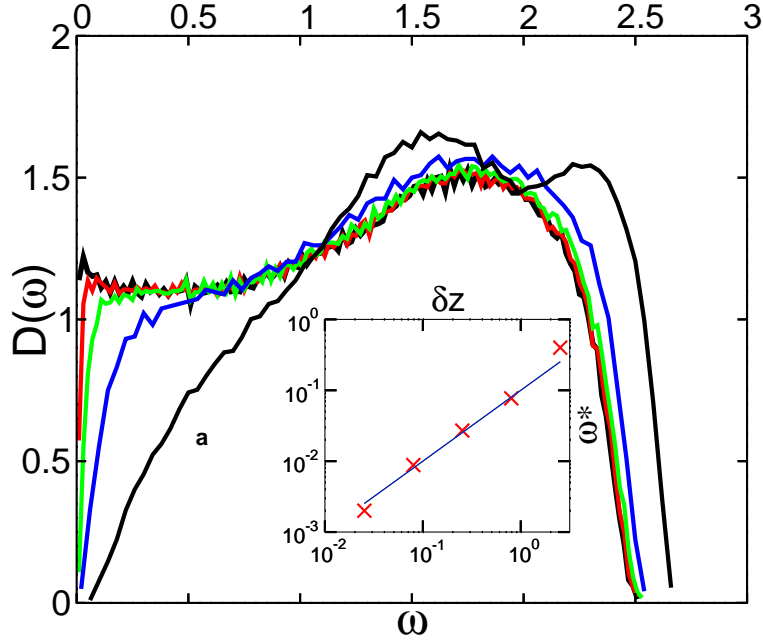}
\caption {$D(\omega)$ {\it vs} angular frequency $\omega$ for the simulation of Ref \cite{J}.  1024 spheres interacting with repulsive harmonic potentials were compressed in a periodic cubic box to volume fraction $\phi$, slightly above the jamming threshold $\phi_c$.  Then the energy for arbitrary small displacements was calculated and the dynamical matrix inferred. The curve labeled $a$ is at a relative volume fraction $\phi - \phi_c = 0.1$.  Proceeding to the left the curves have relative volume fractions $10^{-2}$,  $10^{-3}$,  $10^{-4}$,  $10^{-8}$, respectively.   Inset: Scaling of $\omega^*$ {\it vs} $\delta z$.  $\omega^*$ for each $(\phi-\phi_c)$ is determined from the data in the main panel as the frequency where $D(\omega)$ is half of the plateau value. $\delta z$ vs $(\phi-\phi_c)$ is obtained from the scaling measured in \cite{J}. The line has slope $1$.
}
\label{f1}
\end{figure}

Recently, C.S O'Hern, L.E Silbert  et al. \cite{J,ohern} exhibited a system whose vibrational properties are dramatically different from a conventional solid. They simulate frictionless repulsive particles with short range interactions at zero temperature. The authors consider soft spheres.  For inter-particle distance $r<\sigma$, the particles are in contact and interact with a potential:
\be
\label{opo}
V(r)=\frac{\epsilon}{\alpha} \left(1-\frac{r}{\sigma}\right)^{\alpha} 
 \ee
where $\sigma$ is the particle diameter  and $\epsilon$ a characteristic energy.  For $r>\sigma$  the potential vanishes and particles do not interact.  Henceforth we express all distances in units of $\sigma$, all energies in units of $\epsilon$, and all masses in units of the particle mass, $m$. The simulations were done for $\alpha=2$ (harmonic), $\alpha=5/2$ (Hertzian contacts) and $\alpha=3/2$. When the packing fraction $\phi$ is low, such system is in a gas phase, and the pressure $p$ is zero. At high packing fraction, it forms a solid and has a positive pressure. There is a transition between this two phases where the pressure vanishes: this is the jamming transition. At that point the density of states  behaves as a  {\it constant} instead of the quadratic dependence expected for normal solids, see Fig(\ref{f1}). The authors also study the solid phase when the pressure  decreases toward zero. They find that the jamming transition acts as a {\it critical point}: the microscopic structure, the vibrational modes and the macroscopic elastic properties display scaling behaviors with the pressure $p$ or with $\phi-\phi_c$, where  $\phi_c$ is the packing fraction at the transition. In three dimensions $\phi_c\rightarrow 0.64$ which corresponds to the random close packing \footnote[2]{The parameter $\phi-\phi_c$ is somewhat less natural than the pressure  because $\phi_c$ can vary from sample to sample. The distribution of $\phi_c$ converges to a well-defined value only when the number of particle $N$ diverges. Nevertheless, the parameter $\phi-\phi_c$ has the advantage of being purely geometrical, and  following \cite{J} we should  use it in most cases.}. Concerning the structure, the coordination number $z$, which is the average number of contacts per particles, is found to follow:
\be
\label{ll}
z-z_c\sim(\phi-\phi_c)^{\frac{1}{2}}
\ee
independently of the potential, where $z_c=2d$, and $d$ is the spatial dimension. This singular increase of the coordination was already noticed in \cite{durian}. Another striking observation is the presence of a singularity in the pair correlation function $g(r)$ at the jamming threshold. $g(r)$ has an expected delta function of weight $z_c$ at a distance 1 that represents all particles in contact. But it also displays  the following singularity:
\be
\label{ggg}
g(r)\sim \frac{1}{\sqrt{r-1}}
\ee
which indicates that there are many particles {\it almost} touching. Again this is independent of the potential. This property was observed in other situations \cite{sil}. Note that Eq.(\ref{ll}) and (\ref{ggg}) are related. A small affine compression of the configuration is equivalent to an increase of the particle diameter of an amount $\approx\frac{\phi-\phi_c}{3\phi_c}$. At the jamming threshold this would lead to an increase in the coordination number:
\be
z-z_c \approx 4\pi \int _1^{1+\frac{\phi-\phi_c}{3\phi_c}} g(r) r^2 dr\sim \int _1^{1+\phi-\phi_c} \frac{1}{\sqrt{r-1}} dr\sim (\phi-\phi_c)^{\frac{1}{2}}
\ee
as observed.

As we mentioned, these simulations also reveal unexpected features in the density of states, $D(\omega)$:  (a) As shown in Fig(\ref{f1}), when the system is most fragile, at $p\rightarrow 0$, $D(\omega)$ has a plateau extending down to zero frequency with no sign of the standard $\omega^2$ density of states normally expected for a three-dimensional solid.  (b) As shown in the inset to that figure, as $p$ increases, the plateau erodes progressively at frequencies below a frequency $\omega^*$, which scales in the harmonic case as:
\be
\omega^*\sim \delta z
\ee
(c) The value of $D(\omega)$ in the plateau is unaffected by this compression.  (d) At frequency much lower than $\omega^*$, $D(\omega)$ still increases much faster with $\omega$ than the quadratic Debye dependence. 

Finally, it was conjectured by Alexander \cite{shlomon} that the elastic moduli should scale at the jamming transition. Such scaling properties where observed in emulsions near the jamming transition \cite{mason}, but the presence of noise in the measure of the packing fraction makes the exponent hard to identify. In \cite{J}, the bulk modulus $B$, the shear modulus $G$ and the pressure are found to follow:
\ba
B&\sim&  (\phi-\phi_c)^{\alpha-2}\\
G&\sim& (\phi-\phi_c)^{\alpha-3/2}\\
p&\sim& (\phi-\phi_c)^{\alpha-1}
\ea

These results raise many questions. Among others: (i)  the vibrations of a normal solid are plane waves; what are the vibrations of  a random close packing of elastic spheres? (ii) How  is the behavior of the structure and the density of states related? This critical behavior is a stringent test to such theories. (iii)  How does the microscopic structure, for example the coordination number, depend on the the system history? (iv) What are the different  elastic properties of this system, that behaves almost as a liquid at the transition, as the shear modulus becomes negligible compare to the bulk modulus? For example, how does it react to a local perturbation?

\section{Organization of the thesis}

At the center of our argument lies the concept of {\it soft modes}, or {\it floppy modes}. These are collective modes that conserve the distance at first order between any particles in contact.  They have been discussed in relation to various weakly-connected networks such as covalent glasses \cite{phillips, thorpe}, Alexander's models of soft solids \cite{shlomon}, models of static forces in granular packs\cite{Tom1,moukarzel} and rigidity percolation models, see e.g. \cite{yeye}.  As we will discuss below, they are present when a system is not enough connected.  As a consequence, as Maxwell showed \cite{max},  a system with a low average coordination number $z$ has some soft modes and therefore is not rigid. There is a threshold value $z_c$ where a system can become stable, such a state is called {\it isostatic}.  As we shall discuss later, this is the case at the jamming transition, if rattlers (particles with no contacts) are excluded.  There are no zero-frequency modes except for the trivial translation modes of the system as a whole.  However, if any contact were to be removed, there would appear one soft mode with zero frequency.  Using this idea we will show in what follows  that isostatic states have a constant density of states in any dimensions.   When $z > z_c$, the system still behaves as an isostatic medium at short length scale, which leads to the persistence of a  plateau in the density of states at high frequency. 

The second concept we use is at the heart of the work of Alexander on soft solids \cite{shlomon}.  In continuum elasticity the expansion of the energy for small displacements contains a term proportional to the applied stress (that we shall also call {\it initial stress} term following \cite{shlomon}), as we shall discuss in the next Chapter. It is responsible for the vibrations of strings and drumheads, but also for inelastic instability such as the buckling of thin rods. Alexander pointed out that this term has also strong effects at a microscopic level in weakly-connected solids. For example, it confers rigidity to gels, even though  these do not satisfy the Maxwell criterion for rigidity. We will show that althought this term does not affect much the plane waves, it strongly affects the soft modes. In a repulsive system of spherical particles it lowers their frequency. We shall argue that this can change dramatically the density of states at low frequency, as it will be confirmed by a comparison of simulations where the force in any contact is present, or set to zero. We show that these considerations lead to a inequality between the excess connectivity $\delta z\equiv z-z_c$ and the pressure that guarantees the rigidity of such amorphous solids. This relation between stability and structure will enable us to discuss how the history affects the microscopic structure of the system. In particlar, we shall argue that the preparation of the system used in \cite{J} leads to a {\it marginally stable} state, even when $\phi>\phi_c$. This will account for both the scaling of the coordination, and for the divergence of the first peak in $g(r)$ at the random close packing. 
  
A distinct and surprising property of the system approaching the jamming threshold is the nature of the quasi-plane waves that appear at lower frequency than the excess of modes. The peculiar nature of the transverse waves already appears at zero wave vector, as the shear modulus becomes negligible compared to the bulk modulus near jamming. If the response to a shear stress were be a perfect affine displacement of the system, the corresponding energy would be of the same order of the energy induced by a compression. As this is not the case, this indicates the presence of strong non-affine displacements in the transverse plane waves. To study this problem we shall introduce a formalism that writes the responses of the system in terms of  the force fields that balance the force on every particle. This enables us to derive the scaling of the elastic moduli, and to compute the response to a local perturbation at the jamming threshold. We show that this response extends in the whole system.

We study how these ideas apply to real physical systems, such as granular matter, glasses and dense colloidal suspensions. In granular matter friction is always present. We derive the equation of the soft modes with friction.  The main difference with frictionless particles is that rotational degrees of freedom of grains now matter, but our results on the vibrational modes and on the elastic properties are unchanged. Then we discuss the case of glasses. In these systems the coordination number is not well defined, as there are long range interactions such as Van der Waals forces. We show that if the hierarchy of interactions strengths is large enough, our description of the boson peak still applies. In particular we argue that the boson peak of silica glass corresponds to the slow modes that appear in weakly connected systems. Our argument also rationalizes why the crystal of identical composition and similar density, the crystobalite, has a similar density of state.  We propose testable predictions to check if the same description holds for   Lennard-Jones systems. Finally we study dense hard sphere liquids. Our main achievement is to derive an effective potential that describes the hard sphere interaction when the fast temporal fluctuations are averaged out. Our effective potential is exact at the jamming threshold. It allows to define normal modes and to derive several properties of a hard sphere glass.  In particular it implies that the jamming threshold act as a critical point both in the liquid and in the solid phases. This suggests original relaxation processes.

The thesis is organized as follows. Chapter \ref{c2}  is introductive: we define rigidity and soft modes and discuss how these concept were used to study covalent glasses, force propagation or gels.  In the  Chapter \ref{c3} we use a simple geometric variational argument based on the soft modes to show that isostatic states have a constant density of states. The argument elucidates the nature of these excess-modes.In the Chapter \ref{c4} we compute the density of states when the coordination of the system  $z=z_c+\delta z$ increases with the packing fraction. At that point we neglect the effect of the applied stress on the vibrations. This approximation corresponds to a real physical system: a network of relaxed springs. We show that  such  system behaves as an isostatic state for length scales smaller than $l^*\sim \delta z^{-1}$. This leads to a plateau in the density of states for frequency higher than $\omega^*\sim \delta z$. At lower frequency, we expect that the system behaves as a continuous medium with a Debye behavior, which is consistent with our simulations. We extend this result to the case of tetrahedral networks.  In Chapter \ref{c5} we study the effect  of the applied pressure on $D(\omega)$. We show that althought it does not affect much the plane waves, the applied stress lowers the frequency of  the anomalous modes. We give a simple scaling argument to evaluate this effect, and we discuss its implication for the density of states. Incidentally this also furnishes an inequality between $\delta z$ and the pressure which generalizes the Maxwell criterion for rigidity. We discuss the different length scales that appear in the problem. In Chapter \ref{c6}, we discuss the influence of the cooling rate and the temperature history on the spatial structure and the density of states of the system. We show that the scaling of the coordination and the divergence in $g(r)$ are related to the marginal stability of the system of \cite{J}. Some elastic properties of this tenuous system are computed in Chapter \ref{c7}  , in particular the elastic moduli. Chapter \ref{c8} is devoted to the applications of our arguments to  granular matter and glasses. This approach explains the qualitative shape of the density of state of silica.  In Chapter \ref{c9} we derive an effective  potential for hard spheres, and compute some properties of a hard spheres liquid near the glass transition. To conclude in chapter 11 we discuss the possible applications of these ideas to the low temperature properties of glasses, the glass transition and the rheology of granular matter.

\chapter{Soft Modes and applications}
\label{c2}
\section{Rigidity and soft modes}

\begin{figure}
\centering
\includegraphics[angle=0,width=15cm]{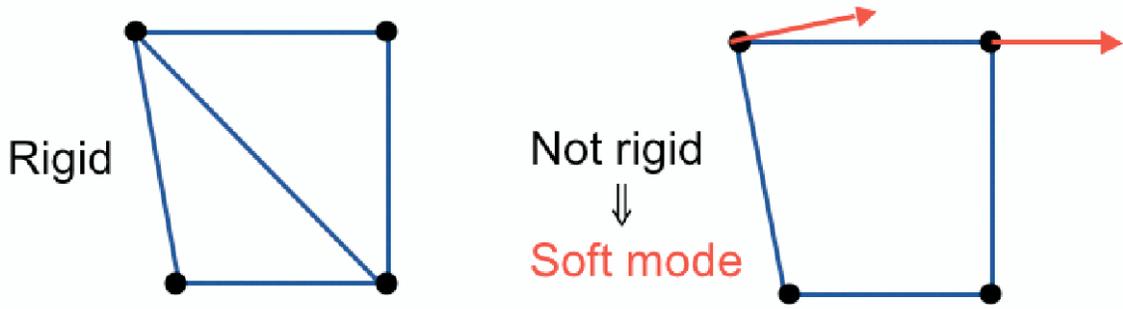}
\caption {Illustration of a rigid and a floppy network made of 4 particles interacting with springs. $N d-d(d+1)/2=5$ contacts are required to rigidify the structure. The red arrow of the floppy network indicates the soft mode that appears when one contact from the rigid system is removed. }
\label{ps}
\end{figure}

More than one century ago, Maxwell \cite{max}, working on the stability of engineering structures, studied the necessary conditions for the rigidity of an assembly of interacting objects. His response is as follows: consider for example   a network of $N$ point particles connected with $N_c$ relaxed springs of stiffness unity in a space of dimension $d$. The expansion of the energy $E$ is:
\ba
\label{2}
\delta E= \frac{1}{2} \sum_{\langle  ij \rangle}[(\delta\vec{R_j}-\delta\vec{R_i}).\vec n_{ij}]^2 
\ea
where the sum is taken on every couple of particles in contact ${\langle  ij \rangle}$, $\vec n_{ij}$ is the unit vector going from $i$ to $j$, and $ \delta\vec{R_i}$ is the displacement of particle $i$.  It is convenient to express Eq.(\ref{2}) in matrix form, by defining the set of displacements $\delta \vec R_1... \delta \vec R_N$ as a $dN$-component vector $|\delta {\bf R}\rangle$.  Then Eq. (\ref{2}) can be written in the form:
\be
\label{m}
\delta E = \langle\delta {\bf R}| {\cal M}|\delta {\bf R}\rangle
\ee
The corresponding matrix ${\cal M}$ is known as the dynamical matrix \cite{Ashcroft}, see footnote\footnote[3]{${\cal M}$ can be
written as an $ N$ by $N$ matrix  whose elements are themselves tensors of rank
$d$, the spatial dimension ${\cal M}_{ij}=-\frac{1}{2}\delta_{\langle ij
\rangle} \vec{n}_{ij}\otimes \vec{n}_{ij} +\frac{1}{2} \delta_{i,j}
\sum_{<l>}\vec{n}_{il} \otimes \vec{n}_{il}$ where $\delta_{\langle ij
\rangle}=1$ when i and j are in contact, the sum is taken on all the contacts $
l$ with $i$.} for an explicit tensorial notation.  The $3N$ eigenvectors of the dynamical matrix are the normal modes of the particle
system, and its eigenvalues are the squared angular frequencies of these modes. A system is rigid if it has no {\it soft mode}, which are the modes with zero energy. Since Eq.(\ref{2}) is a sum of positive terms, such modes satisfy:
\be
\label{3}
(\delta\vec{R_i}-\delta\vec{R_j}).\vec n_{ij}=0 \ \hbox{for all $N_c$ contacts}\ \langle ij \rangle
\ee
This linear equation defines the vector space of displacement fields that conserve the distances at first order between particles in contact. The particles can yield without restoring force if their displacements lie in this vector space. Fig(\ref{ps}) furnishes an example of such mode. Note that Eq.(\ref{3}) is purely geometrical and does not depend on the interaction potential.   

Maxwell noticed that Eq.(\ref{3}) has $N_c$ constraints and $N d-d(d+1)/2$ degrees of freedom if we substract  global translations and rotations. Each equation restricts the $dN-d(d+1)/2$-dimensional space of $|\delta{\bf R}\rangle$ by one dimension.  In general, these dimensions are independent, so that the number of independent soft modes is $dN -d(d+1)/2- N_c$. A rigid system must not have any soft modes, and therefore as at least as many constraints as it has degrees of freedom.  For a large system this yields for the average coordination $z\equiv 2N_c/N$:
\be 
\label{inequality}
z\geq 2d
\ee
This is the Maxwell criterion for rigidity. It is important to note that it is a {\it global} criterion, as it discusses the stability toward collective motions of particles. A local criterion that treat the motions of single particles only leads to $z>d+1$.  As we shall see, some systems live on the bound of Eq.(\ref{inequality}), they are called {\it isostatic}.

If this criterion has been known for quite a long time, it is only in the last decades that the concept of soft mode of non-rigid systems was used to study gels, covalent glasses and force propagation in granular matter. In the present Chapter we discuss these ideas.

\section{Soft modes and force propagation}
  
Recently several theories were proposed to describe the response to forces in sand. Some \cite{cates1,cates2}  have argued that granular matter requires a new constitutive law: they postulate a linear relation between the components of the stress tensor,  the ``null stress law''. This leads to a continuum mechanics different from elasticity,  where forces {\it propagate} along favored directions. In experiments, this description breaks down on large length scales \cite{behringer,reydellet} where an elastic-like behavior is recovered. Nevertheless, as we shall see now, this  theory was further justify  for frictionless grains \cite{Tom1,Tom2,Tom3} using the concept of soft modes.

There is an obvious connection between soft modes and forces: soft modes do not have restoring force. Therefore a non-rigid system can resist to an external force ${\bf |F\rangle }\equiv \{\vec F_i\}$, where $i$ labels the particles, only if  the external force is orthogonal to each soft mode $\beta$, that is:
\be
\label{nono}
\langle {\bf F|\delta R}^\beta \rangle \equiv \sum_i \vec F_i \cdot \vec{\delta R}_i^\beta=0
\ee
If Eq.(\ref{nono}) is not satisfied, the system yields along the soft modes, which are thus the directions of fragility of the system. 

To apply this idea to granular matter, the starting point  is the following remark:  an assembly of frictionless hard spheres (or equivalently elastic spheres at the jamming transition where the pressure vanishes) is exactly isostatic, as was shown in particular in \cite{Tom1,roux, moukarzel} and confirmed in the simulation of \cite{J}.   The argument for hard spheres is as follows: on the one hand, it is a rigid system and therefore must satisfy the bound of (\ref{inequality}). On the other hand, the distance between hard spheres in contact must be equal to the diameter $\sigma=1$ of the spheres:
\be
\label{ddd}
|| \vec{R}_i-\vec{R}_j||=1
\ee
Eq.(\ref{ddd})  brings exactly $N_c$ constraints on the positions of the centers of the particles. Once again, there are $N d-d(d+1)/2$ degrees of freedom for the particle positions.  Therefore one must have $Nd-d(d+1)/2\geq N_c$ which implies $z\leq 2d$ (note that this argument is contradicted in the case of the crystal whose coordination is  larger. In the crystal, the constraints on the particle positions are redundant. Adding an infinitesimal poly-dispersity destroys this effect). Finally these two bounds lead to $z=2d$.
 
The second remark is that an isostatic state is marginally rigid: if one contact is removed, one soft mode appears. This has the following consequence: an isostatic state is very sensitive to boundary conditions. Consider a subsystem of size $L$ in a large isostatic system. Let $N_{ext}\sim L^{d-1}$ be the number of contacts of this subsystem with external beads. The number of contacts inside the subsystem $N_{int}$ is on average  $\langle N_{int} \rangle = Nd-N_{ext}/2$, where the factor $\frac{1}{2}$ shows up because a contact is shared by two particles. This implies that if the $N_{ext}$ contacts were to be removed, the subsystem would not be rigid. On average it will have $N_{ext}/2-d(d+1)/2$ soft modes. Consider now the force field composed of the $N_{ext}$ contact forces applied by the external beads on the subsystem. It must be orthogonal to the $N_{ext}/2-d(d+1)/2$ soft modes. This implies that roughly  half of the external contact forces are free degrees of freedom. If the contact forces are imposed on  half of the boundary of the subsystem,  the contact forces of the other half of the boundary are determined. This is very different from an elastic body where one can impose any stress on the whole boundary and compute the response of the system. In an isostatic system forces {\it propagate} from one side of the system to the other.
In \cite{Tom1}, the authors discuss the nature of the soft modes of isostatic  anisotropic system. Using Eq.(\ref{nono}) they infer a null-stress law among the components of the stress tensor.  This leads to the hyperbolic equations  proposed to describe force propagation in granular matter \cite{cates1,cates2}.

\section{Covalent glasses}

Phillips \cite{phillips} used the soft modes, or ``floppy modes'',  to study the structure of covalent glasses. The counting of degrees of freedom is slightly different from a network of springs where only the stretching of the contacts matters. Covalent interactions also display multi-body forces: the energy of the system depends on the angles formed by the different covalent bonds of an atom. These extra-terms in the energy, the ``bond bending energy'', bring extra-constraints on the soft modes. How many total constraints are there per atom of valence $v$?  As we discussed with Eq.(\ref{3}), the stretching of the $v$ bonds leads to $v/2$ constrains on the soft modes (as there is one constraint per bond and each bond is shared by two atoms). Furthermore, the covalent bonds form $v(v-1)/2$ angles, each of them corresponds to a term in the energy expansion. Therefore the total number of constraints  is $v/2+v(v-1)/2=v^2/2$. 

Consider, following \cite{phillips}, chalcogenide alloys such as Ge$_x$Se$_{1-x}$, where the relative concentration $x$ of Ge can vary from 0 to 1. Ge has a valence of 4 whereas Se has a valence of 2.  When $x=0$ the glass as a polymeric structure. When $x$ increases the connectivity of the covalent network increases too, until it becomes rigid. This takes place when $x \cdot 4^2+ (1-x) \cdot 2^2=2d=6$, that is when $x_c\approx 0.16$. 

\begin{figure}
\centering
\includegraphics[angle=0,width=14cm]{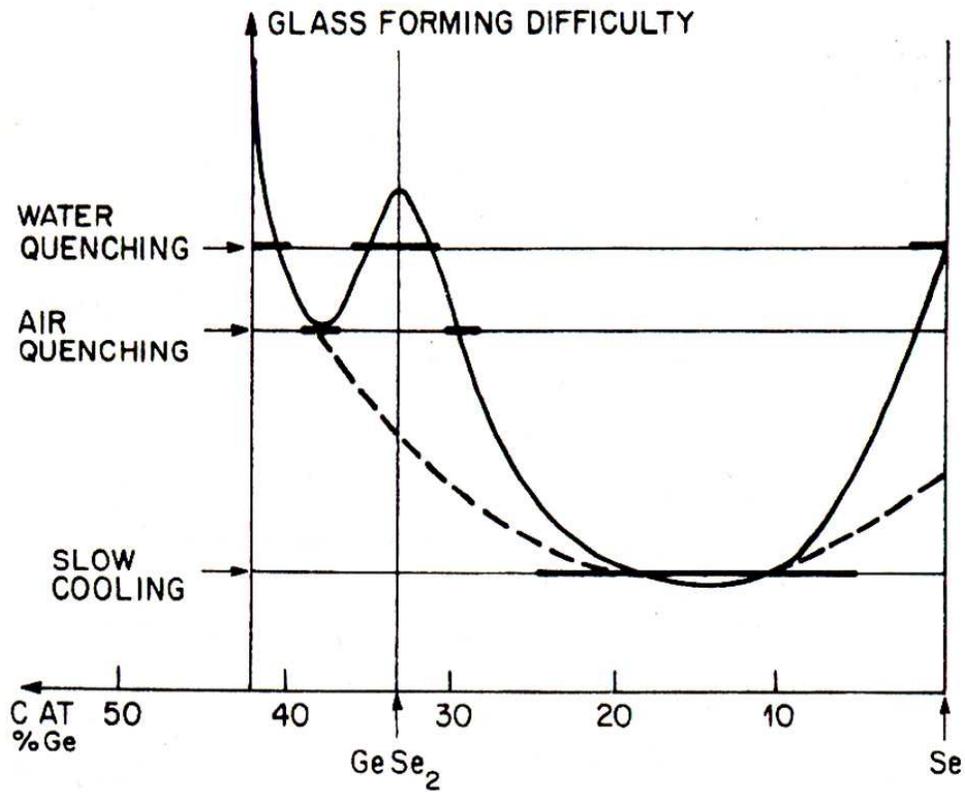}
\caption {The quenching rate (or difficulty of glass formation) was plotted as a function of $x$ in $Ge_x Se_{1-x}$ alloys. Both lines are sketched in to guide the reader's eye. The dashed line corresponds to the network effect discussed in the text. Phillips \cite{phillips}.}
\label{glassformer}
\end{figure}

However, although the covalent network is floppy for smaller $x$,  the glass is still rigid at lower concentration. In particular the shear modulus does not vanish  below $x_c$ \cite{cai}. The reason is that this counting argument neglects the weaker interactions induced by the lone pair electrons, that lead to short range repulsion and long range attraction (Van der Waals interaction).  Nevertheless several experimental results show that the transition at $x=0.16$ affects certain properties of the system. In Ge-As-Se glasses the fragility, that quantifies how the dependence of the viscosity with temperature is different from an Arrhenius law, is maximal at the transition \cite{tat}. Furthermore,  it was argued \cite{phillips} that  the composition of the best glass former should be $x_c$, as it appears experimentally, see Fig(\ref{glassformer}). A qualitative argument is as follows: when $x$ increases toward $x_c$, the covalent network becomes more and more intricate and the viscosity of the system increases. It takes therefore more and more time to nucleate the crystal. On the other hand, if $x$ increases above $x_c$, the covalent network is over-constrained,  the bonds are frustrated and have to store energy. The configuration of the crystal, where the bonds can organize to avoid this frustration, becomes more and more favorable with $x$ in comparison with the amorphous state, and is thus easier to nucleate.

This description is ``mean field'' as it does not consider the possible spatial fluctuations of the coordination. Such fluctuations may lead to a system where rigid, high-coordinated regions coexist with floppy, weakly-coordinated  regions. To study this possibility models were proposed such as rigidity percolation, see e.g. \cite{yeye} and reference therein. In its simplest form, this model considers springs randomly deposited on a lattice. When the concentration of springs increases, there is a transition when a rigid cluster percolates in the system almost fully floppy. Such a cluster is a non-trivial fractal object and contains over-constrained regions. In this form, this model is at infinite temperature, as there are no correlation  among the contacts deposited. It is interesting to note the difference with the transition of jamming that we study in what follows, which is also a transition of rigidity. At the jamming threshold the correlations among particles  are obviously important, as the temperature is not infinite. The system is exactly isostatic, as we shall discuss, and does not contain over-constrained regions.  It is a normal d-dimensional object with only very few holes of size of order unity \footnote[4]{The absence of large voids, or rattlers, has geometrical origins. Because particles are repulsive, the rigid region surrounding a void must be convex. A large void necessitates the creation of a vault. A flat vault is impossible as forces cannot be balanced on any particle of the surface. A weakly curved vault imposes drastic constraints on the distribution of angles of contact between these particles. This suggests that the probability of making a large void decays very fast, presumably at least exponentially in the surface of the void. If friction is present such voids might form more easily.}, the ``rattlers'', which are isolated particles without contact \cite{J}.  

\section{The rigidity of ``soft solids''}

Alexander noticed \cite{shlomon} the following contradiction:  there are solids less connected than required  by the Maxwell criterion,  and that are nevertheless rigid. For example, one may describe a gel as an assembly of reticulated point linked by springs that are the polymers. In general the coordination of such system is less than 6. Why then a gel has a finite shear modulus? The starting point of Alexander's answer is the simple following remark: the stress influences the frequency of the vibrations. At a macroscopic level the presence of a negative stress (that is, when the system is stretched) is responsible for the fast vibrations of strings or drumheads. When a system is compressed, the stress is positive and lowers the frequency of the modes that can even become unstable, as for the buckling of a thin rod.  To discuss the role of stress at a microscopic level, consider particles interacting with a potential $V(r)$. The expansion of the energy leads to:

\be
\label{0}
\delta E=  \sum_{ij} V'(r_{ij}^{eq}) dr_{ij}+ \frac{1}{2} V''(r_{ij}^{eq}) dr_{ij}^2+ O(dr_{ij}^3)
\ee
where the sum is over all pairs of particles, $r_{ij}^{eq}$ is the equilibrium distance between particles $i$ and $j$. In order to get an expansion in the displacement field $\delta \vec{R_i}$  we use:
\be
\label{00}
dr_{ij}= (\delta\vec{R_j}-\delta\vec{R_i}).\vec n_{ij}+ \frac{[(\delta\vec{R_j}-\delta\vec{R_i})^{\bot}]^2}{2 r_{ij}^{eq}}+
O(\delta\vec{R}^3)
\ee
where $ (\delta\vec{R_j}-\delta\vec{R_i})^{\bot}$ indicates the projection of $\delta\vec{R_j}-\delta\vec{R_i}$ on the plane orthogonal to $\vec n_{ij}$.  When used in Eq.(\ref{0}), the linear term in the displacement field disappears (the system is at equilibrium) and we obtain:
\be
\label{000}
\delta E= \{\sum_{ij} V'(r_{ij}^{eq}) \frac{[(\delta\vec{R_j}-\delta\vec{R_i})^{\bot}]^2}{2 r_{ij}^{eq}} \}  \\
+ \frac{1}{2} V''(r_{ij}^{eq}) [(\delta\vec{R_j}-\delta\vec{R_i}).\vec n_{ij}]^2 + O(\delta\vec{R}^3)
\ee
The difference from Eq.(\ref{2}) is the term inside curly brackets. Such term is called the ``initial stress'' term in \cite{shlomon} as it is directly proportional to the forces $V'(r_{ij}^{eq})$. We shall also refer to it as the applied stress term.  If the system has a negative pressure $p$  this term increases the frequency of the modes. As a consequence if $p<0$ all the soft modes of a weakly-connected system gain a finite positive energy: the system is rigid. This occurs in  gels: the osmotic pressure of the solvent is larger than the external pressure. This imposes that the network of reticulated polymers carry a negative pressure to compensate this difference: the polymers are stretched. This rigidifies the system. As a consequence, the shear modulus of gels is directly related to the osmotic pressure of the solvent. 

\chapter{Vibrations of isostatic systems}
\label{c3}

\section{Isostaticity}

When a system of repulsive spheres jams at zero temperature, the system is isostatic \cite{Tom1,moukarzel,roux} (when the rattlers, or particles without contacts, are removed). As we said above, on the one hand it must be rigid and satisfy the bound of (\ref{inequality}). One the other hand, it cannot be more connected: this would imply that the contacts are frustrated as they cannot satisfy Eq.(\ref{ddd}). That is, there would not be enough displacements degrees of freedom to allow particles in contact  to touch each other without interpenetrating, as we discussed for hard spheres in the last Chapter.  Thus the energy of the system, and the pressure, would not vanish at the transition. It must be the case, since an infinitesimal pressure can jam an athermal gas of elastic particles.   

In terms of energy expansion, since the pressure vanishes at the transition, the initial stress in bracket in Eq.(\ref{000}) vanishes. For concreteness in the following Chapters we consider the harmonic potential, corresponding to $\alpha=2$ in Eq.(\ref{opo}). The expansion of the energy is then given by Eq.(\ref{2}). In Chapter \ref{c7} we generalize our findings to other soft sphere potentials and other types of interactions.

An isostatic system is marginally stable: if $q$ contacts are cut, a space of soft modes  of dimension $q$ appears. For our  argument below we need to discuss the extended character of these modes. In general when only one contact $\langle ij \rangle$ is cut in an isostatic system, the corresponding soft mode is not localized near  $\langle ij \rangle$. This comes from the non-locality of the isostatic condition that gives rise to the soft modes; and was confirmed in the isostatic simulations of Ref \cite{Tom1}, which observed that the amplitudes of the soft modes spread out over a nonzero fraction of the particles. This shall be proved by the calculation  of Chapter \ref{c7} that shows that in an isostatic system, the response to a local strain does not decay with the distance from the source. When many contacts are severed, the extended character of the soft modes that appear depends on the geometry of the region being cut. If this region is compact many of the soft modes are localized.  For example cutting all the contacts inside a sphere totally disconnects each inner particle. Most of the soft modes are then the individual translations of these particles and are not extended throughout the system. 

\begin{figure}
\centering
\includegraphics[width=10cm]{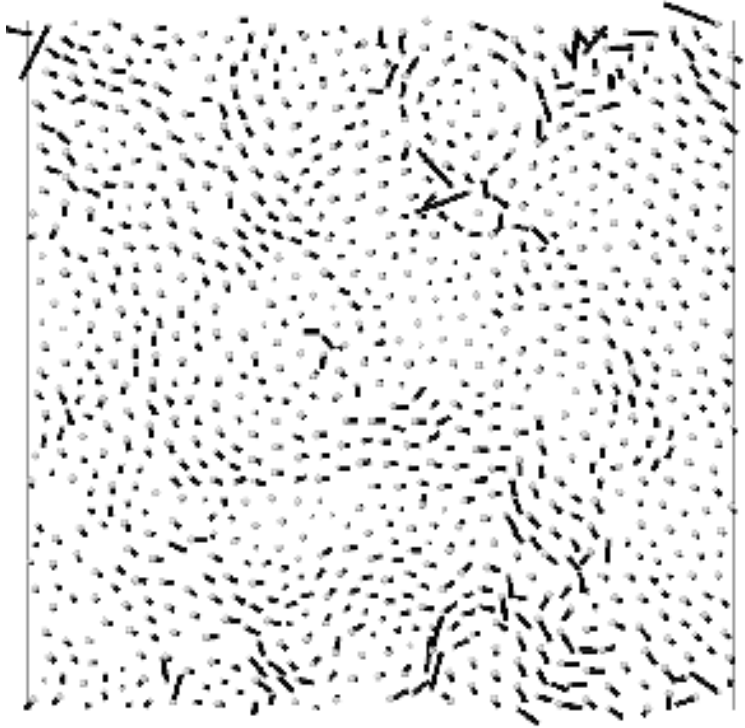}
\caption{One soft mode in two dimensions for $N\approx 1000$ particles.  The relative displacement of the soft mode is represented by a line segment extending from the dot. The mode was created from a previously prepared isostatic configuration, periodic in both directions, following \cite{J}.  20 contacts along the vertical edges were then removed and the soften modes determined.  The mode pictured here  is an arbitrary linear combination of these
modes.}  
\label{softfig}
\end{figure} 

In what follows we will be particularly interested in the case where the region of the cut is a hyper-plane as illustrated in Fig.(\ref{figs}). In this situation occasionally  particles in the vicinity of the hyper-plane can be left with less than $d$ contacts, so that trivial localized soft modes can also appear. However this represents only a finite fraction of the soft modes.  We expect that there is a non-vanishing fraction $q'$ of the total soft modes that are not localized near the hyper-plane. Rather,  as when a single contact is cut, these modes should extend over the whole system, like the mode shown in Fig.(\ref{softfig}).  We shall define extended modes more precisely in the next section. 

\section {Variational procedure}

We aim to show first that the density of states of an isostatic system does not vanish at zero frequency. Since $D(\omega)$ is the total number of modes per unit volume per unit frequency range, we have to show that there are at least of the order of $\omega L^d$ normal modes with frequencies smaller than $\omega$ for any small $\omega$. As we justify later, if proven in a system of size $L$ for $\omega\sim \omega_L\sim 1/L$, this property can be extended to a larger range of $\omega$ independent of $L$. Therefore it is sufficient to show that there are of the order of $L^{d-1}$ normal modes with frequency of the order of $1/L$, instead of the order of one such mode in a continuous solid: the whole translation of the system. To do so we use a variational argument: ${\cal M}$ is a positive symmetric matrix. Therefore if a normalized mode has an energy $\delta E$, we know that the lowest eigenmode has a frequency $\omega_0\equiv \sqrt E_0\leq \sqrt{\delta E}$. Such argument can be extended to a set of modes \footnote[5]{If $m_\alpha$ is the $\alpha$'th lowest eigenvalue of ${\cal M}$ and if $e_\alpha$ is an orthonormal basis such that $\langle e_\alpha|{\cal M}| e_\alpha\rangle \equiv n_\alpha$ then the variational bound of A. Horn [Am. J. Math {\bf 76} 620 (1954)] shows that $\sum_1^q m_\alpha \leq \sum_1^q n_\alpha$.  Since $q n_q \geq \sum_1^q n_\alpha$, and since $\sum_1^q m_\alpha \geq \sum_{q/2}^q m_\alpha \geq (q/2) m_{q/2}$, we have $q n_q \geq (q/2) m_{q/2}$ as claimed.}: if there are $m$ {\it orthonormal} trial modes with energy $\delta E \leq \omega_t^2$, then there are at least $m/2$ {\it normal} modes with frequency smaller than $\sqrt 2\omega_t$. Therefore we are led to find of the order of $L^{d-1}$ trial orthonormal modes with energy of order $ 1/L^2$.

\section{Trial modes}

\begin{figure}
\centering
\includegraphics[width=9cm]{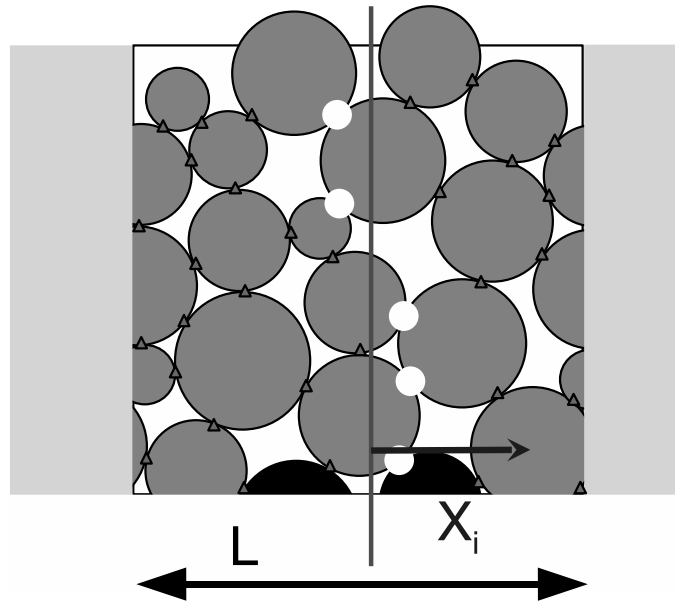}
\caption{Illustration of the boundary contact removal process described in the text.  Eighteen particles are confined in a square box of side $L$ periodically continued horizontally and vertically.  An isostatic packing requires 33 contacts in this two-dimensional system.  An arbitrarily drawn vertical line divides the system.  A contact is removed wherever the line separates the contact from the center of a particle.   Twenty-eight small triangles mark the intact contacts; removed contacts are shown by the five white circles. 
\vspace{2cm}
}
\label{figs}
\end{figure}

For concreteness we consider the three-dimensional cubic $N$-particle system $\cal S$ of Ref \cite{J} with periodic boundary conditions at the jamming threshold.  We label the axes of the cube by x, y, z.  $\cal S$ is isostatic, so that the removal of $n$  contacts allows exactly $n$ displacement modes with no restoring force. Consider for example the system $\cal S'$ built from $\cal S$ by removing the $q\sim L^2$ contacts crossing an arbitrary plane orthogonal to (ox); by convention at $x=0$, see Fig.(\ref{figs}).  $\cal S'$, which has a free boundary condition instead of periodic ones along (ox), contains a space of soft modes of dimension $q$ \footnote[6] {The balance of force can be satisfied in ${\cal S'}$ by imposing external forces on the free boundary. This adds a linear term in the energy expansion that does not affect the normal modes.}, instead of one such mode ---the translation of the whole system--- in a normal solid. As stated above, we suppose that a subspace of dimension  $q'\sim L^2$ of these soft modes contains only extended modes.  We define the {\it extension} of a mode relative to the cut hyper-plane in terms of the amplitudes of the mode at distance $x$ from this hyper-plane. Specifically the extension $e$ of a normalized mode $|\bf \delta R \rangle$ is defined by $\sum_i\sin^2(\frac{x_i \pi}{L}) \langle i|{\bf \delta R} \rangle^2 =e$, where the notation $ \langle i|{\bf \delta R} \rangle$ indicates the displacement of the particle $i$ of the mode considered.  For example, a uniform mode with $ \langle i|{\bf \delta R} \rangle$  constant for all sites has $e=\frac{1}{2}$ independent of $L$. On the other hand, if $ \langle i|{\bf \delta R} \rangle=0$ except for a site $i$ adjacent  to the cut hyper-plane, the $x_i/L\sim L^{-1}$ and $e\sim L^{-2}$. We define the subspace of extended modes by setting a fixed threshold of extension $e_0$ of order 1 and thus including only soft modes $\beta$ for which $e_\beta>e_0$. As we argued at the beginning of the Chapter, we expect that a fraction of the total soft modes are extended. Thus if $q'$ is the dimension of the extended modes vector space, we shall suppose  that $q'/q$ remains finite as $L\rightarrow \infty$; i.e. a fixed fraction of the soft modes remain extended as the system becomes large. The appendix at the end of this Chapter presents our numerical evidence for this behavior.

We use them  to build $q'$ orthonormal trial modes of frequency of the order $1/L$ in the initial system $\cal{S}$. Let us denote $|\bf \delta R_\beta \rangle$ a normalized basis of the vector space of such extended modes, $1\leq \beta \leq q'$.  These modes are not soft in the jammed system $\cal S$ because they deform the previous $q$ contacts located at $x=0$, and therefore cost energy. Nevertheless a set of trial modes, $|\bf \delta R_\beta^* \rangle$, can still be formed by altering the soft modes so that they do not have an appreciable amplitude at the boundary where the contacts were severed.  We seek to alter the soft mode to minimize the distortion at the severed contacts while minimizing the distortion elsewhere. Accordingly, for each soft mode $\beta$ we define the corresponding trial-mode displacement $\langle i|{\bf \delta R}^* \rangle$ to be: 

\be\label{tria}
\langle i|{\bf \delta R}_\beta^* \rangle =C_\beta \sin(\frac{x_i \pi}{L}) \langle i|\bf \delta R_\beta \rangle
\ee
where the normalization constant $C_\beta $  depends on the spatial distribution of the mode $\beta$. If for example $ \langle i|{\bf \delta R} \rangle=0$ except for a site $i$ adjacent  to the cut plane, $C_\beta$ grows without bound as $L\rightarrow \infty$.  In the case of extended modes $C_\beta^{-2}\equiv \sum_{\langle ij \rangle} \sin^2(\frac{x_i \pi}{L})   \langle j|{\bf \delta R}_\beta \rangle^2= e_\beta > e_0$, and  therefore $C_\beta$ is bounded above by  $e_0^{-\frac{1}{2}}$.  The sine factor suppresses the problematic gaps and overlaps at the $q$ contacts near $x=0$ and $x=L$. Formally, the modulation by a sine is a linear mapping. This mapping is invertible if it is restricted to the extended soft modes. Consequently the basis $|\bf \delta R_\beta \rangle$ can always be chosen such that the  $|\bf \delta R_\beta^* \rangle$ are orthogonal.  Furthermore one readily verifies that the $|\bf \delta R_\beta^* \rangle$'s energies are small, because the sine modulation  generates an energy of order $1/L^2$ as expected. Indeed we have from Eq.(\ref{2}):

\be
\delta E =  C_\beta^2 \sum_{\langle ij \rangle} [(\sin(\frac{x_i \pi}{L}) \langle i|{\bf \delta R}_\beta \rangle- \sin(\frac{x_j \pi}{L}) \langle j|{\bf \delta R}_\beta \rangle) \cdot \vec n_{ij}]^2
\ee
Using  Eq.(\ref{3}), and expanding the sine, one obtains:

\ba
\label{kk}
\delta E \approx  C_\beta^2 \sum_{\langle ij \rangle} \cos^2(\frac{x_i \pi}{L}) \frac{\pi^2}{L^2} (\vec n_{ij} \cdot \vec e_x)^2 (\langle j|{\bf \delta R}_\beta \rangle \cdot \vec n_{ij})^2 \\
\leq  e_0^{-1}  (\pi/L)^2 \sum_{\langle ij \rangle}\langle j|{\bf \delta R}_\beta \rangle^2
\ea
where $\vec e_x$ is the unit vector along (ox). The sum on the contacts can be written as a sum on all the particles since only one index is present in each term. Using the normalization of the mode $\beta$ and the fact that the coordination number of a sphere is bounded by a constant $z_{max}$ ($z_{max}=12$ for 3 dimensional spheres \footnote[7]{In a polydisperse system $z_{max}$ could a priori be larger. Nevertheless Eq.(\ref{kk}) is a sum on every contact where the displacement of only one of the two particles appears in each term of the sum. The corresponding particle can be chosen arbitrarily. Chose the smallest particle of each contact. Thus when this sum on every contact is written as a sum on every particles to obtain Eq.(\ref{kkk}), the constant $z_{max}$ still corresponds to the monodisperse case, as a particle cannot have more contacts with particles larger than itself. }), one obtains:

\be
\label{kkk}
\delta E\leq   e_0^{-1}  (\pi/L)^2 z_{max} \equiv \omega_L^2
\ee
One may ask if the present variational argument can be improved, for example by considering geometries of broken contacts different from the surface we considered up to now.  When contacts are cut to create a vector space of  extended soft modes, the soft modes must be modulated with a function that vanishes where the contacts are broken in order to obtain trial modes of low energy. One the one hand, cutting many contacts increases the number of trial modes. On the other hand, if too many contacts are broken, the modulating function must have many ``nodes'' where it vanishes. Consequently this function displays larger gradients and the energies of the trial modes increase. Cutting a surface (or many surfaces, as we shall discuss below) appears to be the best compromise between these two opposing effects. Thus our argument gives a natural limit to the number of  low-frequency states to be expected.

Finally we have found of the order of $L^2$ trial orthonormal modes of frequency bounded by $\omega_L\sim 1/L$, and we can apply the variational argument mentioned above: the density of statesis bounded below by a constant below frequencies of the order $\omega_L$.  In what follows, the trial modes introduced in Eq.(\ref{tria}), which are the soft modes modulated by a sine, shall be called ``anomalous modes''.

\section{Argument extension to a wider frequency range}
\label{pppp}
We may extend this argument to show that the bound on the initial  density of states extends to a plateau encompassing a nonzero fraction of the modes in the system.  If the cubic simulation box were now divided  into $m^3$ sub-cubes of size $L/m$, each sub-cube must have a density of states equal to the same $ D(\omega)$ as was derived above, but extending to frequencies of order $m\omega_L$.  These subsystem modes must be present in the full system as well, therefore the  bound on $D(\omega)$ extends to $[0,m \omega_L]$.  We thus  prove that the same bound on the average density holds down to sizes of the order of a few particles, corresponding to frequencies independent of $L$.  Finally $D(\omega)$ does not vanish when $\omega\rightarrow 0$, as indicates the presence of the observed plateau in the density of states. We note that in $d$ dimensions this argument may be repeated to yield a total number of modes, $L^{d-1}$, below a frequency $\omega_L \approx 1/L$, thus yielding a limiting nonzero density of states in any dimension.

We note that the trial modes of energy $\delta E\sim l^{-1}$ that we introduced by cutting in subsystems of size $l$ are, by construction, localized to distance scale $l$. Nevertheless we expect these trial modes to hybridize with the trial modes of other subsystems, and the corresponding  normal modes not to be localized on such length scale.

\section{Appendix: Spatial  distribution of the soft modes}
\label{geo}

In our argument we have assumed that  when $q\sim L^{d-1}$ contacts are cut along a hyper-plane in an isostatic system, there is  a vector space of dimension $q'=aq$ which contains only extended modes, such that $a$ does not vanish when $L\rightarrow \infty$.  A normalized mode $ |{\bf \delta R}\rangle$ was said to be extended if $\sum_i\sin^2(\frac{x_i \pi}{L}) \langle i|{\bf \delta R }\rangle^2 > e_{0}$, where $e_{0}$ is a constant, and does not depend on $L$. Here we show how to choose  $e_0>0$ so that there is a non-vanishing fraction of extended soft modes. We build the vector space of extended soft modes and  furnish a bound on its dimension.  

Let us consider the linear mapping $\cal  G$ which assigns  to a displacement field $|{\bf \delta R} \rangle$ the displacement field $ \langle i| {\cal G} {\bf\delta R }\rangle= \sin^2(x_i\pi/L) \delta \vec {R }_i$.  For any soft mode $|{\bf \delta R}_\beta \rangle$ one can consider the positive number $a_\beta\equiv {\langle \bf \delta R_\beta}|{\cal G}|{\bf\delta R}_\beta \rangle \equiv \sum_i   \sin^2(x_i\pi/L) \delta \vec {R _\beta}_i^2$. We build the vector space of extended modes by recursion:  at each step we compute the $a_\beta$ for the normalized soft modes, and we eliminate the soft mode with the minimum $a_\beta$. We then  repeat this procedure in the vector space orthogonal to the soft modes eliminated. We stop the procedure when $a_\beta>e_{0}$ for all the soft modes $\beta$ left.  Then all the modes left are extended according to our definition. We just have to show that one can choose $e_{0}>0$ such that when this procedure stops, there are $q'$ modes left, with $q'>aq$ and $a>0$.  In order to show that, we introduce the following overlap function:

\be
\label{lei}
f(x) dx \equiv  q^{-1} \sum_{\beta=1,...,q}\sum_{x_i\in [x, x+ dx]} [\delta \vec{R}_{i,\beta}]^2
\ee

The sum is taken on an orthonormal basis of soft modes $\beta$ and on all the particles whose position has a coordinate $x_i\in [x, x+ dx]$. $f(x)$ is the trace of a projection, and is therefore independent of the orthonormal basis considered. $f(x)$  describes the spatial distribution of the amplitude of the soft modes. The $\delta {\bf R}_\beta$ are normalized and therefore:

\be
\int_0^L f(x)dx=1
\ee

\begin{figure}

\centering
\includegraphics[width=7cm]{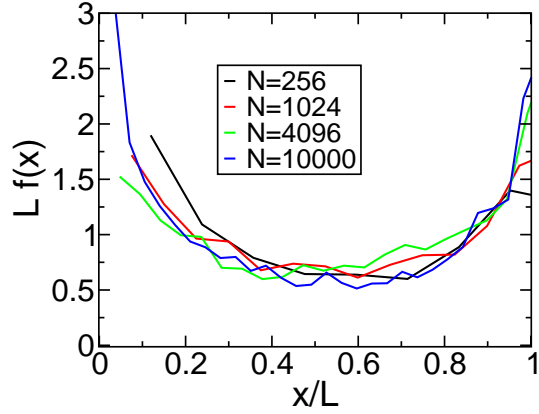}
\caption{The overlap function $f(x)$ as defined in Eq.(\ref{lei}) for different system sizes in three dimensions. Soft modes were created from isostatic configurations as described below Fig. \ref{softfig}.}
\label{f}
\end{figure} 

We have examined soft modes made from  configurations at the jamming transition found numerically in \cite{J}.  The overlap function $f(x)$ was then computed for different system sizes $L$. These are shown in Fig. \ref{f}. It appears from Fig.(\ref{f}) that i) when $f(x)$  is rescaled  with the system size it collapses to a unique curve, and ii) this curve is bounded from below by a constant $c$ ($c \approx 0.6$).  Consequently one can bound the trace of $\cal G$: $tr  {\cal G} =\int_0^L q f(x) \sin^2(x)> q c/2$. On the other hand one has $tr {\cal G} =\sum_{\beta = 1}^q a_\beta$, where the sum is made on the orthonormal basis we just built. Introducing the numbers $a$ and $e_0$  such that there are $q'=aq$ extended modes, and using that the $a_\beta<1$, one can bound this sum and obtain $tr {\cal G}< aq+ (1-a)q e_{0}$.  Choosing for example $e_{0}=c/2>0$,  one finds $a>c/(2-c)$, which is a constant independent of $L$ as claimed.

\chapter{Evolution of the modes with the coordination}
\label{c4}
\section{ An ``isostatic'' length scale}
\label{isos}
\begin{figure}
\centering
\includegraphics[angle=0,width=7cm]{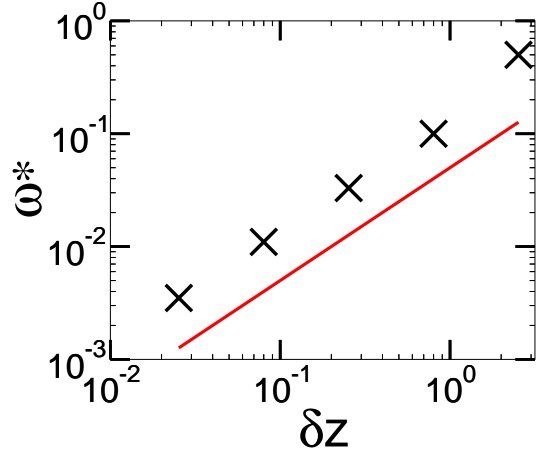}
\caption{ $\omega^*$ as defined in the insert of Fig.1 {\it vs.}  coordination, in the system with relaxed springs. The line as a slope one.}
\label{sans}
\end{figure}

When the system is compressed and moves away from the jamming transition, the simulations showed that the extra-coordination number  $\delta z \equiv z-z_c$ increases. In the simulation, the compression also creates forces on all contacts. In this Chapter we  ignore these forces, and instead only consider the contact network created by compression, but in the absence of applied pressure. Any tension or compression in the contacts is removed. The effect on the energy is to remove the first bracketed term from Eq.(\ref{000}) above, and the expansion of the energy is still given by Eq.(\ref{2}). We note that removing these forces, which add to zero on each particle, does not disturb the equilibrium of the particles or create displacements. In this section we ignore the question of how $\delta z$ depends on the degree of compression. We return to this question in the next section.  Compression causes $\Delta N_c=N \delta z/2 \sim L^3 \delta z$ extra constraints to appear in Eq.(\ref{2}).  Cutting the boundaries of the system, as we did above, relaxes $q\sim L^2$ constraints. For a large system, $q<\Delta N_c$ and Eq.(\ref{2}) is still over-constrained so that no soft modes appear in the system.  However, as the systems become smaller, the excess  $\Delta N_c$ diminishes, and for $L$ smaller than some $l^*\sim \delta z^{-1}$ where $\Delta N_c= q$, the system is again under-constrained, as was already noticed in \cite{Tom1}.  This allows us to build low-frequency modes in subsystems smaller than $l^*$.  These modes appear above a cut-off frequency $\omega^*\sim l^*{}^{-1}$; they are the excess-modes that contribute to the  plateau in $D(\omega)$ above $\omega^*$.   In other words, anomalous modes with characteristic length smaller than $l^*$ are  little affected by the extra contacts, and the density of states is unperturbed above a frequency $\omega^* \sim \delta z$. This scaling is checked numerically in  Fig.\ref{sans}. It is in very good agreement with our prediction up to $\delta z\approx 2$.  

\begin{figure}
\centering
\includegraphics[angle=270,width=7cm]{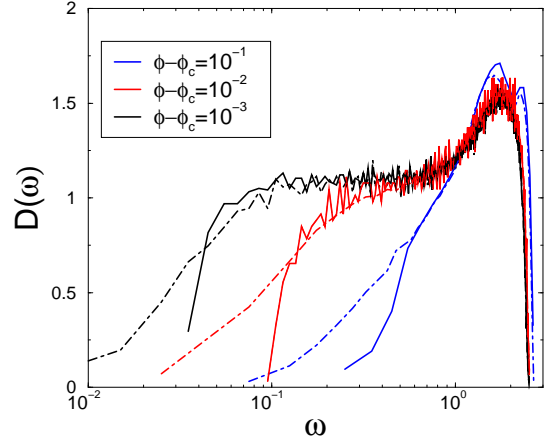}
\caption{Log-linear plot of the density of states for N=1024 for three values of $\phi-\phi_c$ in  the soft spheres system (dotted line) and the system where the applied stress term has been removed (solid line). }
\label{compare}
\end{figure}

At frequency lower than $\omega^*$ we expect the system to behave as a disordered, but not ill-connected, elastic medium, so that  the vibrational modes are  similar to the plane waves of a continuous elastic body. We refer to these modes as ``acoustic modes''. Thus we expect $D(\omega)$ at small $\omega$ to vary as $\omega^2c^{-3}$, where $c(\delta z)$ is the sound speed at the given compression. This $c$ may be inferred from the bulk and shear moduli measured in the simulations; that we shall derive in chapter \ref{c7}. One finds for the transverse velocity $c_t \sim (\delta z)^{\frac{1}{2}}$, and for the longitudinal velocity $c_l\sim \delta z^0$. Thus at low frequency the $D(\omega)$ is dominated by the transverse plane waves and at $\omega=\omega^*$ the acoustic density of states is $\omega^2c_t^3\sim\delta z^2\delta z^{-3/2}\sim \delta z^{\frac{1}{2}}$: the acoustic density of states should be dramatically smaller than the plateau density of states. There is no smooth connection between the two regimes, thus we expect a sharp drop-off in $D(\omega)$ for $\omega<\omega^*$.  Such drop-off is indeed observed, as seen in Fig(\ref{compare}). In fact, because of the finite size of the simulation,  no acoustic modes are apparent at $\omega<\omega^*$ near the transition. 

Thus the behavior of such system near the jamming threshold depends on the frequency $\omega$ at which it is considered. For $\omega>\omega^*$ the system behaves as an isostatic state: the density of states is dominated by anomalous modes. For $\omega<\omega^*$ we expect it to  behave  as a continuous elastic medium with acoustic modes. Since the transverse  and the longitudinal velocities do not scale in the same way,  the wavelengths of the longitudinal and transverse plane waves at $\omega^*$ are two distinct length scales $l_l$ and $l_t$ which follow $l_l\sim c_l \omega^*{}^{-1}$ and $l_t\sim c_t \omega^*{}^{-1}$.  At shorter wave lengths we expect the acoustic modes to be strongly perturbed. Note that since $c_l\sim \delta z^0$, one has  $l_l\sim l^*$. Interestingly, $l_t\sim \delta z^{-\frac{1}{2}}$ is  the smallest system size at which plane waves can be observed:  for smaller systems, the lowest frequency mode is not a plane wave, but an anomalous mode. $l_t$ was observed numerically in \cite{nagel}.

\section{Role of spatial fluctuations of $z$}
Our argument ignores the spatial fluctuations of $\delta z$.  If these fluctuations were spatially uncorrelated they would be Gaussian upon coarse-graining:  then the extra number of contacts $\Delta N_c$ in a subregion of size $L$ would have fluctuations of order  $L^{d/2}$. The scaling of the contact number that appears in our description is $\Delta N_c \sim L^{d-1}$ and is therefore larger than these Gaussian fluctuations for $d>2$. In other terms at the length scale  $l^*$ where soft modes appear, the fluctuations of the number of contacts inside the bulk are negligible in comparison with the number of contacts at the surface. Therefore the extended soft modes that are described here are not sensitive to fluctuations of coordination  in three dimensions near the transition. In \cite{me2} we argued that in two dimensions there are spatial anti-correlations in $z$, and that the fluctuations do not affect the extended soft modes  in two dimensions either. 

Note that these arguments do not preclude the existence of low-frequency localized modes that may appear in regions of small size $l\ll  l^*$, and that could be induced by very weak local coordination, or specific configuration.  The presence of such modes would increase the density of states at low-frequency. There is no evidence for their presence in the simulations of \cite{J}.

\section{Application to tetrahedral networks}

\begin{figure}
\centering
\includegraphics[angle=270,width=7cm]{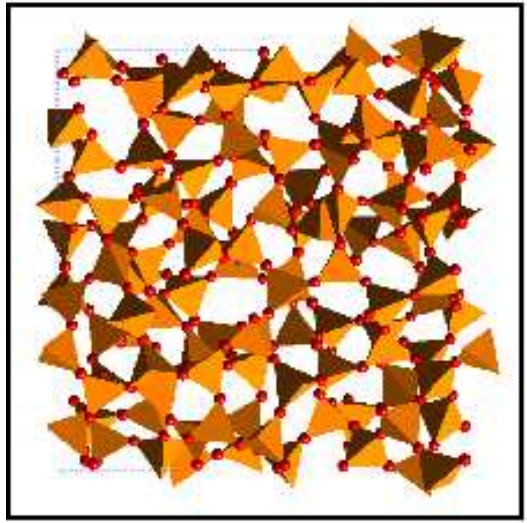}
\caption{Rigid unit modes model applied to silica. Trachenko et al.\cite{kostya}.}
\label{sd}
\end{figure} 

\begin{figure}
\centering
\includegraphics[angle=0,width=7cm]{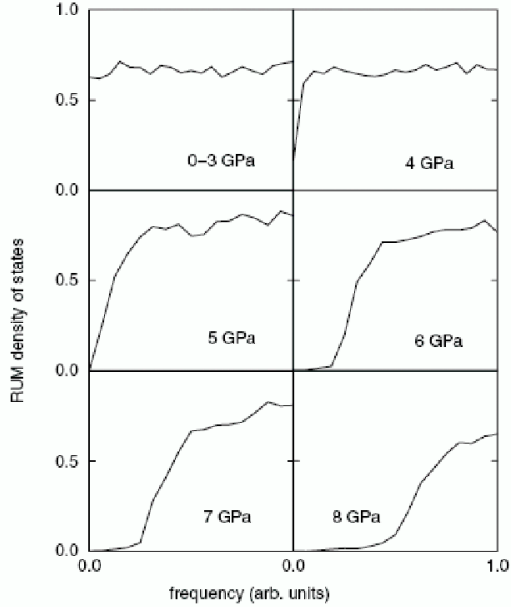}
\caption{Density of  rigid unit modes for silicate at different pressure. Trachenko et al.\cite{dove}. }
\label{dd}
\end{figure}

The nature of slow vibrations -and the possible presence of two-level systems- has been much studied in silica, one of the main glass-forming system.    In this glass (or more generally aluminosilicates) the forces within the tetrahedra $SiO_4$  are much stronger than the forces that act between them \cite{hammonds}: it is easier to rotate two linked tetrahedra than to distort one tetrahedron \footnote[8]{For example the bending energy of Si-o-Si is roughly 10 times smaller than the stretching of the contact Si-o \cite{ff}.}. This suggests to model such glass as an assembly of linked tetrahedra loosely connected at corners: this is the ``rigid unit modes'' model \cite{heine}.  In this model the tetrahedra are characterized by a unique parameter, a stiffness $k$ \footnote[9]{In fact the rigidity of a tetrahedron induced by the covalent bonds should be characterized by 3 parameters corresponding to different deformations of the tetrahedron. If these parameters are of similar magnitude, as one expects for example for silica,  this does not change qualitatively the results discussed here.}. Recently this model was used to study the vibrations of silica  \cite{dove}.  The authors first generate realistic configurations of $SiO_2$ at different pressures using molecular dynamics simulations. At low pressure, they obtain a perfect tetrahedral network. When the pressure becomes large, the coordination of the system increases with the formation of 5-fold defects. Once these microscopic configurations are obtained, the rigid unit model is used and the system is modeled as an assembly of rigid tetrahedra, see Fig.(\ref{sd}). Then, the density of states of such network is computed. The results are shown in Fig.(\ref{dd}). One can note the obvious similarity with the density of states near jamming of Fig.(\ref{f1}). We argue that the cause is identical, and that the excess-modes correspond to the anomalous modes made from the soft modes, rather than to one-dimensional modes as proposed in \cite{dove}. Indeed, a tetrahedral network is isostatic, see e.g. \cite{kostya}. The counting of degrees of freedom can be made as follows: on the one hand each tetrahedron has 6 degrees of freedom (3 rotations and 3 translations). On the other hand, the 4 corners of a tetrahedron bring each 3 constraints shared by 2 tetrahedra, leading to 6 constraints per tetrahedron. Thus the system is isostatic. When the pressure increases the coordination increases too, leading to the erosion of the plateau in the density of states discussed earlier in this Chapter.  Finally, these predictions of the density of states  fail to describe silica glass vibrations at low frequencies, where one cannot neglect the weaker interactions anymore  nor the role of the initial stress that we discuss in the next Chapter.  In Chapter \ref{c8} we evaluate the effect of the weaker interactions. This allows us to propose an explanation for the nature of the boson peak in such glasses.

\chapter{Effect of the initial stress on vibrations}
\label{c5}

In this section we describe how the above simple description of $D(\omega)$ is affected by the presence of applied stress. In general when a system of particles at equilibrium is formed, there are forces between interacting particles. For harmonic soft spheres it leads to a non-vanishing first term in Eq.(\ref{000})  that becomes:
\be
\delta E=\{\frac{1}{4}\sum_{\langle  ij \rangle} (r_{ij}^{eq}-1) [(\delta\vec{R_j}-\delta\vec{R_i})^{\bot}]^2\}+\frac{1}{2} \sum_{\langle  ij \rangle}[(\delta\vec{R_j}-\delta\vec{R_i}).\vec n_{ij}]^2 
\ee
where we used $r_{ij}\approx 1$. This term in bracket is (a) negative for repulsive particles (b) proportional to the transverse relative displacement between particle in contact (c) scales  as the pressure $p$, and  is therefore vanishing at the jamming transition.  The full dynamical matrix ${\cal D}$  can be written:

\be
{\cal D=M+M'}
\ee
where ${\cal M'}$ is written in tensorial notation in footnote \footnote[10]{in three dimension we have ${\cal M'}_{ij}=-\frac{1-r_{ij}}{2r_{ij}} [ \delta_{\langle ij \rangle} (\vec{m}_{ij}\otimes \vec{m}_{ij}+ \vec{k}_{ij}\otimes \vec{k}_{ij}) + \delta_{i,j} \sum_{<l>}(\vec{m}_{ij}\otimes \vec{m}_{ij}+ \vec{k}_{ij}\otimes \vec{k}_{ij})$, where ($\vec{n}_{ij}$,$\vec{m}_{ij}$,$\vec{k}_{ij}$) is an orthonormal basis.}. The spectrum of $\cal D$ has {\it a priori} no simple relation with the spectrum of $\cal M$. Because ${\cal M'}$  is much smaller than ${\cal M}$ near the transition, one can successfully use perturbation theory for the bulk part of the normal modes of  $\cal{M}$. Nevertheless perturbation theory fails at very low frequency, which is of most interest.  In this region the spectrum of ${\cal M}$ contains  the plane waves and the anomalous modes. In what follows we estimate the change of frequency induced by the applied stress on these modes. We show  that the relative correction to the plane wave frequencies is very small, whereas the frequency of the anomalous modes can be appreciably changed. Finally we show that these considerations  lead to a correction of the Maxwell criterion of rigidity.

\section{Applied stress and plane waves}

Consider a plane wave of wave vector  $k$. Since the directions $\vec n_{ij} $ are random, both the relative longitudinal and the transverse displacements are of the same order: $[(\delta\vec{R_i}-\delta\vec{R_j})^\bot]^2\sim[(\delta\vec{R_i}-\delta\vec{R_j}).\vec n_{ij}]^2\sim(\delta\vec{R_i})^2  k^2$. Consequently the relative correction $\Delta E/E$ induced by the applied stress term is very small:

\be
\label{est}
\frac{\Delta E}{E}\approx \frac{\frac{1}{2}\sum_{\langle  ij \rangle} (r_{ij}^{eq}-1)[(\delta\vec{R_j}-\delta\vec{R_i})^{\bot}]^2}{\sum_{\langle  ij \rangle} [(\delta\vec{R_j}-\delta\vec{R_i})\cdot \vec{n}_{ij}]^2}
\ee
since $r_{ij}^{eq}-1$ is proportional to the pressure $p$, while the others factors remain constant  as $p\rightarrow 0$, $\frac{\Delta E}{E}\sim p$, and is thus arbitrarily small near the jamming threshold \footnote[11]{In disordered systems the acoustic modes are not exact plane waves, see e.g.  the recent simulations in Lennard-Jones systems \cite{tanguy1,tanguy2,tanguy3,tanguy4}. As we  discuss below, for transverse plane waves the energy is reduced by a factor $\delta z$. Therefore the relative correction of energy induced by the applied pressure is of the order of  $\frac{p}{\delta z}$, rather than $p$. We have, as shown, $\frac{p}{\delta z} \ll  1$, so that we still expect the correction to be small near the jamming threshold. In principle non-affine displacements could have other interesting effects, such as an increase of the transverse terms amplitude. If so, the effect of applied stress on acoustic modes would be enhanced. }.  Note that when the pressure is high, this effect is non negligible. In particular elastic instabilities can occur, and can be responsible for conformational changes, see  \cite{karki} for such examples in silica crystals.

\section{Applied stress and anomalous modes}
\label{ppp}
\begin{figure}

\centering
\includegraphics[width=5cm]{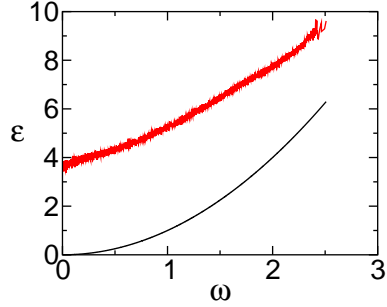}
\caption{Sum of the transverse terms (red curve) $\epsilon=  \frac{1}{2} \sum_{\langle ij\rangle}[(\delta\vec{R_j}-\delta\vec{R_i})^{\bot}]^2$ and longitudinal terms (black curve) $\epsilon=  \frac{1}{2}\sum_{\langle ij\rangle}[(\delta\vec{R_j}-\delta\vec{R_i})\cdot \vec{n}_{ij}]^2$ for each mode of frequency $\omega$ at the jamming threshold in three dimensions. The longitudinal term is equal to the energy of the modes and vanish quadratically at 0 frequency. The transverse term converges toward a constant different from 0. }
\label{trans}
\end{figure}

For anomalous modes the situation is very different:  we expect the transverse relative displacements to be much larger than the longitudinal ones. Indeed soft modes were built by imposing zero longitudinal terms, but there were no constraints on the transverse terms. These are the degrees of freedom that generate the large number of soft modes. The simplest assumption is that the relative transverse displacements are of the order of the displacements themselves, that is $\sum_{\langle ij\rangle}[(\delta\vec{R_j}-\delta\vec{R_i})^{\bot}]^2\sim \sum_i\delta\vec{R_i}^2=1$ for the anomalous modes that appear above $\omega^*$. This estimate can be checked numerically for an isostatic system where the sum of the relative transverse terms is computed for all $\omega$. The sum converges to a constant when $\omega\rightarrow 0$ as assumed, see Fig.(\ref{trans}). 

We can estimate the scaling of the correction in the energy $\Delta E$ induced by the stress term on the anomalous modes: 

\be
\Delta E \sim - \sum_{\langle  ij \rangle} (1-r_{ij}^{eq}) [(\delta\vec{R_j}-\delta\vec{R_i})^{\bot}]^2 \sim -p
\ee
which is an {\it absolute} correction, which can be  non-negligible in comparison with the energy $E$.

\section{Onset of appearance of the anomalous modes}

We can now estimate the lowest frequency of the anomalous modes. The modes that appear at $\omega^*$ in the relaxed springs system have an energy lowered by an amount of order $-p$ in the original system. Applying the variational theorem of the last section to the collection of slow  modes near $\omega^*$ indicates that there must be slow normal modes with a lower energy. That is,  the frequency  $\omega_{AM}$ at which anomalous modes appear verifies:

\be
\label{33}
\omega_{AM}\leq  [(\omega^*)^2-A_2 p]^{\frac{1}{2}} \equiv [A_1 \delta z^2-A_2 p]^{\frac{1}{2}}
\ee
where $A_1$ and $A_2$ are two positive constants.  This indicates that the important parameters of the low frequency excitations are coordination and stress.

\section{Extended Maxwell criterion}

\begin{figure}
\centering
\includegraphics[width=9cm]{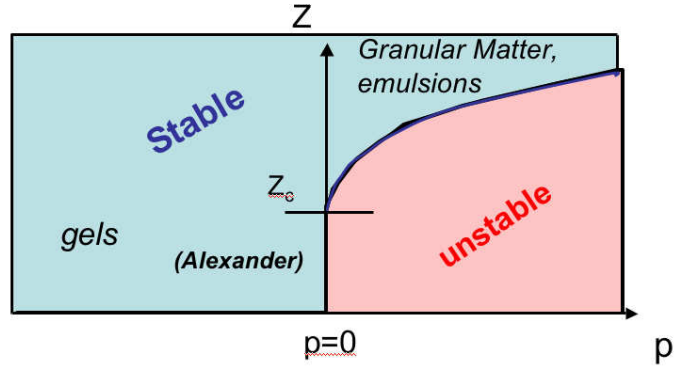}
\caption{Phase diagram of rigidity in terms of coordination and pressure.  When $p>0$, the line separating the stable and unstable regions is defined by Eq.(\ref{22}).  When the pressure is negative, any connected system can be rigid, as it is the case for gels. }
\label{diag}
\end{figure}

From this estimation we can readily obtain a relation among coordination and pressure that guarantees the stability of a system. There should be no negative frequencies in a stable system, therefore $\omega_{AM}>0$.  Thus in an harmonic system the right hand side of Eq.(\ref{33}) must be positive:

\be
\label{22}
\delta z \geq C_0 p^{\frac{1}{2}}\equiv \delta z_{min}
\ee 
 where $C_0$ is a constant. This inequality, which must hold  for any spatial dimension, indicates how a system must be connected to counterbalance the destabilizing effect of the pressure.  A phase diagram of rigidity is represented in Fig.(\ref{diag}). When $p=0$, the minimal coordination $z_c$ corresponds to the isostatic state: this is the Maxwell criterion. As we said earlier for spherical particles $z_c=2d$. As we discuss later $z_c=d+1$ for particles with friction.  When $p>0$, Eq.(\ref{22}) delimits the region of rigid systems: for example granular matter or emulsions  lie above this line. When $p<0$, even systems far less connected than what the Maxwell criterion requires are rigid \cite{shlomon}, as we discussed in Chapter \ref{c2}. These systems contain many soft modes as defined in Eq.(\ref{3}), but there are all stabilized by the positive bracketed term of Eq.(\ref{000}).  Note that a similar phase diagram, with the same singularity of $\delta z$ but a different $z_c$, was recently obtained by a mean field approach \cite{head}.

Finally the relation (\ref{22}) is  verified in the simulations of \cite{J} where $\delta z\sim p^{\frac{1}{2}}$. In the next Chapter we justify why the inequality (\ref{22}) is in fact an equality in the system of \cite{J} and discuss what determines the microscopic structure of this system.

\chapter{Microscopic structure and marginal stability}
\label{c6}

In the previous section we studied how the low-frequency vibrational properties were related to the microscopic structure. The applied pressure has two antagonist effects: on the one hand, it increases the coordination number, which stabilizes the system and increase the frequency of appearance of the anomalous modes. On the other hand, the applied pressure appears in the expansion of the energy and lowers the frequency of the anomalous modes. In what follows we study the relative amplitude of these two effects, or equivalently,   where  such amorphous solids are located in the $(z, p)$ plane of Fig.(\ref{diag}). 

Since these systems are out-of-equilibrium, their microscopic structures depend on the system history.   As we shall see below,   the preparation of the system of \cite{J} leads to {\it marginally stable} systems at any $\phi$.  The two antagonists effects of the pressure compensate \footnote[12]{Assuming an exact compensation of these two terms lead to $\omega_{AM}=0$ in an infinite size system. In Fig.(\ref{compare}) $\omega_{SM}$ is slightly different from zero, as one would expect for a finite size system.}, which leads to  $\omega^*\gg  \omega_{AM}$, as it appears in Fig.(\ref{compare}), and $\delta z\sim p^{\frac{1}{2}}$. In what follows we propose a simple argument to justify such a behavior. We discuss in particular (i) the dynamic that takes place when a liquid of repulsive spheres is hyperquenched and (ii) the decompression  of a jammed solid at zero temperature. This will also enable us to discuss the surprising geometrical property of the random close packing evoked in the introduction: there is a divergence in the pair correlation function $g(r)\sim1/\sqrt{r-1}$ at close contact.  We propose that this divergence is a vestige of the marginal stability that occurs at higher packing fraction.

\section{Infinite quench} 

The simulations of \cite{J} show $\delta z \sim p^{\frac{1}{2}}$, thus saturating the bound of Eq. (\ref{22}), so that there are excess modes extending to frequencies much less than $\omega^*$.  We start by furnishing an example of dynamics that lead to such a situation.  Consider an initial condition where forces are roughly balanced on every particle, but such that the inequality (\ref{22}) is not satisfied. Consequently, this system is not stable:  infinitesimal fluctuations  make the system relax with the collapse of unstable modes. Such dynamics was described by Alexander in \cite{shlomon} as structural buckling events: they are are induced by a positive stress as for the buckling of a rod, but take place in the bulk of an amorphous solid. These events  a priori create both new contacts and decrease the pressure. When the bound of (\ref{22}) is reached, there are no more unstable modes. If the temperature is zero, the dynamics stops. Consequently one obtains a system where Eq.(\ref{22}) is an {\it equality},  therefore (i) this system is weakly connected (ii)  $\omega_{AM}=0$, there are anomalous modes much different from plane waves extending to zero frequency. A similar argument is present in \cite{head}.

In the simulations of Ref \cite{J} the relaxation  procedes as follows.  The system is initially in equilibrium at a high temperature.  Then it is hyperquenched to zero temperature.  At short time scales the dynamics that follows is dominated by the relaxation of the stable, high frequency modes. The main effect  is to restore approximately force balance on every particle. At this point, if inequality (\ref{22}) is satisfied, the dynamics stops. If it is not satisfied, we are in the situation of buckling described above.  The pressure and co-ordination number continue to change until the last unstable mode has been stabilized.  At this point the bound of Eq. (20) is marginally satisfied, and there is no driving force for further relaxation.

It is interesting to discuss further which situations lead to marginal stability. We expect that the situation of marginal stability that follows an infinite cooling rate takes place  for a domain of the parameters of initial conditions $(\phi,T)$, located at high temperature and low density. This domain could end at a finite $\phi$ even when temperature is infinite, as it does in some related systems. This was shown in simulations and  theoretical work on Euclidian random matrix \cite{parisi2} were most of the unstable modes vanish beyond a finite $\phi$ even when $T=\infty$. When the cooling rate is finite, as in experiments, we expect that the relaxation does not stop when all the modes are stable, but that there are activated events that lead to further collapses of anomalous modes. These events  a priori increase the connectivity and decrease the pressure beyond the bound of Eq.(\ref{22}), leading to $\omega_{AM} > 0$.  In agreement with this idea, hyperquenched mineral glasses show a much stronger excess of modes\cite{angell} in comparison with normally cooled glasses, and annealed polymeric glasses expectedly show a smaller excess of modes \cite{duval}.

\section{Decompression } 

The data of \cite{J} were obtained by gradually decreasing the pressure from the above initial state of zero temperature and nonzero pressure.  When pressure is lowered, it is observed that the system remains marginally stable. Here we propose a qualitative interpretation of these findings in the case of harmonic particles \footnote[13]{To extend this argument to other potentials, for example Hertzian contacts, one should assume that $g(1)$ scales at least as $(\phi-\phi_c)^{-\frac{1}{2}}$ when the jamming threshold is approached. This would imply that the number of contacts lost when the system is decompressed is  large enough to generate buckling. This point is related to the evolution of the force distribution of Hertzian particles near the isostatic point. It is a subtle issue, and we are not aware of any numerical results of this sort.}.  In \cite{J}, the decompression is obtained by discrete steps. At each step, the radii of each particle is reduced by a small amount $\epsilon$, while the center of the particles are kept fixed. This corresponds to an affine decompression. This new configuration is not at equilibrium in general. Then the particles are let to relax. The affine decompression has two effects: on the one hand it causes some contacts to open, on the other hand it reduces the pressure.  The opening of these contacts tends to destabilize normal modes and reduce their frequencies, while the reduction in pressure tends to stabilize them.  As we argue below, the destabilizing effect  dominates. Thus, the affine decompression leads the system into the unstable region.  Therefore we expect that when the particles  relax, one recovers the dynamics that follows an infinite quench: modes buckle. As for the infinite quench of temperature discussed above, buckling should occur as long as the relaxation of the stable normal modes, which is faster than the collapse of unstable modes, does not bring back the system into the stable region. If so, the buckling increases the contact number and decreases the pressure until marginal stability is achieved, so that the inequality of Eq. (\ref{22}) is marginally satisfied as the pressure decreases, as observed in the simulations of \cite{J}. 

In what follows we justify the claim that the destabilizing effect of the opening of the contacts dominate the effect of the pressure reduction. When  the particles radii decrease by an amount $\epsilon$, a certain fraction $e\sim g(1)\epsilon$ of contacts opens, where $g(r)$ denotes the radial distribution function.  For harmonic particles we expect $g(1)\sim (\phi-\phi_c)^{-1}$ \footnote[14]{This is related to well-known empirical facts of the force distribution: one has $P(F) dF\sim g(r) dr$. For harmonic particles $dr\sim dF$ and therefore $P(F)\sim g(r)$. When rescaled by $\sim p$, $P(F)$ converges to a master curve with $P(F/\langle F \rangle=0)\neq 0$ \cite{J,ohern}.  This implies that $g(1)\sim p^{-1}\sim (\phi-\phi_c)^{-1}$ \cite{ohern2}.}. Hence, using that $p\sim (\phi-\phi_c)$ for harmonic particles \cite{J} --as we shall demonstrate in the next Chapter--, one obtains $e\sim\frac{\epsilon}{p}$.  On the other hand, the affine decompression lowers the pressure by an amount $\delta p\sim \delta \phi \sim \epsilon$. Thus the system can only afford to lose a fraction $f$ of contacts while remaining stable:   according to Eq.(\ref{22}): $f= \frac{d(\delta z_{min})}{dp}\delta p\sim p^{-\frac{1}{2}} \delta p\sim \frac{\epsilon}{p^{\frac{1}{2}}}\ll  e$. Therefore $f\ll  e$ as claimed.  Hence if an affine reduction of packing fraction $\epsilon$ is imposed, far too many contacts open and the system is unstable. 

To conclude, it was observed in the simulations of \cite{J} that if the steps $\epsilon$ are small, the decompression that takes place in \cite{J} is reversible: cycles of decompression-compression bring the system back to its initial configuration. This empirical fact indicates the absence of discontinuous irreversible events. Thus, the buckling generated by the opening of few contacts  when $\epsilon$ is small enough does not lead to rearrangements of finite amplitude much larger than $\epsilon$.  This indicates that the dynamic of modes collapse increases the coordination by re-closing most of the contacts that open during  the affine decompression (whose particles are separated by distance of at most $\epsilon$), and not by forming new contacts. It is reasonable to think that if several cycles of compression/decompression are made, the system will end up to be reversible.  Why it is  already so at the first decompression is a subtle question that we do not try to justify here.

\section{$g(r)$ at the random close packing}

The probability $g(r)$ of having two particles separated by a vector of length $r$ displays a square root divergence $g(r)\sim (r-1)^{-\frac{1}{2}}$ at the jamming transition. In the introduction we pointed out that this divergence corresponds to the singular increase of the excess coordination $\delta z$ with the packing fraction, as was noted in \cite{J}. Here justify further this correspondence. We show that, if the decompression is assumed to be adiabatic, the singularity in $g(r)$  is a necessary consequence of the marginal stability that characterizes the decompression.  We argue that the pair of particles almost touching at $\phi_c$, responsible for the divergence in $g(r)$, are the vestiges of  the contacts that were present at higher $\phi$  to stabilize the system.  In order to show this,  we first have to count the contacts that open for a given $\phi-\phi_c$. Then we shall estimate the distance between the corresponding pairs of particles at the jamming threshold.

As we discussed in the last section, when the system is decompressed it remains marginally stable: the coordination follows Eq.(\ref{22}), and $z= z_c+\delta z_{min}$. Thus the density of contacts $n(\phi)$ per unit $\phi$ that open  for a given $\phi$ follows exactly in the large $L$ limit:
\be
\label{yyy}
n(\phi)\sim \frac{d(\delta z_{min})}{d\phi}\sim \frac{1}{(\phi-\phi_c)^{\frac{1}{2}}}
\ee

\begin{figure}
\centering
\includegraphics[angle=0,width=9cm]{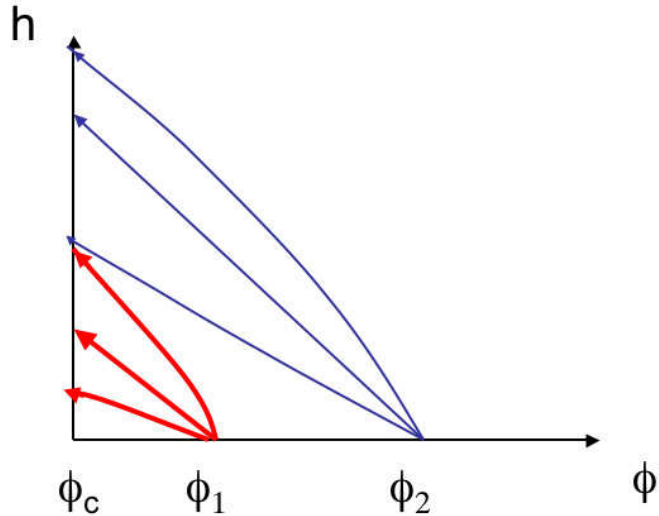}
\caption { Representation of typical variations  of the spacing $h$ between particles whose contacts  opened at  given $\phi>\phi_c$. The arrows are typical trajectories. Those  starting from $\phi_1$ are thicker than those starting from $\phi_2$, indicating that the density of contacts that open is larger as $\phi$ decreases toward $\phi_c$.  $g(r)$ at $\phi_c$ is computed by summing over all trajectories for all $\phi>\phi_c$.}
\label{contribution}
\end{figure}

We now would like to evaluate the distance that separates such particles at the jamming threshold. Let $h_\phi$ be the random variable that corresponds to the spacing $r-1$ at the jamming threshold between two neighboring particles  whose contact opened at a given $\phi$. We want to estimate the fluctuations of $h_\phi$. If the decompression was purely affine $h_\phi$ would be single-valued and given by $h_{aff}(\phi)$:
\be
 h_{aff}(\phi) =\frac{\phi-\phi_c}{d\phi_c}
\ee
As we discussed in the last section  the displacements that follow a decompression are not affine.  For our present argument we need to evaluate the variance  of the distribution of $h_\phi$.  It is directly related to the variance of the non-affine displacements that appear while decompressing. We expect that such non-affine displacements simply lead to a variance of $h_\phi$ of order of its average $h_{aff}(\phi)$  \footnote[15]{More generally, if two particles (in contact or not) are at a distance $r$ of order one at a given $\phi$,  we expect that the fluctuations of the distance that separate them at $\phi_c$ due to non-affine displacements is of order of the affine increase of their distance $\delta r= r \frac{\phi-\phi_c}{d\phi_c}\approx \frac{\phi-\phi_c}{d\phi_c}=h_{aff}(\phi)$. This comes from the following observation: non-affine displacements are induced by the requirement of stability that creates correlations among particles motions. To evaluate such correlations, consider the typical situation discussed in the last section where particles in contact at a given $\phi$ have to stay in contact until $\phi_c$, instead of spreading apart if the displacement was affine. At $\phi_c$  the inter-particle distance is 1, instead of a value $r< 1+h_{aff}(\phi)$. Thus we evaluate  the typical departure from a pure affine displacement to be of order $h_{aff}(\phi)$.}.  Therefore the probability $P_\phi(h)$ that two particles whose contact opened at a given $\phi$ are at a distance $1+h$ at the threshold can be written as:
\be
P_\phi(h)\equiv \frac{1}{\phi-\phi_c} f(\frac{h}{\phi-\phi_c})
\ee
where $f$ is a normalized scaling function $\int f(x) dx=1$. Thus  one can compute $g(r)$ by summing over all the contributions of the contacts that opened at $\phi > \phi_c$, as we represent on Fig.(\ref{contribution}):   
\ba
g(r)\sim \int n(\phi) P_\phi(h=r-1) d\phi\sim \int \frac{1}{(\phi-\phi_c)^{3/2}}  f\left(\frac{r-1}{\phi-\phi_c}\right)d\phi\\
g(r)\sim \frac{1}{(r-1)^{\frac{1}{2}}} \int u^{-\frac{1}{2}} f(u) du
\ea
We do not expect any singularity of $f(u)$ in $u=0$ \footnote[16]{We do not exclude that there is a finite probability $p_0$ for particles that lost their contact at a given $\phi_0$ to recover it at a $\phi_1$ such that $\phi_c<\phi_1<\phi_0$. In our formalism such eventuality will not create a divergence in $f(u)$. The incoming flux of contact would be compensated by a larger rate of contact opening, and the expression (\ref{yyy}) will be corrected by a factor $\frac{1}{1-p_0}$.}, and therefore:
\be
\label{bbb}
g(r)\sim \frac{1}{(r-1)^{\frac{1}{2}}}
\ee

\section{Isostaticity, $g(r)$ and thermodynamics}

In our argument above the divergence in $g(r)$ does not appear through a minimization of a free energy or volume, but rather as a consequence of the specific dynamics that took place to reach the jamming threshold.  Empirically, it turns out that the divergence in $g(r)$ was observed in systems obtained with different dynamics. It was first noticed in molecular dynamics simulations of elastic spheres with friction, where the jammed states were obtained using inelastic collisions \cite{sil}. It was also recently observed --- with possibly a slight difference in the exponent--- in random close packing of hard spheres obtained by increasing continuously the packing fraction from the liquid state \cite{donev}. In this last example thermodynamic equilibrium is maintained until fairly high packing fractions, corresponding to typical inter-particle spacing much smaller than the particles radii \footnote[17]{ For mono-disperse spheres thermodynamic equilibrium can be obtained below typically $\phi\approx 0.58$, which corresponds to an inter-particle spacing $h\approx 0.03$}.  On the other hand in this hard sphere system the range of validity of Eq.(\ref{bbb}) extends to rather large distances, say $r-1\approx 0.3$. Thus, in this system the divergence of $g(r)$ at $\phi_c$ cannot be the vestige of an interaction that would have taken place between pairs of neighbor particles at lower $\phi$, as the system has a finite relaxation time and neighbor particles can move arbitrary far from each other. Instead, this suggests that the divergence of $g(r)$ obtained in \cite{donev} is related to thermodynamic properties of a hard sphere liquid near random close packing. 

In what follows we discuss  why the properties of isostaticity and divergence of $g(r)$, which were introduced in relation with the mechanical stability of the solid phase, may also be connected to the thermodynamic of the liquid phase of hard spheres (see also Chapter \ref{c9}). Our argument is in two steps.  We first argue that the isostaticity and the divergence in $g(r)$ have an interpretation in terms of density phase space of hard sphere packing. We shall show that if an assembly of hard spheres is (i) sub-isostatic, that is with a coordination smaller than $2d$, or  (ii) isostatic with a singularity of $g(r)$ in $r=1$ weaker than Eq.(\ref{bbb}), then it is not a ``good'' local maximum of density: one can build  close configurations which are denser. In a second step, we simply argue that in the liquid phase, the system prefers to lie in configurations close to the denser packing,  which is favorable for entropic reasons. Finally this suggests that when $\phi$ increases toward $\phi_c$, the system becomes isostatic and display  at least a square root singularity in $g(r)$ at $r=1$.

\begin{figure}
\centering
\includegraphics[angle=0,width=9cm]{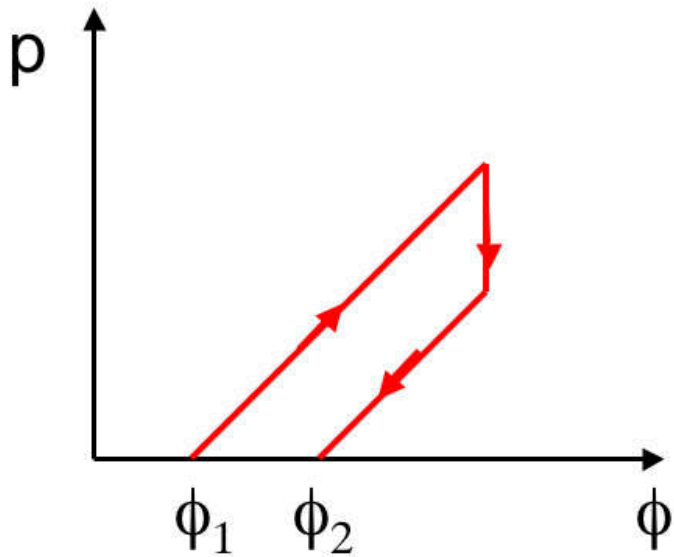}
\caption {Cycle of compression-buckling-decompression starting from a packing fraction $\phi_1$ and ending at a larger packing fraction $\phi_2$.}
\label{p}
\end{figure}

Consider a sub-isostatic configuration of hard sphere in a box. As we discussed in earlier Chapters, this system is not rigid. Therefore, if an infinitesimal pressure is applied at the boundary of the box, the system cannot resist to it and yields. Unstable modes collapse until isostaticity is recovered. Consequently, the volume of the box decreases and one obtains a denser configuration of spheres.  The same kind of argument can be made if one considers an  isostatic configuration of hard spheres, again contained in a box, that do not display  the square root singularity of $g(r)$ in $r=1$.  Let assume temporally that the particles are not hard, but soft,  and interact for example with harmonic interactions. Such system is again unstable toward compression: an affine compression of strain $\epsilon$ increases the coordination by an amount that scales as $\delta z\sim  \int_1^{1+\epsilon} g(r) dr$. If $g(r)$ does not display the square root singularity in $r=1$, one finds $\delta z \ll  \delta z_{min}\sim p^{\frac{1}{2}}\sim \epsilon^{\frac{1}{2}}$. Thus unstable modes collapse until the system is enough coordinated, which lowers the pressure. Then if the pressure is brought  back to 0 adiabatically, one obtains a new isostatic state of higher density, as sketched in Fig.(\ref{p}).

Now we argue that a liquid of  hard spheres lies near the dense jammed states. More precisely, we define the jammed state corresponding to a liquid configuration as the solid obtained by hyperquenching the liquid, that is by increasing rapidly the radii of the particles until a solid is formed. We argue that the densest jammed state are favored, as they offer a larger free volume per particle in the liquid configuration. Consider for example two jammed states of packing fraction $\phi_a<\phi_b$, and a liquid at packing fraction $\phi<\phi_a$.  If the  liquid lies near the jammed state $a$  the free volume per particles varies as $\phi_a-\phi$. Thus if $p_a$ is the probability for the liquid to have $a$ as the jammed state, one can estimate $p_a/p_b\sim (\frac{\phi_a-\phi}{\phi_b-\phi})^N$. Thus the densest jammed states are favored as the packing fraction increases.  For a quantitative treatment of the liquid equilibrium, one should enumerate all the possible jammed states that are accessible to the dynamics ---as there are configurations such as the crystal which are not visited in a reasonable time--- and compute their structure. A similar approach was followed recently in \cite{parisi4}. Our qualitative argument suggests that the isostatic states with the divergence in $g(r)$  evoked above are good candidates to dominate the dynamics near the glass transition, since they are (i) good (in the sense stated above) local maxima of density and  (ii) plausibly numerous, as we exhibited a simple way to create them. If so, $g(r)$ in a random close packing obtained from the liquid phase must display a square root divergence. 

\chapter{Elastic response near the jamming threshold}
\label{c7}

In this Chapter we first derive  a few properties of the response to a local strain, such as the deformation of one contact, near the jamming threshold. We show that at the isostatic limit, such a response  spreads out uniformly in the entire system. This implies that the soft modes of an isostatic state obtained by cutting one contact are extended, as assumed in Chapter \ref{c3}.  When the coordination increases, we show that the energy cost induced by such a local strain grow as $\delta z$.

This suggests that the shear modulus also scales as $\delta z$, as it is observed numerically. We confirm this behavior by computing the response of the system to a global compression and a global shear.  An assembly of repulsive short-range particles near jamming behaves as a gel in the following sense: the shear modulus is much smaller than the bulk modulus, as it was observed numerically \cite{J,makse}. In a gel,  the response to compression is  the one of  liquid of monomers, whereas the response to a shear is related to the stretching of the polymer chains. The cause for the discrepancy between these two moduli is different near jamming. If the responses to compression and shear were purely affine, then the two moduli would scale in the same way.  As we shall show the bulk modulus is simply what one would expect from an affine response of the system.  In the case of a shear, the non-affine displacements lower the energy dramatically, which even vanishes at the transition. This is possible because the low-coordination allows particles to rearrange without much energy cost.  Thus this system is a useful tool to study non-affine displacements, that might play an important role to describe material failure \cite{maloney}, and that were observed in various systems such as granular matter or biological materials, and in simulations of glasses and foams, see e.g. \cite{lubensky} for references.  We discuss very tentatively the possible length scales that characterize such displacements.

\section{Formalism}

In this section we study the relation between forces and soft modes. We show that the equation of  force balance is the dual of the equation that defines the soft modes. This will be used in the next section to compute the response to a strain. 

\subsection{Force propagation}

Many of the properties discussed below concern the response of the system to external forces. It proves convenient to consider our system under the influence of an arbitrary set of forces $\vec{F_i}$ acting on all particles $i$. At equilibrium the sum of the forces on each particle $ i$ is null:
\be
\label{72}
\sum_{\langle j \rangle} f_{ij} \vec{n}_{ij}= \vec{F_i}
\ee
where $ f_{ij}$ is the compression in the contact $<ij>$, the sum is on the particles in contact with particle $i$, and $\vec{F_i}$ the external force applied on $i$. In term of sign convention we recall that $\vec{n}_{ij}$ is the unit vector going from $i$ to $j$. This linear equation can be written:

\be
\label{710}
{\cal T} |{\bf f}\rangle = |{\bf F}\rangle
\ee
$|{\bf f}\rangle$ is  the vector of contact tensions and has $N_c$ components. ${\bf F} $ is the vector of external forces. Its dimension is $ (N d-d(d+1)/2) $. Indeed there are $d$ degrees of freedom for the external force on each particle, which brings the term $Nd$. The term $d(d+1)/2$ corresponds to the constraints on the total torques and forces that must be zero at equilibrium (in what follows our notations shall be lower-case for the contact space of dimension $N_c$, and upper-case for the particles positions space of dimension $Nd-d(d+1)/2$). Therefore ${\cal T} $ is an $ Nd-d(d+1)/2 \times N_c$ matrix. 

In the case where $N_c>Nd-d(d+1)/2 $  equation (\ref{710}) is not sufficient to fix the contact tensions. One must apply elasticity theory to compute the  contact forces for a set of external forces. This is done by writing the relation between forces and distances between particles in contact (which depends on the interaction potential), and by relating the distance among particles to their position. This brings the necessary $N_c- Nd-d(d+1)/2$ extra-constraints. When  $N_c<Nd-d(d+1)/2 $, there is  in general no solution for a given force field, and the system yields under a generic external force field. For such a system no contact forces ${\bf f}$ can satisfy (\ref{710}) unless  ${\bf F}$ is restricted. In the last case, when $ N_c=Nd-\frac{d(d+1)}{2}$, the system is called ``isostatic'', and there are just enough contacts to equilibrate any external force field. Note that in this case Eq.(\ref{710}), which does not depend on  the interaction potential,  is sufficient to determine the contact forces.

\subsection{Duality between force propagation and soft modes}

As we discussed in earlier Chapters, the static equilibrium of the system can be expressed in geometric terms, rather than in terms of the forces as in Eq.(\ref{710}). The change of distance to the first order between particles for a given set of displacements $\delta\vec{R_i}$ is:

\be 
\label{s}
(\delta\vec{R_i}-\delta\vec{R_j}).\vec{n}_{ij}=\delta r_{ij}
\ee
It can be written as:
\be
\label{dis}
{\cal S}|\delta{\bf R}\rangle = |{\bf \delta r}\rangle
\ee
where ${\bf \delta r}\equiv \{\delta r_{ij}\}$ is the set of distance changes for all contacts. If one removes the global translations or rotations from the displacement fields, which obviously do not change any inter-particle distance, ${\cal S}$ becomes an $(Nd-d(d+1)/2)$ by $N_c$ matrix. Its kernel is the space of soft modes we introduced in Chapter \ref{c2}. When $N_c<Nd-d(d+1)/2$, the displacements are undetermined for a given change of inter-particles distances.

Now we  establish the connection among soft modes and forces, which is also present in a similar form in \cite{Tom2,roux}. At equilibrium for any displacement, the net work done by external forces and the contact forces is zero: this assures the stability of the system. Therefore for any acceptable equilibrium force field:
\ba
\label{73}
\sum_{i} \delta\vec{R_i}\cdot \vec{F_i}- \sum_{ij} \delta r_{ij}  f_{ij}=\langle\delta {\bf R}| {\bf F}\rangle - \langle \delta {\bf r} \cdot {\bf f}\rangle=0
\ea
For a soft mode, $\delta r_{ij}=0$ and we are left with:
\ba
\langle \delta {\bf R}| {\bf F}\rangle=  \sum_{i} \vec{\delta R_i}\cdot \vec{F_i}=0
\ea
The soft modes are equivalent to the constraints on the force field one would have obtained by solving Eq.(\ref{72}). Each soft mode represents a direction of fragility of the system,  and the external forces must be  orthogonal to them to avoid yielding. Furthermore, applying the definitions of ${\cal S}$ and ${\cal T}$ in (\ref{73}) we have:
\be
\langle {\bf f}| {\cal S}|\delta {\bf R}\rangle= \langle\delta {\bf R}| {\cal T}|{\bf f}\rangle 
\ee
therefore, introducing the transpose notation, we obtain:
\be
{\cal S=T}^t
\ee

\subsection{Relation with the Dynamical matrix} 

In this Chapter we shall neglect the initial stress term in the energy expansion, as we did in Chapters \ref{c3} and \ref{c4}.  As we discussed in Chapter \ref{c5}, we expect that this term affects only  weakly the plane waves. Thus it should lead to  negligible corrections in the computations of the responses at zero wave vector  such as a shear or a compression. In the harmonic case we discuss, we have then:
\be
\label{7M}
\delta E=\sum_{<ij>} \frac{1}{2} \delta r_{ij}^2=\frac{1}{2}\langle {\cal S} \delta{\bf R}| {\cal S}\delta {\bf R}\rangle 
\ee 
Therefore one finds for the dynamical matrix  defined in Eq.(\ref{m}): 
\be
{\cal M=S}^t{\cal  S}
\ee

\section{Relation between the response to a strain and forces}
\label{iii}

The response to an imposed strain is given by the minimization of the energy with respect to the displacement fields. In this section, using simple linear algebra and the duality between force and soft modes studied above, we show how this minimization can be written as a sum over the vectors in the contact force space that satisfy force balance. This will allow in the next sections to (i)  derive exact results  of the response to a local strain (ii) derive the bulk and shear moduli assuming well-known empirical fact of force properties, that we recall at the end of this section.

As we discussed in earlier Chapters, in the approximation where the initial stress is neglected,  the system is equivalent to a set of point particles interacting with springs. In order to study the elastic behavior of such system, it turns out to be convenient to consider the responses that follow arbitrary changes of rest length of these springs. This is in fact equivalent to imposing dipoles of force. Consider for example two particles $i$ and $j$ in contact, and increase the rest length of their spring by an infinitesimal amount $y_{ij}$. It is equivalent to impose a dipole of force where opposite external forces are imposed on $i$ and $j$ with $\vec{F_j}=-\vec{F_i}=y_{ij}\vec{n}_{ij}$. As we shall see, the response to shear or compression can also be  easily expressed in terms of changes of rest length.

We impose an infinitesimal change of rest length on every contact ${\bf y}=\{y_{ij}\}$. Following Eq.(\ref{7M}) the energy and the displacement field are given by the minimization of:
\be
\label{74}
\delta E= \frac{1}{2} \min _{\{\delta {\bf R}\}} \langle{\cal S}\delta {\bf R}-{\bf y}| {\cal S}\delta {\bf R}-{\bf y}\rangle
\ee
Obviously if ${\cal S}$ was spanning its image space, we would have $\delta E=0$ : one could always find a displacement $|\delta {\bf R}\rangle$ that leads to a change of distances between particles in contact exactly equal to $|{\bf y}\rangle$.  As we said,  ${\cal S}$ is a $N_c\times Nd-d(d+1)/2$ matrix. If $N_c< Nd-d(d+1)/2$,   ${\cal S}$ indeed spans its image space, and the energy associated with any strain $|{\bf y}\rangle$ is zero: the system is floppy.  In the other case, if $N_c> Nd-d(d+1)/2$, there are $ N_c- Nd-d(d+1)/2\equiv N\delta z/2$ relations of dependency among the columns of ${\cal S}$. One can choose a basis of  $ N \delta z/2$ vectors  $|{\bf a}^p\rangle$, with $1\leq p \leq  N\delta z/2$,  in the space of $|\delta {\bf r\rangle}$ such that:
\be
\langle{\bf a}^p| {\cal S}=0
\ee
The $|{\bf a}^p\rangle$ are orthogonal to all the vectors ${\cal S}|\delta {\bf R}\rangle$, for any displacement field  $|\delta {\bf R}\rangle$.  Transposing this relation we have:
\be
{\cal T} |{\bf a}^p\rangle=0
\ee
which indicates that all the vectors in the space of the $|{\bf a}^p\rangle$  satisfy force balance without external force (\ref {72}), but no others. The  $|{\bf a}^p\rangle$  live in the contact-force space, and henceforth we shall denote them $|{\bf a}^p\rangle \equiv |{\bf f}^p\rangle = \{f_{ij}^p\}$. In the following we consider an orthogonal unit basis:
\be
\langle{\bf f}^p| {\bf f}^{p'}\rangle \equiv \sum_{ij} f_{ij}^p f_{ij}^{p'}= \delta_{pp'}
\ee
We can decompose any $|{\bf y}\rangle$ as:
\be
\label{yo}
{\bf y}={\bf y}^{\bot}+\sum_{p=1...N\frac{\delta z}{2}} \langle{\bf f}^p|{\bf y}\rangle {\bf f}^p
\ee
${\bf y}^{\bot}$, the part of $|{\bf y}\rangle$ orthogonal to the $|{\bf f}^p\rangle$, is spanned by the matrix ${\cal S}$, and therefore does not contribute to the energy when the minimization of Eq.(\ref{74}) is performed.  In other words, there is a displacement  field which leads to a strain ${\bf y}^{\bot}$. The energy that results of the minimization of Eq.(\ref{74}) is then simply given by the distance square between $|{\bf y}\rangle$ and the image vector space of  ${\cal S}$, that is $||{\bf y}-{\bf y}^{\bot}||^2$ or:
\ba
\label{rere}
\delta E=\frac{1}{2} \sum_{p=1,..\frac{N\delta z}{2}} \langle{\bf f}^p|{\bf y}\rangle^2
\ea
Furthermore the response to such strain condensates an energy $\delta E_{ij}$ in the contact $<ij>$ that follows:
\be
\label{rerez}
\delta E_{ij}=\frac{1}{2} (y_{ij}-y_{ij}^{\bot})^2= [ \sum_{p=1,..\frac{N\delta z}{2}} \langle{\bf f}^p|{\bf y}\rangle f_{ij}^p]^2
\ee
Expressions (\ref{rere}) and (\ref{rerez}) furnish explicit solutions for the minimization of Eq.(\ref{74}). They do not give access to the displacement  field $\delta {\bf R}$ that follows a given strain (that can be computed in principle by inverting $\cal S$), but to the energy condensed globally or in any contact.  From these expressions we shall derive a few exact properties of the response to a localized strain or force dipole. Furthermore,  they relate the response to a strain to properties of forces, which are well studied empirically. Making simple assumptions on these contact force fields, we shall derive the scaling of the elastic moduli.

\section{Response to a local perturbation}

We impose a local strain corresponding to the compression or the stretching of one contact.  We implement it  by increasing the separation $r_{12}$ of a single contact $12$ chosen to be at the origin by an amount $\epsilon$. This is equivalent to impose a force dipole  $\vec{F_1}=-\vec{F_2}= \epsilon \vec{n}_{12}$. We have:
\ba
y_{ij}&=&0 \hbox{\ if } \ ij \neq 12 \\
y_{12}&=& \epsilon
\ea
It is now straightforward to compute the energy cost of  local strain using Eq.(\ref{rere}). One finds:
\be
\label{77}
\delta E=\frac{1}{2} \sum_{l=1,..N\frac{\delta z}{2}} (f_{12}^l  \epsilon)^2  
\ee
If one sums this expression for every possible contacts $ij$ and use the normalization of the vectors of force contact, one finds exactly $N  \frac{\delta z}{4}\epsilon^2$.  Therefore we have the following exact result for the average energy induced by such deformation:
\be
\label{ta}
\langle \delta E \rangle = \frac{\delta z}{2z}\epsilon^2
\ee
This results can be extended to the case where several contacts are deformed. For example, if a particle swells, all its contacts are compressed by the same factor $\epsilon$, and one finds that the energy still goes as $\delta z$.  In a isotropic continuous elastic medium with a weak shear modulus $G$, the energy resulting from the swelling of a small sphere goes as $G$. Thus we may infer from Eq.(\ref{ta}) that $G\sim \delta z$. This scaling is observed numerically \cite{J}. In the next section we shall justify further this scaling law by computing the response to a global shear.

Eq.(\ref{ta}) shows that when the jamming threshold is approached, deforming a contact (or imposing a force dipole) becomes softer and softer. Eventually, when $\delta z \rightarrow 0$, the deformation of the contact becomes totally soft. The response corresponds then exactly to the soft modes that we described in Chapter \ref{c3} that appear when a contact (here the contact 12) is removed in an isostatic system. A fundamental question was the extension of such modes, that was assumed to be over the whole system at that point.  In what  follows we show that this is indeed the case. 

When the jamming threshold is approached, the coordination diminishes until there is eventually only one term left in the sum of Eq.(\ref{77}).  At that point there is one more contact than  in an isostatic state \footnote[18]{At the jamming threshold, the system is in fact not exactly isostatic: it has one extra contact. It must be the case as there must be a non-zero contact force field to pin the particles. The counting of degrees of freedom of Chapter \ref{c2}, $N_c=Nd-d(d+1)/2$, is in fact incomplete: it must take into account the degree of freedom corresponding to the box size. This subtle point, which does not affect the result of the previous Chapters, is trivial in one dimension, as $N$ particles pinned on a ring have $N$ contacts, and not $N-1$. This was verified numerically in the simulations of \cite{J} in 2 and 3 dimensions as removing $m$ contacts at the threshold creates only $m-1$ soft modes instead of $m$.} and $\delta z=\frac{2}{N}$.   Eq.(\ref{ta}) indicates that the energy resulting from the deformation of one contact is of order $1/N$, which vanishes as the system size increases.  The deformation becomes soft. We can use Eq.(\ref{rerez}), whose sum contains only one force field, to obtain the spatial repartition of the energy:
\be
\delta E_{ij}=\frac{\epsilon^2}{2} (f^1_{12}f^1_{ij})^2 
\ee
$|\bf f^1\rangle$ is the physical set of contact forces that support the system. These contact forces are well studied numerically and experimentally, and we shall discuss more their properties in the next paragraph. At large distance where the contact force are de-correlated $\langle (f^1_{12}f^1_{ij})^2\rangle=\langle (f^1_{ij})^2\rangle^2>0$:  the energy condensed in the contact $ij$ does not vanish when the distance between $ij$ and $12$ diverges. If the displacements that follow such a perturbation were to vanish with the distance from the source, it would be so for the energies condensed in the contacts. As it is not the case, these displacements spread out  in the entire system.

\section{Elastic moduli}

\subsection{Spatial properties of the force fields $|{\bf f}^p\rangle$}

In this paragraph we discuss the properties of the contact force fields that we shall use to derive the scaling of the bulk and the shear moduli. 
Only one vector of the  vector space of the force fields $|{\bf f}^p\rangle$ solutions of Eq.(\ref{710}) without external force is the real set of contact forces  that supports the system. As we discussed at the beginning of this Chapter, it can be computed by solving the whole elastic problem. The solution of such problem is unique. This vector is denoted $|{\bf f}^1\rangle$. The rest of the basis  $|{\bf f}^p\rangle$ with $p \neq 1$ are also solutions of  Eq. (\ref{710}) without external force. Nevertheless, there are not ``physical'' solutions for the interaction potential chosen.  Thus we shall call them $\it virtual$.

 $|{\bf f}^1\rangle$ verifies the following properties: (i) In a system with repulsive interaction,  as we consider here, all the contact forces are compressive and therefore  $ f_{ij}^1>0$ for all contacts. (ii) It is well known from simulations and experiments that the distribution of contact forces is roughly  exponential, or compressed exponential (see for example \cite{J} for simulations in the frictionless case). This implies that the fluctuations of the contact forces are of order of the average value, leading to $\langle f_{ij}^1\rangle^2\sim \langle (f_{ij}^1)^2\rangle=1/N_c$ for a normalized force field.  Thus we may introduce a constant $c_0$ such that:
\be
\label{coco}
\langle f_{ij}^1 \rangle = c_0 \frac{1}{\sqrt N_c}
\ee

Now we turn to the properties of  the virtual forces $|{\bf f}^p\rangle$: (i) There are no physical constraints on the sign of the contacts forces for the virtual vectors. Furthermore, the ${|\bf f}^p\rangle$ must be orthogonal to  $|{\bf f}^1\rangle$, whose signs of contact forces are strictly positive, and where the fluctuations in the contact compression is small. Therefore  the virtual force fields have  roughly as many compressive as tensile contacts.

\subsection{Implementation of global strain}

In our framework it turns out to be convenient to study the response to shear or compression as there are generally implemented in simulations. When periodic boundary conditions are used, an affine strain is first imposed on the system. Then the particles are let to relax. In general the affine strain is obtained by changing the boundary condition. Consider a 2-dimensional system with periodic boundary: it is a torus. For example a shear strain can be implemented by increasing one of the principal radii  of the torus and decreasing the other. Then the distance between particles in contact increases or decreases depending on the direction of the contact. In fact, this procedure of change of boundary conditions is formally equivalent to a change of the metric of the system. If the metric is changed from identity $ I$  to the constant metric $ G= I+U $, the length of a vector $\vec{\delta l_0}$ becomes $\delta l$, such that $\delta l^2= \vec{\delta l_0} \cdot G \cdot \vec{\delta l_0}$.  From this expression one deduces the change of distance between two particles is  given by the formula: 
\ba
\delta r_{ij}=\vec{n}_{ij}\cdot U \cdot \vec{R_{ij}}
\ea
Near jamming $\vec{R_{ij}}\approx \vec{n}_{ij}$ and therefore $\delta r_{ij}\approx\vec{n}_{ij}\cdot U \cdot \vec{n}_{ij}$. Formally,  such a change of metric is strictly equivalent to a change of the rest length of the interactions with $y_{ij}=\vec{n}_{ij}\cdot U \cdot \vec{n}_{ij}$. Incidentally  Eq.(\ref{rere}) can be used to compute the energy of such strain.

\subsection{Compression}
For a compression $U=-\epsilon  I$ where $ I $ is the identity matrix and $\epsilon$ is the magnitude of the strain. Eq.(\ref{rere}) becomes:
\be
\label{75}
\delta E=\frac{1}{2} (\sum_{ij}-\epsilon f_{ij}^1)^2 + \frac{1}{2} \sum_{p=2,..\frac{N\delta z}{2}} (-\epsilon \sum_{ij} f_{ij}^p )^2 
\ee
In the first sum all the terms have the same signs for a purely repulsive system, and this term leads to the strongest contribution. We have:

\be
\delta E\geq (\sum_{ij}-\epsilon f_{ij}^1)^2 = \epsilon^2 (N_c \langle f_{ij}\rangle)^2= \epsilon^2 c_0^2 N_c  
\ee

On the other hand, $\delta E$ is certainly smaller than an affine compression  whose energy also goes as $\epsilon^2 N$.  Therefore we find that:
\ba
\delta E \sim N \epsilon^2\\
B \equiv \frac{\delta E}{N \epsilon ^2}\sim \delta z^0
\ea
The bulk modulus of an harmonic system  jumps from 0 in the ``gas'' phase toward a constant when the system becomes jammed, as verified in the simulations. From this result follows that $p\sim (\phi-\phi_c)$. Note that this result holds only for purely repulsive systems. If there are as many tensile and compressive contacts, one recovers for the bulk modulus the result valid for the shear modulus, that we derive in the next section.

\subsection{Shear}
If a pure a shear strain is imposed, the tensor $U$ is  traceless. Let be $\epsilon$ the largest eigenvalue (in absolute value). The change of distance of two particles in contact due to shear $\delta r_{ij}= \vec{n}_{ij} \cdot U \cdot \vec{n}_{ij}$. It is a number of zero average if the system is isotropic, and fluctuates between  $+\epsilon$ and $-\epsilon$ depending on the orientation of $\vec{R}_{ij}$.  Eq.(\ref{rere}) becomes:
\be
\delta E= \frac{1}{2} \sum_{p=1,..\frac{N\delta z}{2}} ( \sum_{ij} f_{ij}^p \delta r_{ij})^2 
\ee
Each term in the summation gives on average:
\ba
\label{pg}
\langle(\sum_{ij} f_{ij}^p \delta r_{ij})^2\rangle = \sum_{ij} \langle (f_{ij}^p)^2 \delta r_{ij}^2\rangle+\sum_{mn\neq ij}  \langle f_{ij}^p f_{mn}^p\delta r_{ij} \delta r_{mn}\rangle \\
= \sum_{ij} \langle(f_{ij}^p)^2\rangle\langle \delta r_{ij}^2\rangle+\sum_{mn\neq ij} \langle f_{ij}^p f_{mn}^p\delta r_{ij} \delta r_{mn}\rangle
\ea

In principle they can be spatial correlations in the contact force amplitudes that leads to $\langle f_{ij}^p f_{mn}^p\delta r_{ij} \delta r_{mn}\rangle \neq 0$  even if $mn\neq ij$. Nevertheless we expect these terms to be negligible, as we argue at the end of the paragraph. Concerning the diagonal terms, one has $\delta r_{ij}^2 \approx \epsilon^2$ while $\sum (f_{ij}^p)^2=1$ by construction. Thus each term in the $p$ summation is of order $\epsilon^2 \cdot 1$, and:
\be
\delta E \sim N\delta z \epsilon^2
\ee 
which implies:
\be
\label{xixi}
G \equiv \frac{\delta E}{N \epsilon^2}\sim \delta z
\ee
This is in agreement with the observation of \cite{J}.

We come back to the subtle question of the possible presence of spatial force correlations.  In the simulation of soft spheres of \cite{J}, the spatial correlations of the real force field $|{\bf f}^1\rangle$   vanish for distance larger than roughly two particle diameters at the jamming transition.  This property is restituted by simple models of force propagation such as the scalar $q$-model  \cite{q}. In this model, the system is decomposed in layers, and forces propagate  downward from the top of the system to the bottom.  Each particle as 2d contacts, with d particles of the upper layer and d particles of  the layer below. The force field is builded recursively, following a local rule: when a particle receives from an upper layer a total force amplitude $f=f_1+...+f_d$, this force is distributed randomly to the d contacts below. This model mimics the fact that in an isostatic system external forces  imposed at the top of a system propagate downward, as we discussed in Chapter \ref{c2}. The randomness of the force splitting sketches the randomness of the local configuration of contacts. This model presumably captures some of the relevant physics, as it leads both to a rather realistic force distribution,  and does not display any spatial correlation in the force amplitude. Extending this model, we argue that when $\delta z>0$ the typical virtual force fields do not display long-range correlations. Indeed following the same line of thought a typical virtual force fields can also be builded recursively with a similar local rule,  the only difference being that some particles have more than 2d contacts. Thus the level of randomness increases as there are sometimes more ways to split the force between the different contacts below, accounting for the fact  that there are many possible virtual force fields. Such an increase of randomness will certainly not create long-range correlation, and this simple model justifies our assumption that such correlations are negligible. Note that this model does not preclude that a few force fields can display long-range correlations, but simply supports that such correlations do not occur for the bulk part of the set of force fields. In particular, we expect that when $\delta z>0$, the real force field $|{\bf f}^1\rangle$ displays long range correlations. Indeed this force field is solution of the whole elastic problem, and cannot be builded from any local rule. We expect that at large distances $r$, the correlation in force amplitude would follow $\langle f^1(0) f^2(r)\rangle \sim r^{-d}$ as in a continuous elastic medium with random disorder \cite{lubensky}. Accounting for this particular force field in Eq.(\ref{pg}) leads to a relative correction of $G$ that goes as $\frac{\ln(N)}{N}$ which vanishes at the thermodynamic limit. Finally, note that some subset of these virtual force fields were studied in simulations \cite{jocco1}, and that no long range correlations were noticed \cite{jocco2}.

\section{Discussion: non-affine displacements and length scales.}

An affine shear costs an energy comparable to a compression. Thus the non-affine displacements that follow  a pure shear diminish the cost in energy by a ratio that diverges at the jamming transition. Such non-affine displacements $|\delta \bf{R}_{n.a.}\rangle$ are simply the displacements corresponding to the minimization of Eq.(\ref{74}). It follows that  $|{\bf y}^\bot\rangle={\cal S}  |\delta \bf{R}_{n.a.}\rangle$. $\cal S$ can be made invertible if it is restricted to its image space, and we may write:
\be
\label{lin}
|\delta \bf{R}_{n.a.}\rangle= |{\cal S}^{-1}{\bf y}^\bot\rangle
\ee
A surprising recent observation was made in  the Lennard-Jones simulations of \cite{tanguy1,tanguy2}:  a rather large  length scale ($\sim 30$ particle sizes) appears in the correlations of the non-affine displacements.  Eq.(\ref{lin}) is a linear equation. Any strain $|\bf y\rangle$ can be decomposed as the sum of individual contact deformations. Therefore the non-linear displacements that follow a shear or a compression can be written as  the sum of  the responses to a contact deformation. The formalism of the present Chapter does not give any access to the spatial behavior of the response to a local deformation when $\delta z>0$ . On the other hand the results on the vibrational modes of Chapter \ref{c4} might bring interesting insights.  Since any deformation can be decomposed  on the vibrational modes, we expect the characteristic  lengths $l^*$ and $l_t$ to  characterize the non-affine displacements that follow a global strain. It would be useful to study these questions numerically near the jamming threshold.

\chapter{Granular matter and Glasses}
\label{c8}

In the previous  Chapters we discussed some properties of weakly-compressed harmonic soft spheres. Our main results are that (i)  anomalous modes appear at low-frequency. They  are related to the soft modes of floppy subsystems and are characterized by a length scale $l^*$ (ii) there is a frequency scale $\omega^*$ below which the system does not behave as a continuous medium, but as in an isostatic state. In this Chapter we study the applicability of these results to more realistic systems. We first discuss granular matter, then glasses. The sections contained in this Chapter, apart the last one, can be read independently since they deal with distinct matters.

At least two modifications are necessary to apply our results to granular matter: (i) the presence of friction. Up to now we studied mostly radial interactions, and covalent bonds  in Chapters \ref{c2} and \ref{c3}.   In  section \ref{s81} we shall see that there is no conceptual difference when friction is present, apart from  a change in the equations that define the soft modes. (ii)  the potential used: In three dimensions grains do not  interact harmonically. Modeling the contact with a Hertzian potential is more realistic, even thought it might not be perfect \cite{goddard}. It corresponds to $\alpha=5/2$ in Eq.(\ref{opo}). In  section \ref{s82} we compute the density of states for such non-harmonic interactions. Our results agrees with the numerical findings of \cite{J}.

Then we discuss the vibrations of glasses. All glasses present attractive forces, such as Van der Waals interactions. Therefore each particle a priori interacts with all other ones, and the coordination number is not-well defined. In section \ref{s83} we show how one can deal with this problem and compute the density of states in simple cases where there is a strong hierarchy in the contact stiffness. Our results apply for systems as  silica, where the covalent interactions inside the tetrahedron are the  strongest as discussed in Chapter \ref{c4}. In section \ref{s84} we consider systems without any clear-cut hierarchy in the distribution of  contact stiffness, such as Lennard-Jones systems. The question is to know under which conditions our results on the vibrations of weakly-connected system can apply in such situations.  We propose an improved variational argument which uses the distribution of the contact stiffnesses to evaluate the density of states. This leads to testable predictions on the nature of the excess-modes in such systems. In particular, this should enable to decide  whether or not the length scales that appears in the vibrations, the responses to a point force or the non-affine displacements in some Lennard-Jones simulations \cite{tanguy1,tanguy2,tanguy3,tanguy4}  corresponds to the length scales we introduced in this thesis. 

Let us specify that in this Chapter we shall not consider the effect of initial stress: this is a separated issue, which can be  treated independently, as shown in Chapter \ref{c5}.

\section{Particles with friction}
\label{s81}

In this section we show how our results can be extended to particles with friction. We shall assume that all the contacts lie {\it inside} the Coulomb cone: particles in contact do not slide irreversibly for an infinitesimal perturbation. This means that there is no plastic deformation at the surface of contact: the contacts act as if they were ``welded''.  Simulations showed that the validity of this assumption depends on the system preparation. The distribution of the contact tangential force was observed to vanish on the edges of the Coulomb cone in \cite{sil}. In \cite{zhang}, this was also observed if the dynamics that leads to jamming is fast. Otherwise, a finite fraction of the contacts was observed to lie on the Coulomb cone.   For concreteness we consider the two dimensional case where the notations are simpler. The same ideas apply in higher dimensions. We consider elastic  discs, whose  repulsive interaction is harmonic: $\alpha=2$. We add a term of friction,  function  of the shearing of the contacts. If two particles  are adjacent, and one of them rotates an angle $\delta \theta$ while the other is rigidly pinned, the energy stored can be written in the form:
\be
\label{pet}
\delta E =\gamma \delta\theta^2
\ee
$\gamma$ describe the energy associated with the shearing of a contact. This can be easily generalized in 3-dimensions \cite{mudlin}.    

If a small displacement $|{\bf \delta R}\rangle$ and a small rotation field ${\delta \cal \theta}=\{\delta \theta_i\}$ are imposed, the shear  between two particles in contact is characterized  by the following angle:
\be
\delta \theta_{ij}= \theta_i+\theta_j+(\delta\vec{R_j}-\delta\vec{R_i}).\vec{n}_{ij}^\bot
\ee
where $\vec{n}_{ij}^\bot$ is obtained by rotating  $\vec{n}_{ij}$ by $+\frac{\pi}{2}$. Therefore, using (\ref{pet}), we can write for the expansion of the energy:

\be
\label{exp}
\delta E = \sum_{ij}   \frac{1}{2} [(\delta\vec{R_j}-\delta\vec{R_i}).\vec{n}_{ij}]^2\\
+  \gamma  [\theta_i+\theta_j+(\delta\vec{R_j}-\delta\vec{R_i}).\vec{n}_{ij}^\bot]^2 +O(\delta\vec{R}^3)
\ee
 
For each contact $ij$ there is a compression force $f_{ij}$ and a tangential force $f_{ij}^\bot$. The system is at equilibrium when the force and the momentum are balanced on every particle:
\ba
\label{oyo}
\sum_{j}f_{ij} \vec{n}_{ij}+f_{ij}^\bot \vec{n}_{ij}^\bot= \vec{F}_i \hbox{ \ \ for all i}  \\
\sum_{j} f_{ij}^\bot= {\cal M}_i \hbox{\ \  for all i}
\ea
where ${\cal M}_i$ is the external momentum applied on particle $i$. This set of linear equation has $2 N_c$ degrees of freedom and $  3 N -3$ constraints in two dimensions. Therefore a jammed system with friction that can sustain any generic external force field must be such that $2 N_c > 3 N-3$. At the isostatic point the coordination number is given by:
\be
z_c= 2 N_c/N \rightarrow 3
\ee
In three dimensions one finds similarly $z_c=4$. The soft modes are modes of null energy.  In two dimensions Eq.(\ref{exp}) gives:
\ba
\label{gggg}
(\delta\vec{R_j}-\delta\vec{R_i}).\vec{n}_{ij}=0   \\
\theta_i+\theta_j+(\delta\vec{R_j}-\delta\vec{R_i}).\vec{n}_{ij}^\bot=0
\ea
Again this is a linear system. Now there are $ 3 N -3$ degrees of freedom, and $2N_c$ constraints. Following the procedure of Chapter \ref{c3} and \ref{c4} one can build anomalous modes that appear at a frequency $\omega^*\sim \delta z$, where $\delta z=z-z_c$, $z_c$ being the isostatic limit with friction. This result was recently observed numerically in \cite{somfai2}. Furthermore,  it is easy to show that this system (\ref{gggg}) is, as in the frictionless case, the dual of (\ref{oyo}). The results on the dipole of force and the scaling  of the elastic moduli of Chapter \ref{c7} can then be recovered.

Note that contrarily to the frictionless case,  this system does not need to be isostatic when the pressure vanishes. As we discussed in Chapter \ref{c2} and \ref{c3}, in frictionless systems  the two conditions that (i) the system must be rigid and (ii) the particles cannot interpenetrate lead to the unique solution for the coordination number: $z=z_c$.  In the frictional case, these two antagonist requirements do not lead to a unique solution. The geometrical requirement that forbids the interpenetration of particles  does not change, whereas  the condition of rigidity becomes less demanding, as we just showed. In two dimensions one finds $3\leq z\leq4$, and in three dimensions $4\leq z \leq6$. Consequently the coordination of stiff frictional particles depends on the preparation of the system \cite{zhang,sil}. In three dimensions when the friction coefficient $\mu \approx 1$ the simulations of \cite{sil} leads to $z\approx 4.5$ that implies $\delta z\approx 0.5$. We guess that this corresponds to a $l^*$ of few tens of particle sizes, which could be probed by computing the vibrational modes of this system.

\section{Extension to non-harmonic contacts}
\label{s82}
Up to now we considered harmonic interactions. Here we discuss the generalization of our argument to other contact potentials. Ref.\cite{J} explored several other potentials, in particular the Hertzian contact potential describing the compressive energy of two elastic bodies.  It corresponds to $\alpha = 5/2$ in Eq. (\ref{opo}).  Ref.\cite{J} observed a plateau in the density of states whose height $D_0$ scales as $p^{-1/6}$.  They also observed a cutoff frequency $\omega^*$ varying as $p^{\frac{1}{2}}$.  In the Hertzian case the quadratic energy of Eq. (\ref{2}) becomes:

\be
\delta E= \frac{1}{2} \sum_{\langle  ij \rangle} (1-r_{ij})^{\frac{1}{2}}[(\delta\vec{R_j}-\delta\vec{R_i}).\vec n_{ij}]^2 
\ee
The new factor $(1-r_{ij})^{\frac{1}{2}}$ amounts to a spring constant 
$k_{ij}$ that depends on compression.  The contact force $f_{ij} = 
\partial \delta E/\partial r_{ij}$ evidently varies as 
$(1-r_{ij})^{3/2}$. In what follows we start by neglecting the fluctuations that exist between the stiffnesses of the contacts. This treatment is sufficient to recover the scaling results of \cite{J}. Then we discuss how such fluctuations can be taken into account to improve the bound on the density of states derived in Chapter \ref{c3}.

The new factor  $V''(r_{ij})=(1-r_{ij})^{\frac{1}{2}}$  rescales the energy.  To account for this overall 
effect, we replace $(1-r_{ij})^{\frac{1}{2}}$ by its average 
$\langle (1-r_{ij})^{\frac{1}{2}}\rangle$.  Expressed in terms of contact 
forces, this factor is proportional to 
$\langle f_{ij}^{1/3}\rangle$.  Again replacing $f_{ij}$ by its 
average, the factor becomes $\langle f_{ij}\rangle^{1/3}$.  This 
average is related to the pressure $p$, via $ p\approx 
\langle f_{ij}\rangle$.  Thus in this approximation the overall effect is to rescale the 
energy by a factor $k(p) \sim p^{1/3}$. 
\be
\delta E= \frac{k(p)}{2} \sum_{\langle  ij \rangle} [(\delta\vec{R_j}-\delta\vec{R_i}).\vec n_{ij}]^2 
\ee

Apart from this prefactor, the energy and the dynamical matrix 
are identical to the harmonic case treated above.  Thus each normal 
mode  frequency gains a factor $k^{\frac{1}{2}} \sim p^{1/6}$.  In the 
harmonic case the crossover frequency follows $\omega^* \sim \delta z$.  In 
the Hertzian case, it too gains a factor $k^{\frac{1}{2}}$, so that $\omega^* 
\sim k^{\frac{1}{2}} \delta z$.  When the initial stress is taken into account, the bound on the lowest-frequency anomalous modes $\omega_{AM}$ still has the
form  
\be
\label{34}
\omega_{AM}{}^2 \leq  \omega^*{}^2 -A_2 p
\ee
For a marginally-stable system we still have $\omega_{AM}\approx 0$ which leads to an unaltered relationship between
$\omega^*$ and $p$, namely $\omega^* \sim p^{\frac{1}{2}}$.  Comparing with our previous estimate of $\omega^*$ yields $\delta z\sim p^{1/3}$. Furthermore, the plateau density of states $D_0$ has the dimensions an inverse frequency, and
 thus gains a factor $p^{-1/6}$.  Since the 
harmonic $D_0$ had no dependence on $p$, the Hertzian $D_0(p)$ also 
should vary as $p^{-1/6}$.  The scaling behaviors seen in \cite{J} agree 
with these expectations.  These arguments may be
applied to general values of the interaction exponent $\alpha$. 

Additional effects could in principle alter the low-frequency modes in the Hertzian case.  When harmonic springs are replaced by Hertzian springs, the contacts supporting different forces have different stiffnesses. The variational argument derived in Chapters \ref{c3} and \ref{c4} does not consider such fluctuations. If it is applied as is to the more general case with fluctuations, the energy of the corresponding anomalous modes defined in Eq.(\ref{tria})  simply gains a factor $\langle k \rangle$, the average stiffness. This can be deduced from Eq.(\ref{kk}) by neglecting the correlations between the soft modes displacements and the stiffnesses amplitudes\footnote[19]{We expect such correlations to be very small, as the amplitudes of the stiffnesses do not enter in the soft modes equation.}. Thus, this variational argument corresponds to the simple derivation of the previous paragraph where quantities are replaced by their average. Here we propose to use the stiffness fluctuations to improve this variational argument by a numerical factor.  Instead of modulating the soft modes by $\sin(\frac{x_i \pi}{L})$ to obtain the anomalous modes of Eq.(\ref{tria}), we introduce a more general phase $\psi(i)$ and  modulate the soft modes with $\sin(\psi(i))$. Then we minimize the average energy of these anomalous modes with respect to $\psi$, imposing $\psi=0$ and $\psi=\pi$ on the boundaries $x=0$ and $x=L$.  The expression of the energy corresponds to Eq.(\ref{kk}) with  $\sin(\psi(i))$ replacing  $\sin(\frac{x_i \pi}{L})$ and each term of the sum  multiplied by a stiffness $k_{ij}$. When averaged on the soft modes amplitude, one obtains the average energy of the anomalous modes  $\langle \delta E \rangle \sim\sum_{ij} k_{ij} \cos^2(\psi(i)) [\psi(i)-\psi(j)]^2$. This has to be minimized with respect to the variables $\psi(i)$. We expect the phase $\psi$ to vary slowly in space, and not to differ dramatically from our previous solution $\phi= \pi x/L$.  Hence for a large system we may consider that the cosines terms that appear in the  energy sum are constant locally. Then the local minimization of the energy corresponds to the problem of conductivity in a random network of resistors where $\delta E=\sum_{ij} r_{ij}{}^{-1}  [U(i)-U(j)]^2$, where $r_{ij}$ is the resistor between the vertex $i$ and $j$, and $U(i)$ is the Coulomb potential in $i$. Therefore $k_{ij}$ corresponds to the inverse of a resistance $r_{ij}$, and $\psi(i)$ to the potential $U(i)$.  Effective medium theory  \cite{choy} furnishes a good approximation of the conductivity, and therefore of the energy $\delta E$, of such a random system. For example  if the network of contacts is sketched by a square or cubic  lattice, one obtains for the effective stiffness the following equation:
\be
\label{eff}
\langle \frac{k_{eff}-k}{(d-1) k_{eff}+ k_{ij}} \rangle=0
\ee
where the average is taken on the distribution of stiffness.  Because of the convexity of the inverse function, one obtains that $k_{eff} \leq \langle k \rangle$.  Expectedly, the energy in this improved variational argument is smaller than in our previous result, since it is proportional to  $k_{eff}$ instead of $\langle k \rangle$.  For the distributions of stiffnesses we are considering here \footnote[20]{As we discussed in Chapter \ref{c7}, the potential does not enter in the computation of the force field at the isostatic threshold. Thus if the configurations obtained at  threshold are similar for the different potentials $\alpha$, as it seems to be the case, the same force distribution is obtained for any $\alpha$. Such force distribution is known to be roughly exponential. From this we can deduce the distribution of stiffnesses $P(k)\sim k^{\frac{1}{\alpha-2}}e^{-k^\frac{\alpha-1}{\alpha-2}}$.}  where we expect no delta function in $k=0$, nor any fat tail at large $k$, $k_{eff}$ as computed in Eq.(\ref{eff}) and $\langle k \rangle$ are of the same order of magnitude. Thus the bound on the density of states is improved by a numerical factor when this improved argument is used.

\section{$D(\omega)$ in systems with various interaction types}
\label{s83}
In this section we study the density of states of systems where several types of interactions are at play, and  where the amplitude of the strongest interaction is much larger than the others. In particular we think about  tetrahedral covalent networks such as silica. The strong interaction corresponds to the deformation of the tetrahedra, which is much larger than the bending of the Si-o-Si  bonds or the Van der Waals attractions, as we discussed in Chapter \ref{c3}.   In Chapter \ref{c3} and \ref{c4} we showed how to compute the density of states of weakly-connected networks with only one type of interaction:  no weak interactions were present. The density of states is then described by a plateau that appears above a frequency $\omega^*\sim \delta z$. In this section we show how to treat the weak interactions by simple perturbation. We simply consider the anomalous modes that would appear at $\omega^*$ if the weak interactions were not there. Then we compute the change in energy induced by the weak interactions on these anomalous modes. If these modes are on average shifted by an energy $\zeta$, the plateau in the density of states appears at a frequency $\sqrt{\omega^*{}^2+\zeta}$. In what follows we show how to evaluate $\zeta$ and discuss some examples.

\subsection{Model}
Consider for concreteness  a system with radial interactions, where ``strong'' interactions with  stiffness close to unity form a rigid network of coordination $z>z_c$, and where the weak interactions have stiffnesses of order $\eta\ll  1$. Consider the lowest-frequency anomalous modes that appear at $\omega^*$ in the rigid network of coordination $z$.  The energy correction $\zeta$ induced by the weak interactions on such mode is:

\be
\zeta= \frac{1}{2} \sum_{\langle ij\rangle}\eta_{ij}[(\delta \vec{R}_i-\delta\vec {R}_j)\cdot \vec {n}_{ij}]^2
\ee
where the sum is taken on all pairs of  particles interacting with the weak interaction of stiffness $\eta_{ij}$. Here  we consider that all pairs of particles $<ij>$ that appears in this sum are not in contact through the strong interaction, since such terms would simply re-normalize the strong stiffnesses, and can be taken into account when $\omega^*$ is computed in the first place.  To estimate $\langle \zeta \rangle$ we need to compute $\langle [(\delta \vec{R}_i-\delta\vec {R}_j)\cdot \vec {n}_{ij}]^2\rangle$. If the displacement of particles $i$ and $j$ were uncorrelated, one would get for a normalized mode $ \langle [(\delta \vec{R}_i-\delta\vec {R}_j)\cdot \vec {n}_{ij}]^2\rangle=\frac{2}{d N}$. We expect this to be the case when the distance $r_{ij}$ between the two particles is large. When the distance between the pair $ij$ diminishes, the displacement  of particles $i$ and $j$ becomes correlated. Thus we may write $\langle [(\delta \vec{R}_i-\delta\vec {R}_j)\cdot \vec {n}_{ij}]^2\rangle =\frac{2 c(r_{ij})}{d N}$, where $c(r_{ij})<1$ characterizes the correlations at distance $r_{ij}$. Since we are only considering pairs of particles which are not in contact for the rigid network, we expect that  for the anomalous modes of such a network considered here, the correlation of the displacement of particles $i$ and $j$ is ``weak", that is $c(r_{ij})$ is bounded below by a constant $c_0$ of order one. This is supported by our numerical result of Fig.(\ref{trans}), that shows that if two displacement degrees of freedom are not fixed by the soft mode condition, then the correlation between these displacements is weak, even when these two particles are close. At the level of approximation we are considering here we shall simply assume that $c(r_{ij})=1$. Then we obtain:

\be
\label{ee}
\zeta= \sum_{ij}\eta_{ij} \frac{1}{dN}\equiv\frac{1}{d}\int \rho(\eta) d\eta
\ee
where we introduced the number $\rho(\eta)$ of pairs with stiffness $\eta$ per particle per unit stiffness.  

\subsection{Density of states of square and cubic lattices}

\begin{figure}
\centering
\includegraphics[angle=0,width=10cm]{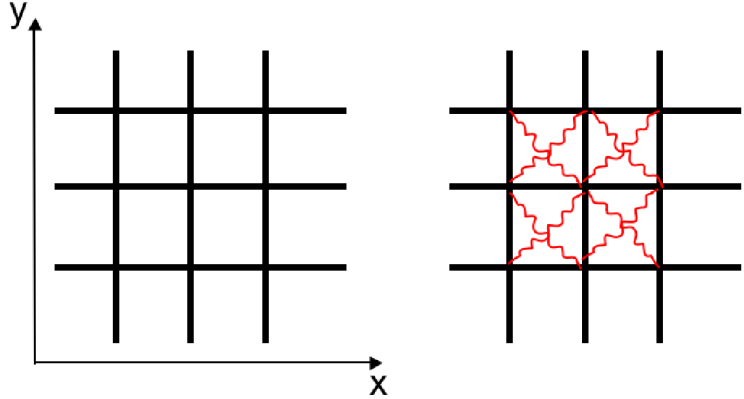}
\caption{Left: Square lattice with interaction between first neighbor of stiffness 1. Right: the same lattice where interactions of stiffness $\eta\ll 1$ have been added between second neighbors.}
\label{square}
\end{figure}

In what follows we discuss in particular the situation where the rigid network is isostatic $z=z_c$. Our present argument shows that a plateau appears in the density of states at $\omega\sim \sqrt \zeta$. To illustrate this idea we consider the example of the square (or cubic) lattice of point particles where first neighbors interact with spring at rest, see Fig.(\ref{square}). Our argument of Chapter \ref{c3}, which does not depend on disorder but only on coordination, applies to this system. The square and  the cubic lattice are isostatic, therefore there density of states is constant. In this particular case the soft modes are trivial one-dimensional object: they are the independent translations of the columns. Thus for these systems a simpler  way to prove that the density of states is constant, instead of going through the argument of Chapter \ref{c3}, is  to consider such lattices as an assembly of disconnected one dimensional lattices, which obviously have a constant density of states. One may ask what happens to the density of states of such system if a weak interaction is added to couple these independent lines.  As an example, we consider that the second neighbors are now coupled: we add springs with small stiffness $\eta$ as drawn in Fig.(\ref{square}).   $\rho$ is then a delta function at stiffness $\eta$, of amplitude 2 if one considers the square lattice  (there are two weak interactions per particles). Therefore our rough evaluation for the appearance of the plateau of Eq.(\ref{ee}) yields $\omega=\sqrt \eta$. Since this system is crystalline, the normal modes are plane waves and the density of states can  be computed exactly. The density of states display a cross over between a Debye regime and a plateau. A direct estimation of the cross-over energy gives $\delta E\approx2\eta$ which leads to $\omega=\sqrt {2\eta}$ \footnote[21]{The frequency of a plane wave of wave vector $\vec \kappa$ can be written $\omega\sim a(\vec{\kappa}/\kappa) \sin(\kappa)$, where the function  $a(\vec{\kappa}/\kappa)$ only depends on the direction of the wave vector. If the sine is approximated by an affine function on obtains $\omega\sim a(\vec{\kappa}/\kappa) \kappa$.   Then it is straightforward to show, by summing on all directions  $\vec{\kappa}/\kappa$, that in this approximation the density of states is linear (as expected for a 2-dimensional crystal) until a frequency $\omega_0$, defined as the lowest frequency at which some plane waves reach the  edge of the Brillouin zone. These plane waves have the smallest  $a(\vec{\kappa}/\kappa)$. In this  example they correspond to the transverse waves whose wave vector is in the direction of the axis $x$ or $y$ as indicated in Fig.(\ref{square}). When the wave vector reach the boundary of the Brillouin zone, the energy of such waves can be computed. One finds $E_0=2\eta$.}. A cross-over energy is arbitrarily defined, here we simply note that our two estimates are similar. Finally, note that our argument indicates that the cubic lattice are very floppy if the second neighbor interactions are small in comparison  with those of the first neighbors. This happens if the interactions potential decays  rapidly. In practice, there are no simple elements that crystallize in a simple cubic crystal \cite{kittel} because such structures are too floppy. For example it is not possible to crystallize a 12-6 Lennard-Jonnes in a square lattice even at zero pressure and zero temperature, because such a structure is mechanically unstable \footnote[22]{Imposing that the pressure is zero brings a condition between the forces carried by the first neighbors and by the second neighbors. In a 12-6 Lennard-Jonnes it can be satisfied only if the second neighbors are located behind the inflection point of the potential. Therefore the second neighbors have a negative stiffness, which destabilizes the marginally stable network constituted by  the first neighbors.}.  In general simple cubic crystals have charged particles, as in NaCl, and the second neighbors interactions are not negligible.

\subsection{The boson peak of silica}

Silica is the most common glass,  much studied in experiments and in simulations. It is also known to have one of the strongest  excess of low-frequency modes, or boson peak, see \cite{nakayama}  for a review of empirical results and models. In this paragraph we propose an explanation for its density of states at low-frequency, and for the nature of the excess-modes. As we discussed in Chapter \ref{c4}, the strongest  interactions in silica lies inside  the tetrahedron SiO$_4$. If the other interactions are neglected, one can model the glass  as an assembly  of linked tetrahedra of appropriate stiffness loosely connected at corners,  the ``rigid unit modes'' model \cite{heine}. This description is supported by recent Hyper-Raman scattering experimental results that show that the low-frequency excess modes correspond to the motion and rotation of stiff tetrahedra \cite{hehlen}. As we discussed in Chapter \ref{c4} such a tetrahedral network has a constant density of states at low frequency.  Following the results of the last section, one may evaluate the effect of the weaker interactions (in particular  the bending of the Si-o-Si bond and the  Van der Waals interaction) on the density of states by computing the energy shift $\zeta$ that they induce on the anomalous modes. $\zeta$ can in principle be estimated from the knowledge of the distribution of the stiffnesses of the weak interactions $\rho$. Using the stiffness of the Si-O-Si bending interaction obtained {\it ab initio} \cite{ff}, and the molecular mass, we obtain a frequency 1.4 Thz, which estimates  the order of magnitude of $\sqrt \zeta$. According to the previous arguments of the present section,  we expect thus  silica glass to display a plateau in its density of states that should appear at a frequency of the order of 1 Thz. This is indeed what is observed in simulations: silica glass present a well-defined plateau in the density of states, which appears at a frequency corresponding to the boson peak, see e.g.  Fig.15 of \cite{kob2} or  Fig.(\ref{cry}) for numerical results.  

Our argument does not involve disorder. Thus it must also apply for the crystals of the same composition and similar densities such as the $\alpha$ and $\beta$-cristobalite, since these crystalline structures are formed, as silica,  by  SiO$_4$ tetrahedra connected at the corners.  $\beta$-cristobalite has the structure of the diamond, in which the tetrahedra correspond to the 4 carbons bonded  to a central carbon, whereas  $\alpha$-cristobalite has a tetragonal structure. Empirically a boson peak is observed in all these materials \cite{orenbarre}. Numerically, a plateau indeed appears in $D(\omega)$ at roughly the same frequency in the cristobalite $\alpha$ and $\beta$ \cite{kostya} and in the glass, as shown in Fig.(\ref{cry}).

 More generally, there are crystals showing an excess  density of states at frequencies that correspond to the typical boson peak frequency in glasses \cite{nakayama,leadbetter,caplin,bilir}. This implies that the disorder is not a necessary condition to obtain excess modes. This is a crucial point as  in many theories of the boson peak disorder is the only relevant parameter. Such theory certainly cannot explain the density of states of silica, very similar to the cristobalites.  Rather, we argue here that the key feature that determines the density of states is weak coordination.
 
Note that if disorder is not relevant to compute the density of states, it affects the nature of the vibrational modes.  For example properties such as transport are very different between  silica glass, and the cristobalites. Thus the peculiarity of the amorphous state  lies in the {\it nature} of the excess-modes,  not in the density of states \cite{nakayama}. It is useful to note the  parallel between cristobalite and  silica glass on the one hand and cubic lattice and the jamming threshold of elastic spheres on the other hand.  In both cases the amorphous solid and the crystal have a similar density of states, but the anomalous modes in the amorphous phase are not plane waves. Disorder strongly affects the anomalous modes, and makes them very heterogeneous,
as it appears in Fig{\ref{softfig}. 

\begin{figure}
\centering
\includegraphics[angle=0,width=13cm]{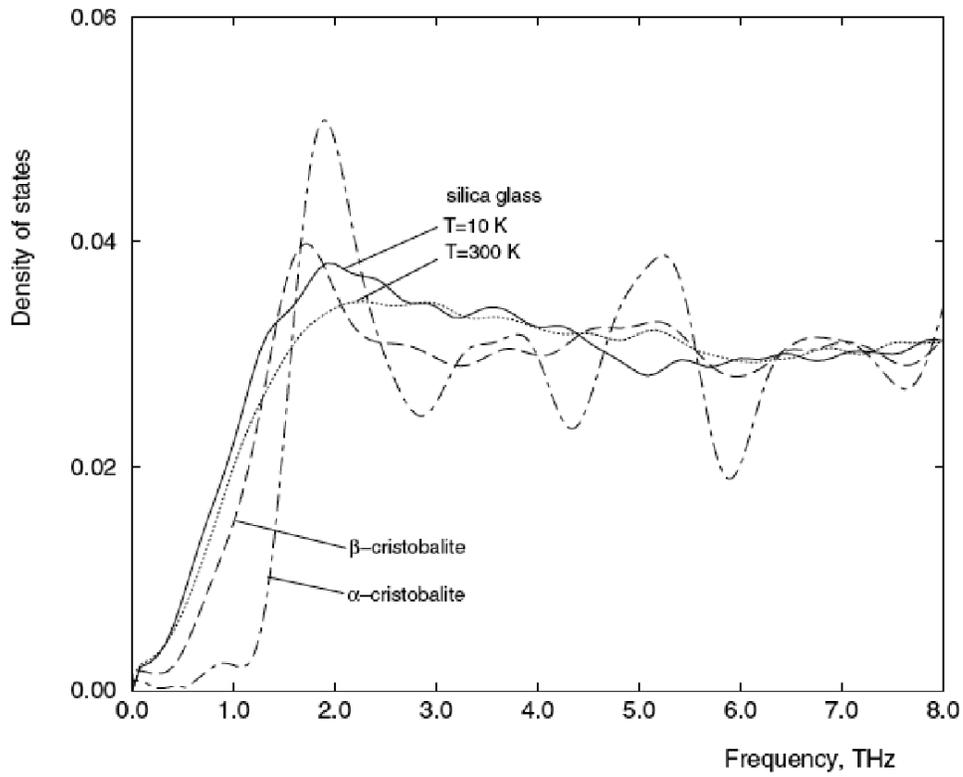}
\caption{Density of states of silica glass (at temperatures of 10 and 300 Kelvins), $\alpha$-cristobalite and $\beta$-cristobalite. This figure is taken from the simulations of \cite{kostya}.  }
\label{cry}
\end{figure} 

There is an interesting geometrical parallel  between a tetrahedral network and an assembly of spherical particles with friction. The equation that defines the soft modes of these two systems are the same, and corresponds to the extension of Eq.(\ref{gggg}) in three dimensions.   This comes from the fact that when friction is present, the contacts can be considered as  ``welded'', as we discussed in \ref{s81}. Similarly in a tetrahedral network the corners of each tetrahedra in contact are attached. In both cases the soft modes corresponds to the rigid motion  of the particles  that keep the contact point  welded. 

Finally, we considered systems in which the  strong interactions form a rigid network, that is $\delta z> 0$. Our finding is  that the density of states can be understood in terms of the anomalous modes that we  introduced in Eq.(\ref{tria}). Note that if the concept of such anomalous mode is new, soft modes have been used in several different fields as we discussed in  Chapter \ref{c2}. When the network of strong interactions is not rigid, $\delta z<0$, and the stiff network has got some  soft modes. Consequently there is delta function in the density of states at  $\omega=0$ if the weak interactions are neglected. When the weak interactions are taken into account, Thorpe and collaborators  \cite{t} argued that these soft modes shift to higher frequency (in our notation by an amount $\sqrt \zeta$), leading to a peak in the density of states. The existence of such peak was  observed in neutron scattering data of Ge$_x$Se$_{1-x}$ glasses.  When the composition of this glass is changed as $x$ decreases, the covalent network becomes less and less connected. At $x=0.24$ it is isostatic. The peak amplitude was observed to diminish as $x$ decreases toward 0.24 \cite{kami,t}, as predicted.

\section{Lennard-Jones systems}
\label{s84}

In this section we discuss the vibrations of systems where (i) coordination is not well defined a priori and (ii) there is not a clear-cut hierarchy in the amplitudes of interactions. We have  in mind Lennard-Jones systems, or purely repulsive system with a potential of the form $V(r)=r^{-\beta}$.  In particular, we discuss under which conditions   such systems should display the anomalous modes we introduced in Chapters \ref{c3}. One important  motivation is the recent results on Lennard-Jones systems \cite{tanguy1,tanguy2,tanguy3,tanguy4} where excess-modes are observed, and where a characteristic length scale appears, in particular in the vibrations of the system and in the response to a local force. It is tempting  to associate the excess-modes of \cite{tanguy1} to the anomalous modes, and the length scale observed by the authors   to the lengths we introduced in the previous Chapters. In this section we propose an improved variational argument, in the spirit of the  last section, to evaluate the density of states of such systems. We shall argue that this variational method is efficient if the interaction potential decays rapidly. In this case anomalous modes appear characterized by a length scale. We propose a numerical test to check whether or not the excess-modes found in the Lennard-Jones system of \cite{tanguy1} correspond to the anomalous modes we discussed. 

For concreteness we consider a system with radial interactions, and introduce the density of stiffness $P(k)$, which denotes the number of pairs of particles whose interaction has a stiffness $k$ per particle per unit stiffness. In the harmonic soft sphere system that we considered in the previous Chapters $P(k)=\frac{z}{2} \delta(k=1)$.  In the previous section, $P(k)$ could be decomposed in two well-separated distributions, one around $k=1$ of amplitude $z/2$ for the strong interactions, and one at much weaker $k$ for the weak ones. Now we consider a general distribution $P(k)$.  We propose the following family of variational arguments to define an effective coordination and evaluate the density of states. Let us introduce an artificial cut off $\lambda>0$, such that the coordination number $z(\lambda)=z_c+\lambda$.  We use this cut-off to decompose the interactions in two groups: the $z(\lambda)/2$ strongest ones, which form a network of `` strong'' interactions of coordination $z(\lambda)$, and the rest. This allows us to repeat the argument of the previous section: we first compute the energy $\omega^*{}^2(\lambda)$ of the low-frequency anomalous modes of the ``strong" network. Then we evaluate the correction in energy $\zeta(\lambda)$ induced by the rest of the interaction on these anomalous modes, as computed in the last section. Finally, one obtains the energy $E(\lambda)=\omega^*{}^2(\lambda)+\zeta(\lambda)$ which gives a bound of the energy of appearance of the excess-modes. Then the standard procedure consists in minimizing  $E(\lambda)$ with respect to $\lambda$. If $\lambda_0$ minimizes this quantity, then $\sqrt{E(\lambda_0)}$ gives the best estimate of the frequency of the anomalous modes. Furthermore, one can define an effective coordination $z=z_c+\lambda_0$. The length that characterizes   the anomalous modes follows, according to Chapter \ref{c4}, $l^*\sim \lambda_0^{-1}$. In what follows we  compute $E(\lambda)$ and $\lambda_0$, discuss  the quality of the estimate obtained by this argument, and compute other quantities that should enable to test if the excess-modes of Lennard-Jones systems are correctly described in terms of anomalous modes.

We first evaluate $\omega^*{}^2(\lambda)$.  In the harmonic case where all the stiffnesses are 1, we had $\omega^*{}^2= A_1 \delta z$, where $A_1$ can be deduced numerically from the data of \cite{J}. From Fig.\ref{sans} one has in three dimensions $A_1\approx 0.12$.  In the network of coordination $z(\lambda)$, the amplitude of the stiffnesses fluctuates. To present this argument in its simplest form, we neglect these fluctuations, and replace the stiffness of each contact by the average stiffness as we discussed in the first part of section \ref{s82}. We also showed there how a more accurate argument could be made, and we shall come back to this issue later. From our approximation follows that $\delta E\sim \langle k \rangle$, and in our notation  we obtain $\omega^*{}^2(\lambda)=A_1^2 \langle k\rangle_\lambda \lambda^2$, where  $ \langle k \rangle_\lambda=\frac{2}{z(\lambda)}\int_{k(\lambda)}^\infty k P(k) dk$.   $k(\lambda)$ is the weakest stiffness of the network of ``strong'' interactions, defined as:
\be
\label{kl}
\int_{k(\lambda)}^\infty P(k) dk= \frac{z(\lambda)}{2}
\ee. 

We now compute  $E(\lambda)$ and $\lambda_0$. From Eq.(\ref{ee}) we get for the correction in energy $\zeta$:  $\zeta(\lambda)=\frac{1}{d} \int_0^{k(\lambda)}P(k) k dk$, and finally:
\be
E(\lambda)=A_1^2 \langle k\rangle_\lambda \lambda^2+\frac{1}{d} \int_0^{k(\lambda)} P(k) k dk
\ee
The effective extra-coordination $\lambda_0$ is defined as $\frac{dE(\lambda)}{d\lambda}|_{\lambda=\lambda_0}=0$. This equation defines a non-trivial minimum in general, since (i) $\frac{dE(\lambda)}{d\lambda}|_{\lambda=0}=-  k(0)<0$ as required by mechanical stability, and (ii) if the potential decays reasonably  fast with distance, $E(\lambda)\sim \langle k\rangle_\lambda \lambda^2 \sim \lambda \rightarrow \infty$ as $\lambda\rightarrow \infty$. Using Eq.(\ref{kl}), the minimization of $E(\lambda)$  leads to:
\be
\label{zero}
2\lambda A_1^2 \langle k\rangle_{\lambda_0}+A_1^2 \lambda_0^2 \frac{d\langle k\rangle_\lambda}{d\lambda}=\frac{1}{2d} k(\lambda_0)
\ee
A necessary condition for this variational argument to be relevant is that the solution $\lambda_0$ of Eq.(\ref{zero}) is small (at least smaller than the end of validity of the scaling of Fig.\ref{sans}, say  $\lambda_0<2.5$ in three dimensions). Assuming that it is the case (and checking it self-consistently later), one may neglect the term quadratic in $\lambda^2$ in Eq.(\ref{zero}) to find:
\be
\lambda_0=\frac{1}{4dA_1^2}  \frac{k(\lambda_0)}{\langle k\rangle_{\lambda_0}}
\ee
In three dimension using the value of $A_1$ one gets $\lambda_0\approx 5  \frac{k(\lambda_0)}{\langle k\rangle_{\lambda_0}}$. Thus a first requirement for this variational argument to apply  is that $k(\lambda_0)$ must be several times smaller than $\langle k \rangle_{\lambda_0}$. A second requirement is that the bound of the energy $E(\lambda_0)$ must be relatively small in comparison with the Debye energy $E_D=\omega_D^2$: if this bound is too large, our variational argument does not lead to a correct estimate of the density of states, and the vibrational modes are not well described by anomalous modes. This implies that the correction in energy $\zeta(\lambda)$ must not be too large. It is clear that these two conditions  are satisfied when the interaction potential decays fast enough. Consider for example a potential of the form $V(r)=r^{-\beta}$.  If the potential decays slowly, $\beta>0$ is small, then the two requirements we just evoked are not satisfied. In three dimensions for example there are on average 12 neighbors surrounding a particle. If $\beta$ is small they all interact with a similar stiffness with the particle lying at the center, and one finds that $k(\lambda_0)$ is close to $\langle k \rangle_{\lambda_0}$, and not several times smaller. Furthermore the interactions with the second, third, etc... neighbors are not negligible either, and $\zeta$ is large. Thus our variational argument is not relevant in this case, and we expect the vibrations of such system to correspond to plane waves, and not to display excess-modes.  If $\beta$ is large enough, then the two requirements are satisfied if the system is amorphous. To show that we assume that $g(r)$ does not evolve much with the potential considered, as it is generally observed with radial interactions in the glass phase. Then, when $\beta$ becomes large, the hierarchy among the stiffnesses diverges: if one particles has 2 neighbors at distance $r_1$ and $r_2$ with $r_1<r_2$, the ratio of the stiffnesses of these two interactions is $\frac{k_1}{k_2}=(\frac{r_2}{r_1})^{\beta+2}$ which diverges when $\beta$ increases. It follows that $\zeta$ becomes negligible, and that $\lambda_0\sim \frac{k(\lambda_0)}{\langle k\rangle_{\lambda_0}}\rightarrow 0$ as $\beta\rightarrow \infty$. 

To decide whether or not the anomalous modes that appear in this variational argument are responsible for the excess-modes observed in Lennard-Jones systems, such as those of \cite{tanguy1},  one may consider a more precise observable that characterizes  the anomalous modes. The quadratic energy of a mode  is the sum the energy condensed in every contact $\delta E(k_{ij})= \frac{1}{2}k_{ij}[(\delta \vec{R}_i-\delta\vec {R}_j)\cdot \vec {n}_{ij}]^2$. Consider the average energy condensed in a contact of stiffness $k$, that we denote $\langle \delta E(k)\rangle$,  where the average is taken on all the contacts of similar stiffness $k$. According to our variational argument,  for the lowest-frequency  anomalous modes $\langle \delta E(k)\rangle$ varies as follow: if $k<k(\lambda_0)$, the displacement of the particles in contact are not correlated  and  $\langle \delta E(k)\rangle=\frac{1}{d} k$. If $k>k(\lambda_0)$, the interaction belongs to the rigid network. Then the relative longitudinal displacement between particles in contact corresponds to the modulation by a sine on a length scale  $l^*\sim \lambda_0^{-1}$, therefore $\langle \delta E(k)\rangle\sim k \lambda_0^2$.  Thus the curve $\frac{\langle \delta E(k)\rangle}{k}$ is a step function, which jumps at $k=k(\lambda_o)$, as represented in Fig.(\ref{step}). In the next paragraph we shall argue that this result is not exact and that the step is not sharp, but smooth. In any case,  if the excess-modes observed in Lennard-Jones systems are related  to the anomalous modes describe here, they should exhibit this cross-over. This could be verified numerically. Furthermore, from this curve  $k(\lambda_0)$ can be characterized: it is  the stiffness at which the cross-over takes place. From $k(\lambda_0)$ one can compute the effective coordination $\lambda_0$ if the distribution of stiffness $P(k)$ is known. 

\begin{figure}
\centering
\includegraphics[angle=0,width=13cm]{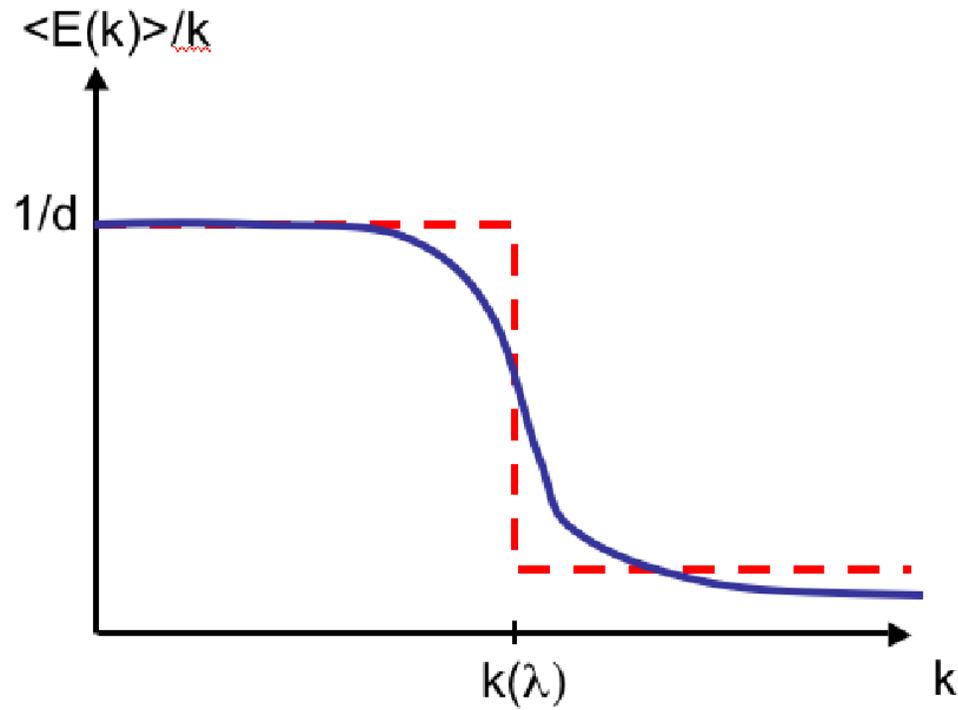}
\caption{$\langle \delta E(k)\rangle/k$ as defined in the text {\it vs} the stiffness $k$ for an anomalous mode. The red dashed curve corresponds to the variational argument presented in the text. The black continuous curve is what we expect qualitatively when the variational argument is improved. }
\label{step}
\end{figure} 
 
To conclude, we expect that the present variational argument could be improved in several ways. For example the fluctuations of stiffnesses could be taken into account to estimate $\omega^*$, as discussed  at the end of section \ref{s82}. This leads to corrections on the quantity $\langle \delta E(k)\rangle/k$ for $k>k(\lambda_0)$ that can be estimated using effective medium theory. One finds for the anomalous modes that  the quantity  $\langle \delta E(k)\rangle/k$ is not constant for $k>k(\lambda_0)$, but decreases with $k$, as represented in the full line curve of Fig.(\ref{step}). Such effects smooth the step function drawn in Fig.(\ref{step}).

\chapter{Rigidity of  hard sphere liquids near the jamming threshold}
\label{c9}

There is something mystifying about the properties of a hard sphere
system. On the one hand it is a simple system: no energy is involved,
the only rule is that particles cannot  interpenetrate,  and only
entropy matters. On the other hand, it displays a rich phenomenology. When
the packing fraction $\phi$ is slowly increased from the liquid phase,  it
crystallizes. If the $\phi$ is increased rapidly,
crystallization is avoided, and a glass transition is observed. The
 time $\tau_\alpha$ that characterizes the de-correlations of the density fluctuations at some vector $q$
 grows rapidly. For  a 3-dimensional monodisperse system $\tau_\alpha$ becomes
 inaccessible numerically above  $\phi_0\approx 0.59$. Nevertheless the packing fraction can be increased further until the pressure diverges, when the distance between particles vanish: it takes place at $\phi_c\approx 0.64$, the random close packing.  At  packing fractions between $\phi_0$ and $\phi_c$
 the structure of the system is frozen, apart from the fast rattling of the particles around their average position.
 
 The mode coupling theory furnishes predictions for the relaxation of the density fluctuations near the glass transition in rather good agreement with empirical data \cite{barrat}. Nevertheless there is no clear understanding of the spatial nature of the events that relax the system, neither of  the heterogeneous dynamics that has been observed near $\phi_0$. A necessary first step to study such questions is  to understand  the cause of the rigidity of a hard sphere system at times scales $t\ll  \tau_{\alpha}$. In the conventional picture \cite{sjorgen}, the freezing of the liquid at times  $t\ll \tau_{\alpha} $ is interpreted  with the ``cage'' effect. As the density increases, the cages formed by the neighboring particles tighten, and the characteristic time for a particle to escape its cage increases. This description considers the stability toward the motion of one single particle. It is dangerous since, as we  discussed in the previous Chapters,  the  stability
 against {\it collective} motions of particles is more demanding than the
 stability against individual particles displacements. For example  in $d$ dimensions $d+1$ particles are enough to
 pin one particle. Nevertheless, a system with a coordination number $d+1$ is unstable to cooperative motion, as shown by the Maxwell criterion.

 In this Chapter we study the
 microscopic cause of the glass rigidity. In particular we study the hard sphere glass at packing fraction $\phi$ near
 $\phi_c$, and we derive some elastic properties of the glass phase.
 Following the ideas of the previous chapters, we show
 that the solid-like behavior at $t\ll  \tau_\alpha$  requires the
 formation of a rigid structure  with a sufficiently large coordination number. 
We argue that such structure corresponds to the network that carries momentum between the particles, which was  introduced recently to study granular flows \cite{bubble}.   Our main achievement is to compute an effective potential between particles in contact through this network.  We show theoretically, after averaging over the fast fluctuations or rattling of the particles, that the hard sphere potential becomes logarithmic. This result is exact at $\phi_c$, and is in very good agreement with a direct numerical check.  This allows us to define normal modes and to apply  the results we found for soft  sphere solids near the jamming threshold.  In particular, the extended Maxwell criterion applies, which yields  an inequality for the network coordination. We confirm this result  numerically. We compute the scaling of the high-frequency elastic moduli near $\phi_c$. More generally, this approach shows that the jamming threshold acts as a critical point both in the solid and in the glassy liquid phase. This suggests original relaxation processes in the super-cooled phase.

\section{Coordination number and force}

For concreteness we consider a hard sphere system without dissipation where particles collide elastically. We show in the next section how to generalize our finding to the dissipative case where particles follow Brownian motion. We consider packing fractions $\phi$ such that  the typical collision time between two neighbors $\tau_c$ is much smaller than $\tau_\alpha$. We introduce an arbitrary time $t_1$, much larger than the collision time, and much shorter than any time scales at which the structural relaxation occurs. Two particles are said to be in contact at a time $t$ if they collide at least once in the time interval $[t-t_1/2,t+t_1/2]$. This enables us to define the coordination number $z$ as $z=2N_c/N$, where $N_c$ is the total number of contacts. Then, we define the contact force $\vec{f}_{ij}$ as the total momentum per unit time exchanged between the two particles:

\be
\label{imp}
\vec{f}_{ij}=\frac{1}{t_1}\sum_{n=1}^{n=n_{col}} \Delta \vec{P}_n  
\ee
where the sum is made on the total number of collisions $n_{col}$ between $i$ and $j$ that took place in the time interval $[t-t_1/2,t+t_1/2]$, and $\Delta \vec{P}_n$ is the momentum exchanged at the nth shock. These definitions were first introduced in a  work on dense granular flows \cite{bubble}, and were used recently on hard sphere systems \cite{donev,donev2}. Coordination and contact forces depend a priori on an arbitrary parameter $t_1$. In the high-packing fraction system we studied, we did not observe any relevant dependence of these objects with $t_1$ as long as $\tau_c\ll t_1\ll  \tau_\alpha$ \footnote[22]{If $t_1$ is too small some contacts can disappear, which can lead to the appearance of unstable modes when they are computed following the procedure bellow, even thought the system is stable}.

In Fig(\ref{forcefield}) we show a two-dimensional  example of the
 contact force field obtained with such procedure at packing fraction
 $\phi$ very close to $\phi_c$.  Note that the forces are roughly balanced on every particle, as it
 must be the case  on time scales over which the structure is stable.
  
 To obtain  high packing fraction configurations such as the one of
 Fig.(\ref{forcefield}), we proceed as follows: we consider the
 two-dimensional  polydisperse \footnote[23]{Half of the particles
 have a diameter unity, the other half have a diameter 1.4.}
 configurations of \cite{J} at the jamming threshold at packing
 fraction $\phi_c\approx 0.83$. At this packing fraction the particles are in contact. Then we reduce all the particles diameter
 of a relative amount $\epsilon$.  This leads to configuration of
 packing fraction $\phi=\phi_c(1- \epsilon)^2$.  We assign a random velocity to every particle. A Newtownian dynamics  is then computed using an event-driven simulation.
 As we shall discuss later, such a protocol does not lead to a system at thermal equilibrium. Nevertheless, we are not interested in thermodynamic properties, and  in practice  systems with such high packing fraction are never equilibrated in any reasonable time. We rather aim to study the conditions that guarantee mechanical stability.  Such condition should be fulfilled whether the system is at thermal equilibrium or not, as long as it is stable on reasonably long time scales.
 
 Note that since there is no energy involved in such system, the temperature only fixes the time unit. In what follows we impose for the average square velocity $\langle v^2 \rangle=1$.

\begin{figure}[htbp]
\begin{center}
\rotatebox{0}{\resizebox{5.5cm}{!}{\includegraphics{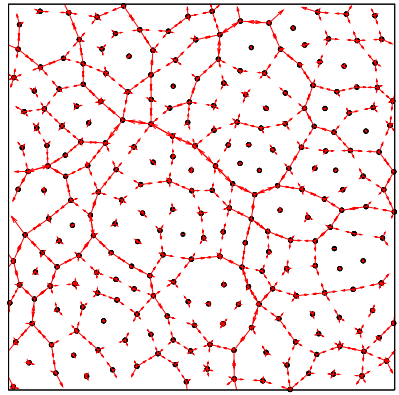}}}
\caption{Contact forces for $N=256$,
  $\epsilon=10^{-5}$ and $t_1 = 10^5$. Black points represent 
  particles.  Contact forces are sketch by arrows which start from
  the particle center, and whose length is proportional to the
  force amplitude. Note that the forces are balanced on every particle, as it
 must be the case on  time scales where the structure is stable.
For similar force networks see \cite{donev2}. }
\label{forcefield}
\end{center}
\end{figure}

\section{Effective potential}

In the previous Chapters we studied the rigidity of amorphous solids by
considering their vibrational modes.   It is a priori problematic to
use the same analysis to study the rigidity of   contact networks
such as the one of Fig.(\ref{forcefield}): the hard sphere potential
is discontinuous, and the energy cannot be expanded as in
Eq.(\ref{0}). Nevertheless we shall go around this difficulty by
deriving a smooth effective potential.  The trick is to average on the
fast fluctuations that take place on characteristic time smaller than
some $t_1\gg \tau_c$.  Consider two particles $i$ and $j$  in contact
separated by a  spacing $ r_{ij}-r_i-r_j$ of typical value $h$, where
$r_i$ and $r_j$ are the radii of the particles, and $r_{ij}$ the
distance between $i$ and $j$. Such a spacing fluctuates in time between
 $0$  (when the particles are colliding) and a few $h$, so that the instantaneous value of this spacing does not give much information about the contact $ij$. Nevertheless, if $r_{ij}-r_i-r_j$ is averaged on large time intervals $t_1\gg \tau_c$, the spacing $h_{ij}\equiv \langle r_{ij}-r_i-r_j \rangle$ converges to a well defined value. We shall define the averaged position $\vec{R}^{av}(t)$ of particle $i$ as:
\be
\vec{R}_i^{av}(t)=\frac{1}{t_1} \int_{t-t_\frac{1}{2}}^{t+t_\frac{1}{2}} \vec{R}_i(t')dt'.
\ee

In what follows we estimate the contact force $f_{ij}$ exchanged between two particles $i$ and $j$ whose spacing is $h_{ij}\approx ||\vec{R}_i^{av}-\vec{R}_j^{av}|| -r_i-r_j$. We furnish thermodynamic arguments, that apply both  to Newtonian and Brownian dynamics. We start by considering a one-dimensional system with beads of diameter 1. The partition function $Z$ is:
 \be
 {\cal Z}=\prod_i\int_{h_i=0}^{h_i=\infty} dh_i e^{-p h_i}
 \ee
 where $h_i$ is the spacing between particle $i$ and $i+1$. If an external force dipole $p_i=-p_{i+1}\equiv p_1$ is applied on $i$ and $i+1$, the partition function becomes:
  \be
 {\cal Z}=\prod_{j\neq i}\int_{h_j=0}^{h_j=\infty} dh_j e^{-p h_j} \int_{h_i=0}^{h_i=\infty} dh_i e^{-(p+p_1) h_i}
 \ee
 From the partition function one can compute the average spacing $\langle h_i\rangle=\frac{1}{p+p_1}$.  Since the contact force $f_i$ in the contact $i,i+1$ is $f_i=p_1+p$, one finds:
 \be
 \label{ff}
 f=\frac{1}{h}
 \ee
 
This result can be extended to an isostatic state of any spatial dimension. A particularity of the isostatic state is that the number of displacements degrees of freedom is precisely equal to the number of contact. Hence the configuration of the system can be described by the set of distances between particles in contact. If the system is at equilibrium in a meta-stable state where the  contact forces field $|{\bf f}\rangle = \{f_{ij}\}$ is well-defined, the partition function can be written:
 \be
 \label{part}
 {\cal Z}=\prod_{\langle ij \rangle} \int_{h_{ij}=0}^{h_{ij}=\infty} dh_{ij} e^{-f_{ij} h_{ij}}
 \ee
The argument valid for the one-dimensional  line of particles is valid here, and one obtains $\langle h_{ij}\rangle=f_{ij}^{-1}$.  This shows that in an isostatic assembly of colliding hard spheres, when the particles rattling are averaged, the hard sphere potential converge to an effective  potential.  At time $t\gg \tau_c$ the system behaves as an assembly of particles of positions $|{\bf R}^{av}\rangle=\{\vec{R}^{av}_i\}$ interacting with the  potential $V_{ij}(r)$:
\ba
\label{pot}
V_{ij}(r)&=&\infty \ \ \ \ \ \ \ \ \ \ \ \ \ \ \ \ \ \ \ \ \ \hbox{ if} \ \ \ \ r<r_i+r_j  \nonumber \\
V_{ij}(r) &\sim& -ln (r-r_i-r_j)\ \ \ \  \hbox{if $i$ and $j$ are in ``contact''} \nonumber  \\
V_{ij}(r) &=& 0 \ \ \ \ \ \ \ \ \ \ \ \ \ \ \  \ \ \ \ \ \ \ \ \hbox{if $i$ and $j$ are not in ``contact"}
\ea

The relation force/distance  is checked numerically in Fig. (\ref{ff1}) at packing fraction close to $\phi_c$. At such packing fractions the system is nearly isostatic, as we shall see in the next section. The exponent found is in very good agreement with Eq.(\ref{ff}).    Fig. (\ref{ff1}) also shows that for large $t_1$ the dispersion of the contact forces around  their average value described by Eq.(\ref{ff}) is extremely small. This indicates that the only relevant parameter that characterizes the contact force amplitude is the spacing $h$, as predicted by Eq.(\ref{pot}). 

%*****caro{il faut diminuer le vspace-> la figure est sur le text!}
\begin{figure}[htbp]
\begin{center}
\rotatebox{0}{\resizebox{4.5cm}{!}{\includegraphics{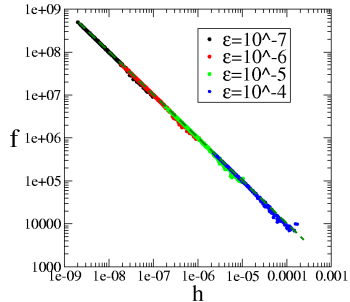}}}
%\rotatebox{-90}{\resizebox{5.5cm}{!}{\includegraphics{graf-forca-hist--semLEG--3nov.ps}}}
\caption{{\small Log-log plot of the  contact force amplitude versus the
 spacing $h=r-r_i-r_j$ for various $\epsilon$ in a system of  $N=256$
 particles. Each dot represents the  pair of number ($f_{ij}$, $\langle h_{ij}\rangle$)
 associated with the contact $ij$. Dots collapse on the dotted theoretical curve defined by Eq.(\ref{ff}). }}
\label{ff1}
\end{center}
\end{figure}

When the coordination $z$ increases, the $h_{ij}$ are not independent variables anymore. We aim to evaluate the corrections to Eq.(\ref{pot}) when the system is at a finite distance from isostaticity, that is with $\delta z>0$. Isolating the contact $ij$ we may write the partition function as follows:
 \be
 \label{part1}
 {\cal Z}= \int _{h_{ij}=0}^{h_{ij}=\infty} dh_{ij}e^{-f_{ij} h_{ij}}  e^{-{\cal F}(h_{ij})}
 \ee
where ${\cal F}(h_{ij})$ is the free energy of the entire system conditionally to the value of the spacing $h_{ij}$. It can be written as ${\cal F}(h_{ij})={\cal F}_0+\Delta {\cal F}(h_{ij})$, where ${\cal F}_0$ does not depend on $h_{ij}$.  To evaluate  $\Delta {\cal F}(h_{ij})$, we consider as a zero order approximation that Eq.(\ref{pot}) is true.  Then $\Delta {\cal F}(h_{ij})$ corresponds to the energy cost induced by a local strain of amplitude $h_{ij} - \langle h_{ij}\rangle= h_{ij}-f_{ij}^{-1}$ in an elastic system where particles interact with the potential (\ref{pot}). In Chapter \ref{c7} we computed the response to  a local strain of amplitude $e$. The corresponding energy cost varies with the contact considered, and for small strain its amplitude follows $\delta E\sim  \delta z B e^2$, where $B$ is the bulk modulus. For an interaction given by Eq.(\ref{pot}), following chapter \ref{c7} one finds $B\sim \langle V'' (r)\rangle \sim \langle (r-r_i-r_j)^{-2}\rangle \sim p^2$, so that $\Delta {\cal F}(h_{ij})\sim \delta z p^2 (h_{ij}-f_{ij}^{-1})^2$. Thus we may write $\Delta {\cal F}(h_{ij})\equiv  \delta z C_{ij} f_{ij}^2 (h_{ij}-f_{ij}^{-1})^2$, where $C_{ij}$ is positive, of order one, and can a priori depend on the contact considered.  Using this expression in Eq.(\ref{part1}), and expanding $Z$ to first order in $\delta z$, we can compute the corrections to $\langle h_{ij} \rangle$. One finds $\langle h_{ij} \rangle=\frac{1}{f_{ij}} (1-2 C_{ij} \delta z)$, so that the force-displacement relation satisfies:
\be
\label{corec}
f_{ij}=\frac{1}{h_{ij}} (1-2 C_{ij} \delta z)
\ee
This estimates the corrections to the potential of Eq.(\ref{pot}),  which vanish when the system becomes isostatic. Thus, one non trivial consequence of these corrections is to weaken the force for a given inter-particle distance.  To test this effect we compute numerically $C(\delta z)\equiv \langle f_{ij} \langle h_{ij}\rangle\rangle_{ij}-1$, where $\langle\rangle_{ij}$ denotes the average over all contacts.  The results are represented in Fig.(\ref{correction_f}).  Small corrections are indeed found, which are in good agreement with our prediction $C(\delta z)\sim -\delta z$.
 In what follows we are mainly interested in scaling relations near $\phi_c$, 
 where corrections of the order $\delta z$ do not matter. Thus we shall neglect them, and consider that the effective interaction is constant and given
 by Eq.(\ref{pot}).
 
\vspace{2cm}
\begin{figure}[htbp]
\begin{center}
%\rotatebox{-90}{\resizebox{5.5cm}{!}{\includegraphics{graf-correcao_vs_dz-artigo--25ago.ps}}}
\rotatebox{0}{\resizebox{4.5cm}{!}{\includegraphics{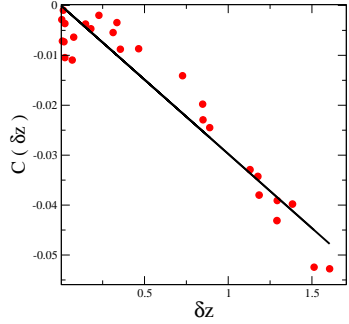}}}
\caption{{\small Average correction $ C(\delta z)$  as defined in the text {\it vs.} excess coordination $\delta   z$ for  various $\phi$. The line is a linear fit  consistent with our predictions at small $\delta z$. Corrections are small,  of the order of 3-4 percents when $\delta z=1$.}}
\label{correction_f}
\end{center}
\end{figure}

Note that  Eq.(\ref{pot}) can be recovered with a simple scaling argument. We may evaluate the collision frequency $\nu$ as $\nu\sim \frac{v}{h_{ij}}$, where $v$ is the average amplitude velocity. The typical momentum exchanges during one collision is, taking $m=1$ for the particles  mass, $||\Delta \vec{P}||\sim v$. Therefore following Eq.(\ref{imp}) we recover that the force follows $f_{ij}\sim \frac{\langle v^2\rangle}{h_{ij}}\sim \frac{1}{h_{ij}}$.

In what follows we shall only assume that the potential $V(r)$ is differentiable and that the force $f(h)$ scales as $h^{-1} $. The corrections estimated above in Eq.(\ref{corec}) do not affect these results, therefore we neglect them and use Eq.(\ref{pot}). As we shall see, this implies that $\phi_c$ acts also as a critical point in the liquid phase.  In particular, Eq.(\ref{pot}) enables us to define the stiffness $k_{ij}$ of a contact $ij$ as:
\be
\label{kqk}
k_{ij}=V''(r)\sim \frac{1}{(r-r_i-r_j)^2}\sim f_{ij}^2
\ee
This allows to define a dynamical matrix and vibrational modes once the average particles positions and the contacts are known. As we discussed above, the potential of Eq.(\ref{pot}) has an entropic nature. Thus such vibrational modes describe the local curvatures of the entropic landscape of the system. In what follows we show that imposing the stability of these modes leads to constraint on the coordination of the force network. Then we discuss the elastic property of hard sphere glass, derive the elastic moduli and discuss the length scales that appear in the response of such systems. Finally we discuss how these modes may be related to the structural relaxation.

\section{Stability of hard sphere systems}

If the  network of contact of a hard sphere system is weakly
 connected, we can apply the results of Chapters \ref{c3}, \ref{c4}
 and \ref{c5}. In particular, following Eq.(\ref{33}) and Eq.(\ref{34}) and the paragraph above it, we obtain that
 such system presents anomalous modes that appear at a frequency
 $\omega_{AM}^2=A_1 B(p) \delta z^2- A_2 p$, where $B(p)$ is the bulk modulus of the system.  The bulk modulus scales as the average stiffness of the contacts, and following Eq.(\ref{kqk}) we obtain $B(p)\sim p^2$. As we discussed in Chapter \ref{c5}, a rigid system does not display unstable modes. Therefore we obtain that there must be a constant $C_0$ such that:
\be
\label{dz}
\delta z \geq C_0 p^{-\frac{1}{2}}
\ee 
which is another realization of the extended Maxwell criterion we derived in Chapter \ref{c5}. Note that near the jamming threshold, the typical inter-particle
 spacing $\epsilon$ goes as  $\epsilon\sim \phi_c -\phi$ and thus $p\sim \epsilon^{-1}\sim (\phi_c-\phi)^{-1}$, as was already computed with different methods \cite{donev,parisi4}. Thus Eq.(\ref{dz}) indicates that  $\delta z$ must scale  as $\delta z\sim (\phi_c-\phi)^{\beta}$ with $\beta\leq\frac{1}{2}$, as  is the case for a soft sphere system {\it above} $\phi_c$.

\subsection{Stability of the hexagonal and the square crystals}

\begin{figure}
\centering
\includegraphics[angle=0,width=8cm]{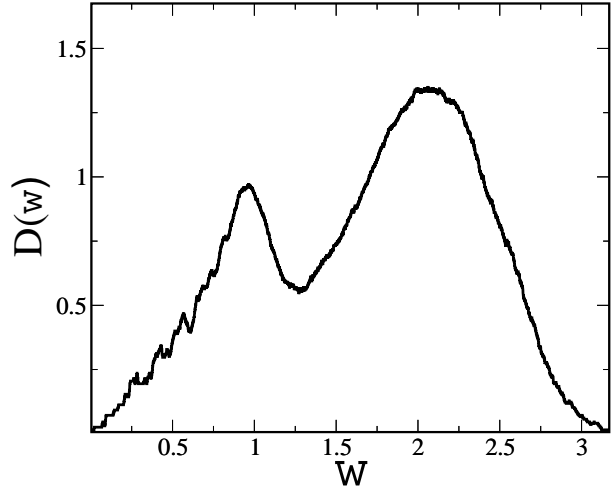}
\caption{Density of vibrational modes $D(\omega)$ versus frequency in
  a hexagonal crystal of 1024 particles for $\epsilon=10^{-4}$. The
  frequencies are rescaled by $\epsilon^{-1}$. The particle positions
  were average in time  $t_1 = 4 \times 10^4 $ to obtain $|{\bf
    R}^{av}\rangle$.  Eq.(\ref{pot})  was used to compute the dynamical matrix, from which the vibrational frequencies were inferred.}**
\label{d1}
\end{figure}

To test Eq.(\ref{dz}) we perform tests on three different systems in
 two dimensions. We start by considering an hexagonal monodisperse
 crystal and a monodisperse square crystal. We consider  these two
 systems at their maximum packing fraction where hard spheres are in
 contact. Then we reduce the particles diameter by a relative amount
 $\epsilon$, and start the dynamics. According to Eq.(\ref{dz}) these systems must behave very differently. In the
 hexagonal crystal, the coordination is 6, therefore $\delta
 z=2\gg p^{-\frac{1}{2}}\sim \epsilon^{\frac{1}{2}}$ for small $\epsilon$. Therefore
 the condition of Eq.(\ref{dz}) is satisfied  and we expect the system
 to be stable. On the other hand, in the square crystal, the number of
 first neighbors is 4, $\delta z=0$,  and the system cannot satisfy
 Eq.(\ref{dz}) without large structural rearrangements for any
 $\epsilon$. Thus such  a system cannot be rigid.  These predictions are
 verified numerically. For small $\epsilon$, the hexagonal crystal
 displays no structural changes, whereas the square crystal collapses
 rapidly. Such collapse leads to an hexagonal configuration, and is generated by the buckling of unstable modes. We show the corresponding displacements in Fig.(\ref{dc}). The instability can be also observed in the free energy landscape by computing the vibrational modes of the corresponding systems. Fig.(\ref{d1}) shows the density of vibrational modes of the hexagonal crystal, which varies linearly at low frequency, as expected for a two-dimensional crystal. No unstable mode are observed. In the square crystal, see Fig.(\ref{d2}), the density does not vanish at $\omega\rightarrow 0$, as expected for an isostatic system. Furthermore, we observe unstable modes, as implied by Eq.(\ref{dz}).

\begin{figure}[htbp]
\begin{center}
\mbox{
{\rotatebox{0}{\resizebox{3.1cm}{!}{\includegraphics{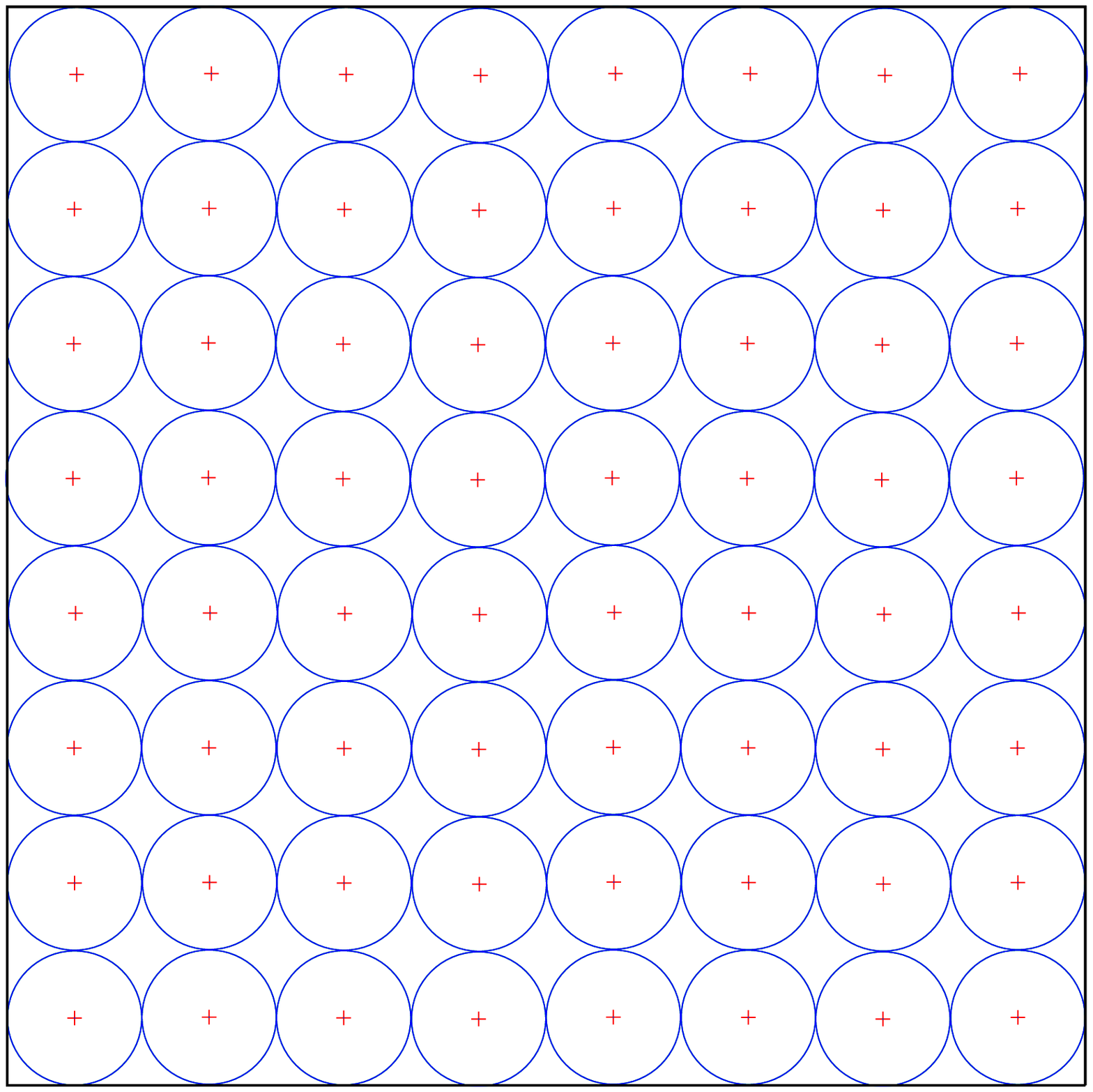}}}}\quad

{\rotatebox{0}{\resizebox{3.1cm}{!}{\includegraphics{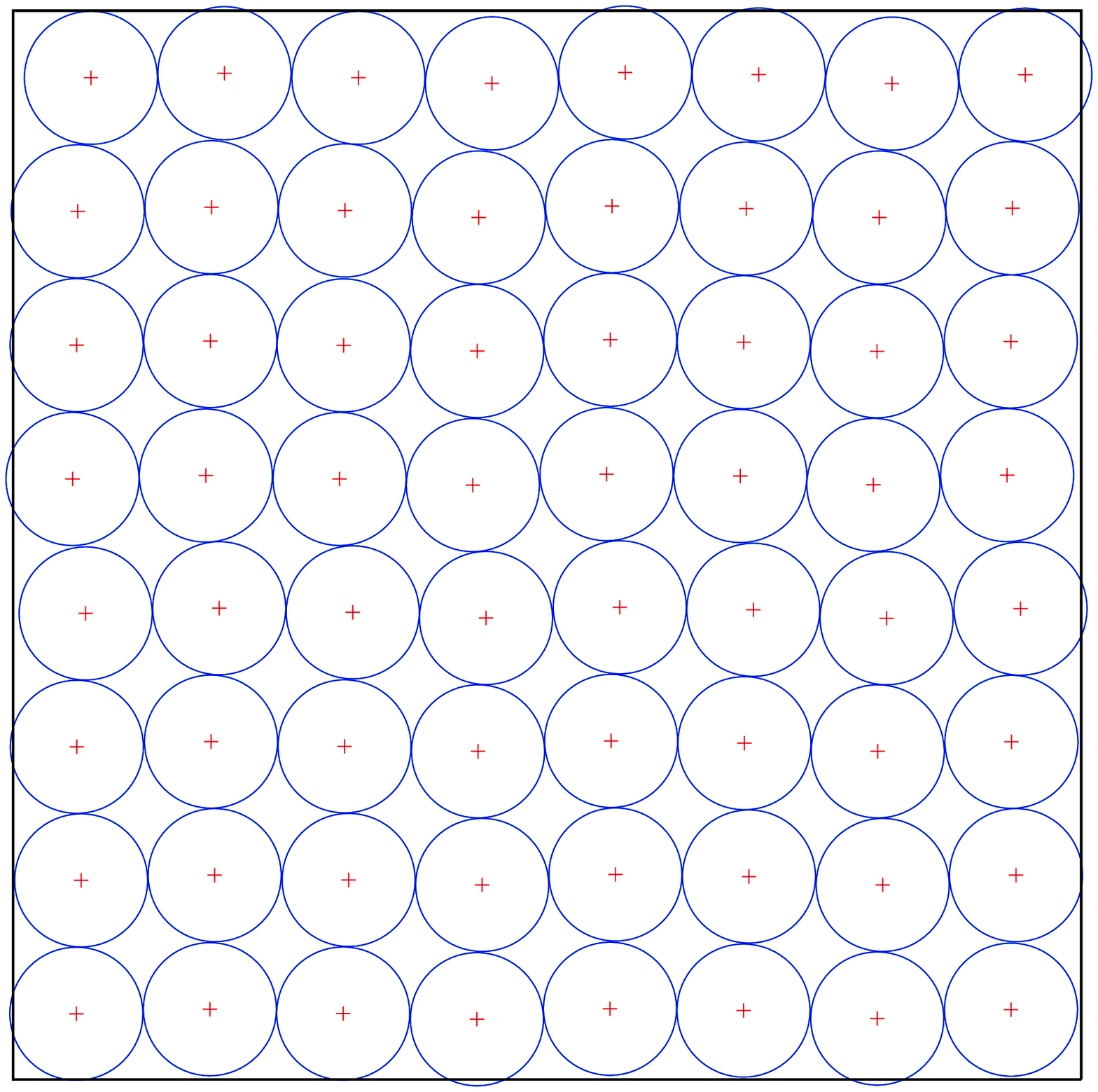}}}}\quad

{\rotatebox{0}{\resizebox{3.1cm}{!}{\includegraphics{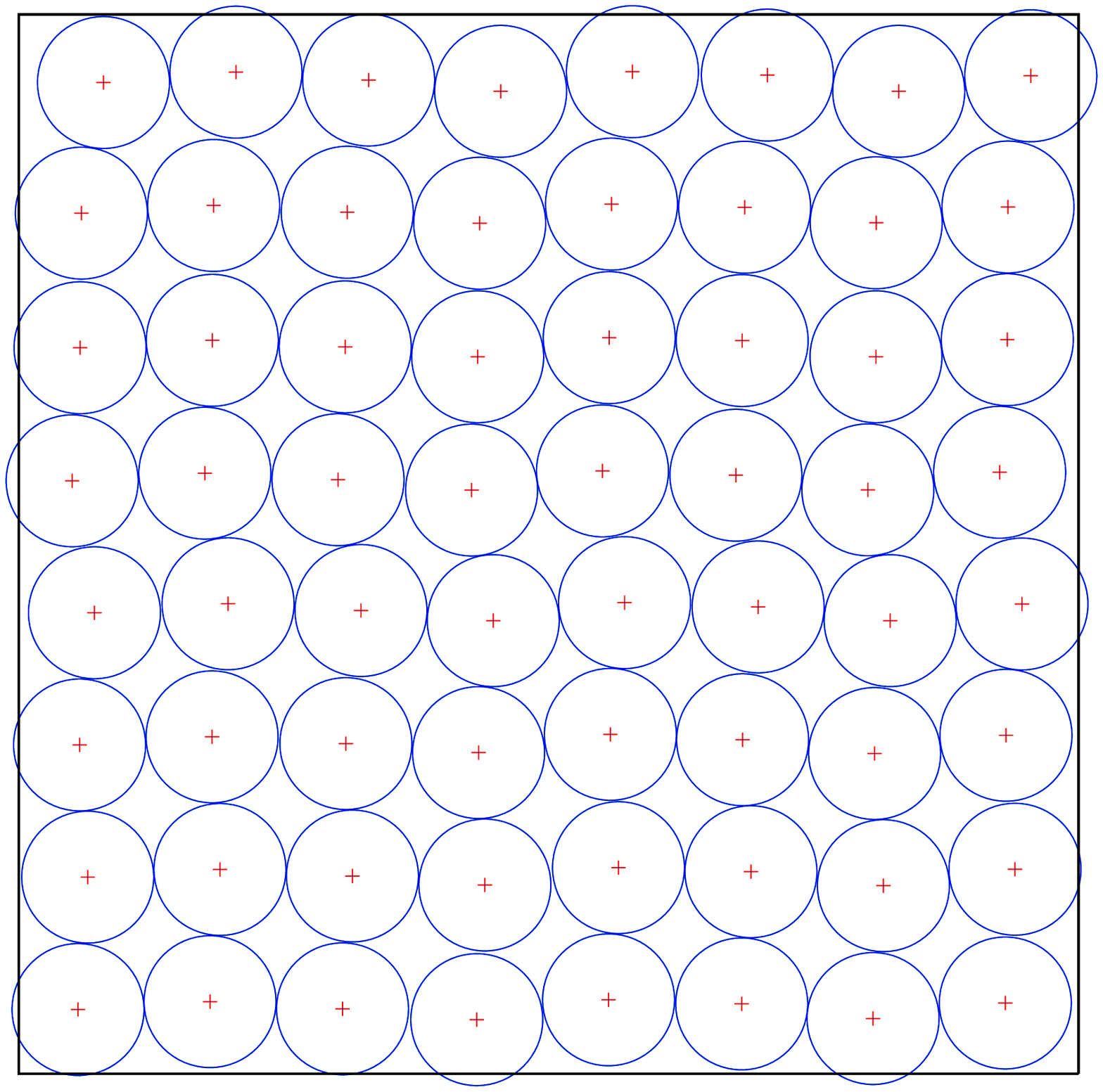}}}}\quad

{\rotatebox{0}{\resizebox{3.1cm}{!}{\includegraphics{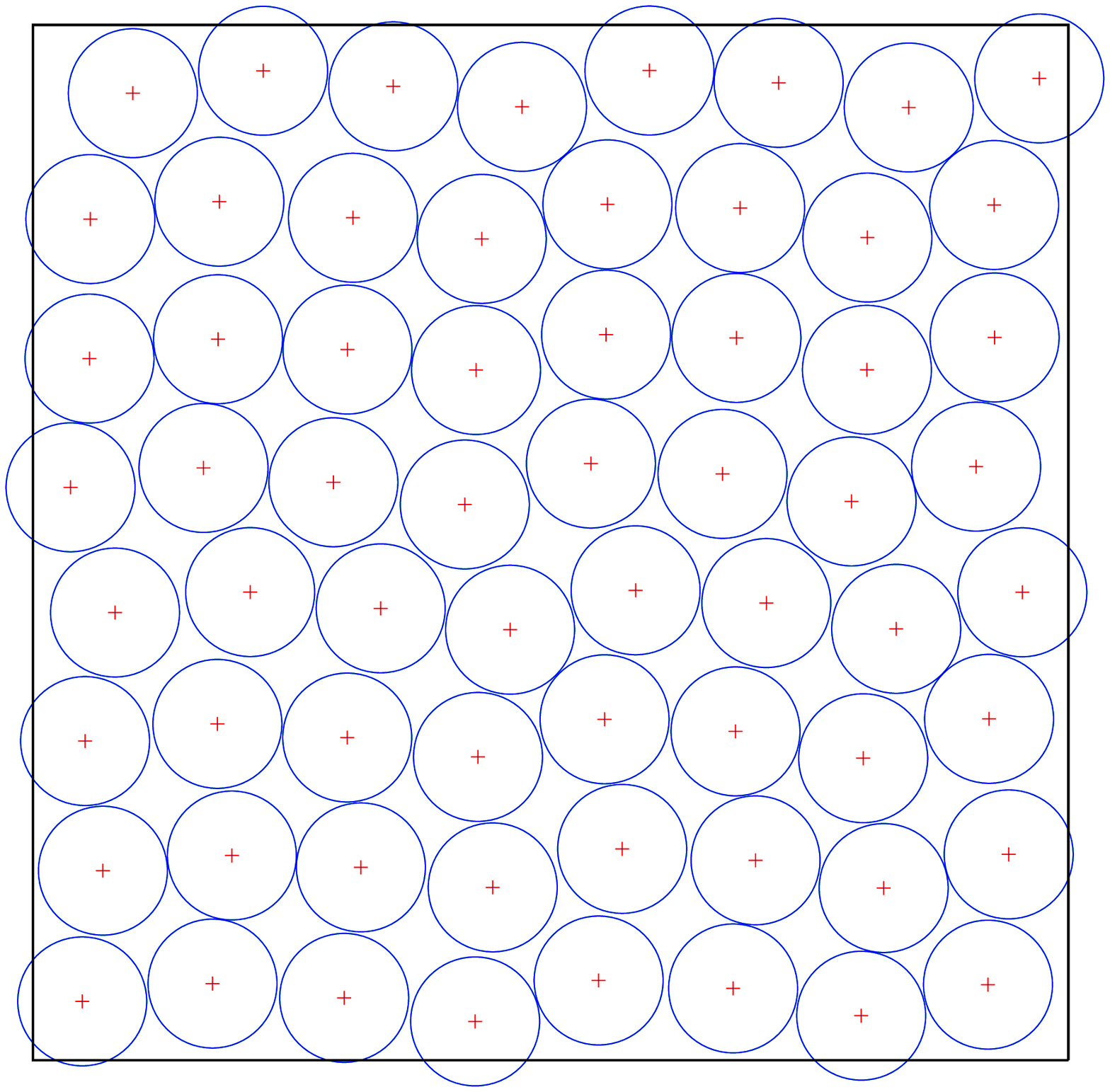}}}}\quad

{\rotatebox{0}{\resizebox{3.1cm}{!}{\includegraphics{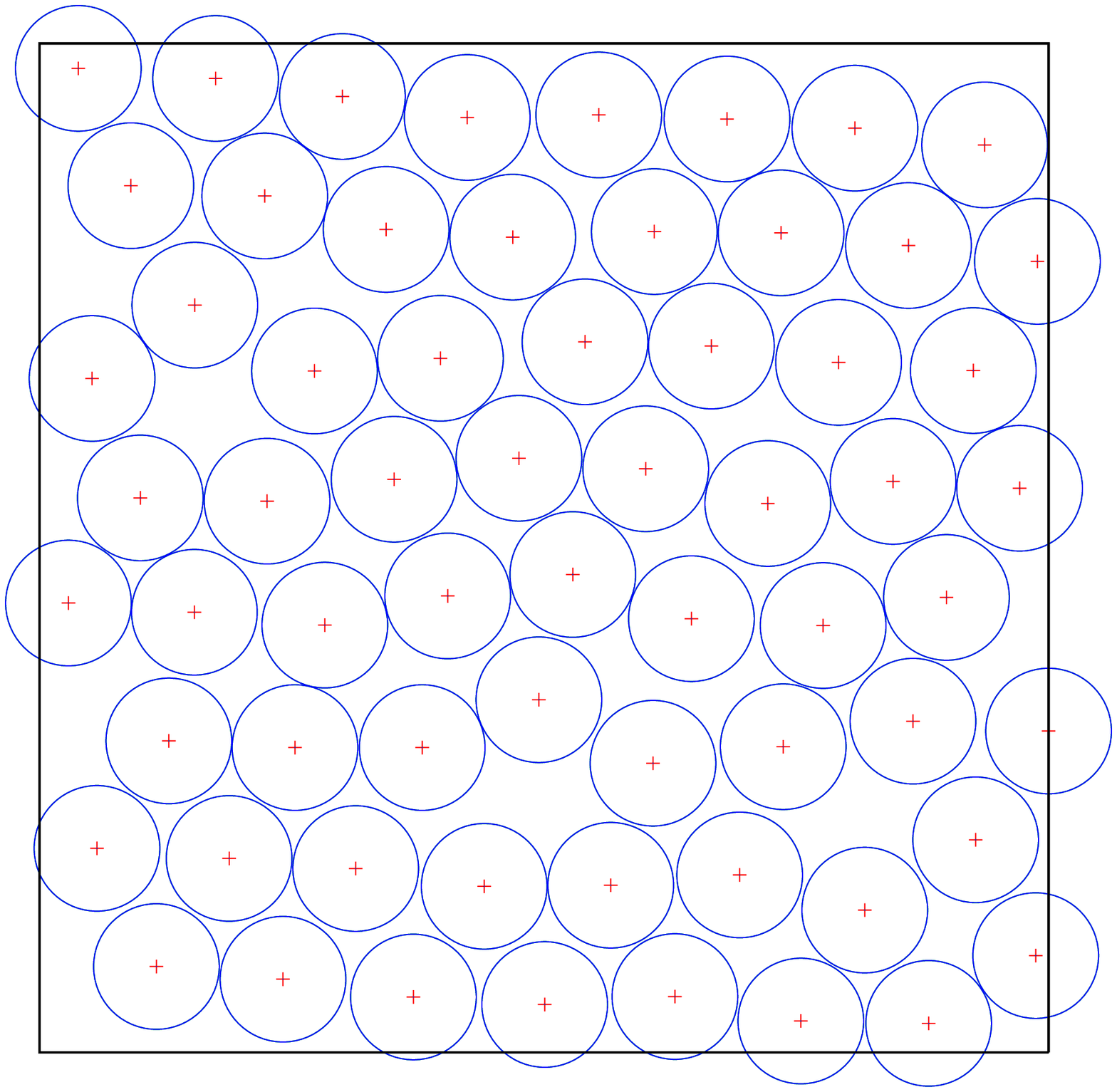}}}}}
\caption{ Buckling of a square lattice. }
\label{dc}
\end{center}
\end{figure}

\begin{figure}
\centering
\includegraphics[angle=0,width=6cm]{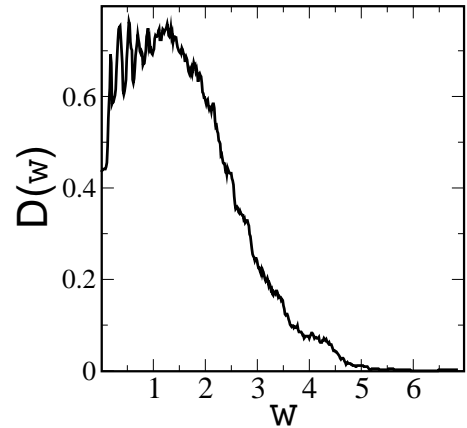}
\caption{ Same caption as  Fig.(\ref{d1}) for a N=1024 square crystal.  80 unstable modes were observed, lying in the frequencies  range $\omega\in[  -0.015,  0]$. There are very degenerated and therefore not appropriate for a plot, thus we do not represent them.    }
\vspace{2cm}
\label{d2}
\end{figure}

\subsection{Stability of hard sphere systems  near the jamming threshold}

\begin{figure}
\centering
\includegraphics[angle=0,width=6cm]{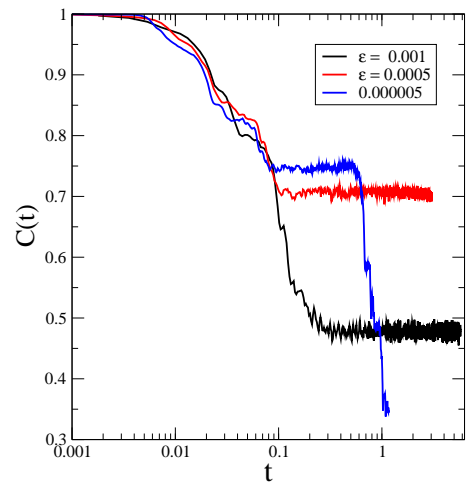}
\caption{ Examples of $C(t)=C(q=1,t)$ for $q = \pi / r_1$   versus real time for
  different $\epsilon$  in a $N=256$ particles  system. $t=0$ corresponds to the initial time where random velocities are assigned to the particles.}
\label{c(t)}
\end{figure}

We perform the same test using the  polydisperse configurations of
\cite{J} at the jamming threshold. We study the structural stability of
the system by computing the self correlation function
$C(\vec{q},t)=\langle \exp[i\vec{q}\cdot (\vec{R}_j(t)-\vec{R}_j(0))]
\rangle$ where the average is taken on every particle $j$. In what follows
we consider  wave vectors $\vec{q}$ of norm $\frac{\pi}{r_1}$,
where $r_1$ is the radius of the smallest particle.  Typical
curves for different $\epsilon$ are represented in
Fig.(\ref{c(t)}). For $\epsilon$ smaller that roughly $\epsilon=0.05$, the system ages and dynamics is intermittent. There are long
 periods of time where the structure is stable, that appear as
 plateaus  in the self correlation function. Such periods ends with  crashes, where the self correlation function drops of a large amount in a very short time. These events must correspond to sudden collective rearrangements involving a large number of particles.  During such crashes,  we observe that the coordination increases, and that  the pressure drops. We also observed that the correlation function of the force network  $H(t)= \langle f_{ij}(\tau) f_{ij}(\tau+t)\rangle$ is constant during the plateau, and drops when a crash occurs. 
 
 Such crashes, or ``earthquakes'', were reported in other glassy systems. They occur  when a Lennard-Jones liquid is rapidly quenched at temperature much lower than the glass transition  \cite{aging}. In this case the crashes typically involve 100 particles. Such crashes were also observed in colloidal pastes using dynamic light scattering \cite{cipelletti}, and in the dielectric response of laponite \cite{ciliberti}.   It would be obviously interesting to understand what is the nature of these crashes, and what triggers the sudden collapse of apparently mechanically stable structure. These are fundamental questions of the glass transition. In the present Chapter we study the stable structures that appear before and after the crashes, when the system is quiet. Understanding why such structures are rigid is certainly a necessary step to find out why and how  they can yield.

We define the quiet periods as the  plateaus of $C(t)$. To  study the rigidity of the dense hard sphere assemblies during these periods, we compute the vibrational modes. If the rattlers are removed \footnote[24]{The rattlers do not participate to the rigidity of the structure. They can be  identified as the distance with their neighbors is much larger than for the rest of the inter-particle distances, so that the frequency of the shocks they have with their neighbors is much smaller than the other particles.}, we find that, the system is mechanically stable: there are no unstable modes. The density of states for $\epsilon=10^{-4}$ is shown in Fig.(\ref{d3}). The main difference with the other stable structure that we studied above--- the hexagonal crystal--- is the presence of  a large excess of modes at low frequency.  A large amount of modes are nearly unstable, as we expect near the jamming threshold where we must have $D(\omega)\rightarrow \omega^0$ when the frequency vanishes.  Note that the density of states does not have a flat plateau as the harmonic soft spheres at the isostatic point, despite that the coordination network is the same when $\epsilon$ is small. The main difference between these systems is the stiffness disparity:  in the amorphous hard sphere system, the disparity of the  contact force leads to a disparity in the stiffness since $k\sim f^2$. If this disparity is removed by keeping the same contact network, but by imposing a constant stiffness in all contacts, we find that a flat plateau is recovered.

The absence of unstable modes enables us to test Eq.(\ref{dz}), that must be satisfied as the system is stable.To  check this relation we computed numerically both the coordination and the pressure for  various packing
fractions, and for various stable periods that appear along the aging
regime.  As shown in Fig.(\ref{PvsZ}), the data are consistent with an {\it equality} of the inequality (\ref{dz}).
This suggests that  a hard sphere glass is only {\it marginally} stable, as it is the case for soft spheres slowly decompressed toward $\phi_c$.

\begin{figure}
\centering
\includegraphics[angle=0,width=8cm]{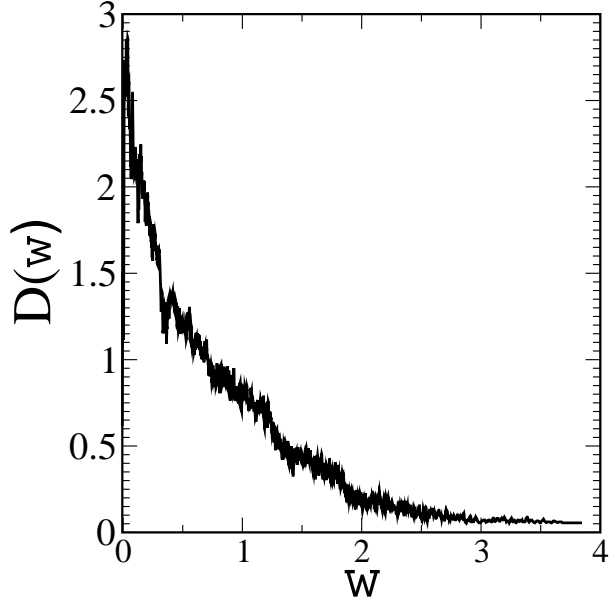}
\caption{ Density of vibrational modes $D(\omega)$ versus frequency in a amorphous hard sphere glass of 256 particles for $\epsilon=10^{-4}$. The frequencies are rescaled by $\epsilon^{-1}$. $D(\omega)$ was computed in a quiet period preceding   the first crash. No unstable modes were observed. }
\label{d3}
\end{figure}

\begin{figure}[htbp]
\begin{center}
\rotatebox{0}{\resizebox{5.5cm}{!}{\includegraphics{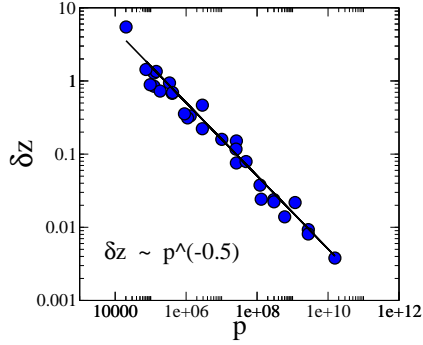}}}
\caption{{\small Log-log plot of $\delta z$  versus the average contact force $\langle f\rangle\sim p$. The data were obtained for different $\epsilon$ and different time periods.
The black line corresponds to the equality of the inequality (\ref{dz}). } }
\label{PvsZ}
\end{center}
\end{figure}

\section{Elastic property of the hard sphere glass}

We now use this approach to derive the elastic behavior of the hard sphere glass near $\phi_c$. The results of Chapter \ref{c7} apply. In this system the elastic moduli have purely entropic causes: they describe how the number of configurations is reduced under a global strain.   As we already discussed, the scaling of the bulk modulus $B$ is:
\be
B\sim p^2\sim (\phi_c-\phi)^{-2}
\ee
In repulsive systems  near $\phi_c$ the shear modulus scales as the bulk modulus times the excess coordination. Thus the extension of Eq.(\ref{xixi}) to non harmonic contacts yields:
\be  
G\sim B \delta z\sim \delta z p^{2}
\ee
If one assumes that the glass is marginally stable, that is to say that the system lies on the bound of inequality (\ref{dz}), one obtains:
\be
\label{gg}
G\sim  p^{3/2}
\ee
As in practice such glass is not at equilibrium near $\phi_c$, the hypothesis of marginal stability certainly depends on the preparation of the system. It is reasonable to think, following the analogy with the soft spheres, that if the system is slowly compacted toward $\phi_c$ the dynamics would not bring the system far away from marginal stability.  Thus we expect $\delta z\sim p^{-\frac{1}{2}}$ to be a good approximation and the shear modulus $G$ not to differ to much from the scaling law of Eq.(\ref{gg}).

Furthermore, this system present  excess modes: there are the anomalous modes we discussed in the previous Chapters.  Near the jamming
 threshold, when the pressure diverges, we expect $D(\omega)\rightarrow \omega^0$. When $\phi$ decreases,  the density of states is characterized  by some frequency $\omega^*\sim B^{\frac{1}{2}}(p) \delta z\sim p^{-\frac{1}{2}} \delta z$.  Consequently a hard sphere glass is characterized the length $l^*\sim \delta z^{-1}$  which characterizes the anomalous modes present at low-frequency.

\section{Discussion}

In this Chapter  we studied some properties of the hard-sphere glass near the close packing at $\phi_c$.   We argued that the mechanical stability at time $t\ll \tau_\alpha$ relies on the emergence of a  rigid force network. The key point  was to compute the properties of the force networks by averaging on fast fluctuations, the particles rattling. This allows one to derive an effective potential, exact at the isostatic limit, and to define rigorously the vibrational modes in hard sphere systems.  They characterize the entropic landscape expansion around a configuration. This allows one to recover the main results valid for  soft sphere systems near the jamming threshold. In particular the stability of the vibrational modes imposes a bound on the coordination of the contact force network.  More generally,  we showed that $\phi_c$ acts as a critical point: the elastic moduli and  the coordination  $z$  scale as $\phi\rightarrow \phi_c$. The system responses are characterized by  $l^*$ that diverges at $\phi_c$.

As critical points and diverging length scales are in general
associated with diverging time scales, the presence
of  a critical behavior at $\phi_c$ may be the cause for the glass transition  observed
in hard sphere systems. In particular, our work suggests that the slowing-down of the dynamics 
is due to the appearance of an {\it extended} rigid network of interactions, rather than to any local properties such as the tightening of cages. 
This suggests an alternative perspective for the relaxation in super-cooled liquid, at equilibrium or in an aging regime such as the one of  Fig.(\ref{c(t)}). The particularity of the systems that crash, such as the square crystal or the amorphous state near isostaticity, is the large amount of modes around zero frequency. As we discussed in  previous Chapters such modes are very different from plane waves, as they have rapid spatial fluctuations, which makes them sensitive to the applied stress term responsible for buckling. Furthermore they can be localized on length scales larger than some $l^*$ without  changing in their frequency. These modes indicate a nearby elastic instability, and may well play a role in the rigidity loss. Furthermore, whatever rearrangements occur when the system relaxes, suddenly or not, the corresponding displacements must be similar to the soft modes, as there are, rougly speaking, the only modes where particles can avoid to inter-penetrate. Thus it is tempting to associate the structural relaxation with the structural buckling of the weakest frequency modes, that could be induced for example by pressure or coordination fluctuations. If so, we expect the relaxation to occur on length scales of order $l^*\sim p^{\frac{1}{2}}$, since for smaller subsystems the frequency of the modes increases, and the elastic instability goes away. Thus our work suggests an original perspective on the possible cause of the heterogeneous  dynamics that occur in super-cooled liquids \cite{Ediger,Weeks}.  Many models of the glass transition lead to heterogeneous dynamics  \cite{twbbb}. In most of them, a local rule describes the motion  of one or few particles, and leads to cooperative dynamics. Our work suggests an alternative view: the  rule that allows the particle motion is {\it itself} non local, for the simple reason that rigidity is not global property, as was already understood by Maxwell more than one century ago.

\chapter{Conclusion}

\section{Summary}

The elastic properties of an assembly of short-range, repulsive particles display  critical behaviors near the jamming transition. In particular the elastic moduli, the frequency of the excess-modes, and the coordination number scale. In the first part of this thesis we derived the corresponding exponents. Then we showed how some of these results  can apply to various systems, such as silica glass or colloidal particles. 

The starting point is the following variational argument: if a rigid system of coordination $z$ is cut in sufficiently small subsystems of size $l$, these subsystems are not rigid. This is true as long as the subsystems size is smaller than some length $l^*\sim \delta z^{-1}$ that diverges at the jamming threshold. Smaller subsystems contain  modes of zero frequency, the soft modes. From these modes one can build what we called the ``anomalous modes'' which have a frequency of order $1/l$ in the original system. This gives for the dependence of the onset of excess-modes $\omega^*\sim \delta z$, as observed numerically. The system can be described as a continuous elastic medium only for $\omega\leq\omega^*$. At larger frequencies, it behaves as an isostatic state. 

Then, by arguing that the soft modes have large transverse relative displacements, we showed that the anomalous modes were much more sensitive to the applied stress than the conventional acoustic modes. This has two direct consequences:  in a purely repulsive system, anomalous modes can appear at frequencies much smaller than $\omega^*$. Furthermore, in rigid solids  the coordination  number must be sufficiently large to guarantee the  stability of the anomalous modes. We find  $\delta z\geq C_0 (p/B)^{\frac{1}{2}}\sim (\phi-\phi_c)^{\frac{1}{2}}$, where $p$ is the pressure and $B$ the bulk modulus. This generalizes the Maxwell criterion for rigidity $\delta z\geq 0$ valid when applied stress is absent.  It follows that, if the jamming threshold $\phi_c$ is reached adiabatically, the pair correlation $g(r)$ measured at $\phi_c$ must display a divergence  $g(r)\sim \frac{1}{(r-1)^\gamma}$ with $\gamma \geq \frac{1}{2}$. This divergence is the vestige of the excess contacts $\delta z$ necessary to maintain the rigidity of the structure at larger $\phi$.  

These arguments furnish the scalings of both the onset of excess modes and the coordination number, but they do not enable us to compute the elastic moduli. To do so, we used the fact that the linear equation that defines the soft modes is the dual of the force balance equation. This allows us to derive an original formalism  relating the response to a strain  to the contact force fields that satisfy force balance on each particle.  This enables us to compute the shear and the bulk modulus. We found  that for  a repulsive system the ratio $G/B$ vanishes at the jamming transition.  We also obtain exact results on the response to a local strain. The energy  cost of such deformation vanishes at the transition as the shear modulus. At the transition, this response extends in the entire system. Therefore this is also true for the soft modes that appear when one contact is cut in an isostatic state, an assumption that was  essential to compute the frequency of the anomalous modes.

In a second part we studied the applications of these concepts to glasses, granular matter and colloids. All glasses have an excess of low-frequency modes, the boson peak. It is especially strong in silica, one of the best glass former.  In silica the strongest interactions are those responsible for the rigidity of the SiO$_4$ tetrahedra.  If the weaker interactions, such as Van de Waals,  are neglected, the tetrahedral network obtained is isostatic.  Following our argument, such a network must  have a constant density of states at low frequency, as it is observed numerically.  When the weaker interactions are turned on, the anomalous modes responsible for the plateau shift to higher frequencies.  This improved variational argument predict the appearance of a plateau in the density of states. Such plateau is indeed observed in the simulations of silica, and appears around 1 THz, the boson peak frequency.  As the key parameter of our description is coordination rather than disorder, this argument also justifies the similarity between the density of states of silica and  the one of the corresponding crystals, which also present a plateau at roughly the same frequency.   This is certainly a strong point, as most of the existing boson peak theories are based on disorder only, and cannot explain  the excess modes that show up in these crystals.  Finally we proposed to extend these ideas to Lennard-Jones systems.

Generally critical points display similar behavior on both sides of the transition.  We argued that it is also the case for the jamming transition. We showed that when a hard sphere liquid approaches the jamming transition, the contact force network, that now characterizes how particles exchange momentum, is very similar to the one of elastic spheres above $\phi_c$. The key point was to show that when the dynamics of an isostatic system is averaged on short time scales, the interactions among particles can be described by a  logarithmic effective potential. We evaluated the corrections of this potential when the coordination increases. This allows to compute the normal modes of such systems. As a consequence  the results of elastic spheres solids apply to hard sphere liquids. In particular the extended Maxwell  criterion must be satisfied near $\phi_c$, when the system is rigid on short time scales. This approach  also yields the elastic moduli that characterize the system for $t\ll \tau_\alpha$, the relaxation time. This description supports that the relaxation in super-cooled liquid should not be described in terms of the motion of individual particles, even if models with such local rules can lead to non-trivial cooperative dynamics. Rather, it suggests that the elementary motions to consider are themselves collective, and are related to the anomalous modes  introduced above.

\section{Perspectives}

\subsection{Low temperature glass properties}

The transport properties of glasses, but also of granular matter \cite{somfai,liu1}, are not understood \cite{AndyAnderson}. In particular there is a plateau in the thermal conductivity around 10 K, temperatures at which the heat transport can be dramatically smaller than in the crystal phase \cite{bergman}.  These temperatures correspond to frequencies in the Thz range, where the boson peak appears. In repulsive, short-range systems we showed that the cause for the boson peak is the following: above some frequency $\omega^*$ the system behaves as an isostatic state, whose density of states is much larger than the one of the crystal at low frequency. We argued that these ideas are more general and  apply for example to silica and granular matter.  This suggests that the transport properties  of glasses in the Thz range correspond to those of a system at the jamming threshold. Then the question is to understand how the anomalous modes we introduced contribute to the transport. Accordingly, it is necessary to compute their spatial power spectrum $E(q)$. For a normal mode of displacement $\{\delta \vec{R}_i\}$, $E(q)$ is  defined as:
\ba
\label{the}
E(q)=\frac{1}{q^2} \sum_{j,l} (\vec{q}\cdot \delta \vec{R}_j) (\vec{q}\cdot \delta \vec{R}_l) e^{i \vec{q}\cdot (\vec{R}_j-\vec{R}_l)} \\
\sim \frac{N}{q^2} \int dr^d \langle (\vec{q}\cdot  \delta \vec{R}(r)) (\vec{q}\cdot \delta \vec{R}(0)) \rangle e^{i \vec{q} \cdot \vec{r}}
\ea
For an acoustic mode of frequency $\omega$, $E(q)$ presents a peak for $q=c^{-1} \omega$, where $c$ is the longitudinal velocity of sound. If this peak has a finite width $\Delta q$, the scattering length of the acoustic mode follows $l\approx \Delta q^{-1}$.  This length enters in the computation of the thermal conductivity: the contribution of a mode to the heat transport goes as $l \cdot c$ \cite{W.Phillips}. The empirical data suggests that at frequency smaller than the boson peak, the acoustic modes have a scattering length  $l\approx 150 \lambda$, where $\lambda$ is the wavelength of the corresponding mode \cite{AndyAnderson}. We expect that the anomalous modes that appear at higher frequencies will transport much less than that, because their spatial correlations $\langle \delta \vec{R}(0)\cdot \delta \vec{R}(x)\rangle$, which enter in Eq.(\ref{the}), are presumably very small. This quantity is similar to the spatial correlations of the soft modes. If these correlations were zero, both the soft modes and the anomalous modes  would be equally distributed on all wave vectors. Then the anomalous modes would have a purely diffusive behavior, corresponding to a scattering length $l=1$ particle size.  Thus they would almost not contribute to the transport at all. It is apparent  from Fig.(\ref{softfig}) that the spatial  correlations of the soft modes are indeed small. This observation may furnish an explanation for the observed bad quality of the transport at the boson peak frequency. It  is also consistent with the empirical observation that the boson peak frequency depends only weakly on wave vectors $q$ if at all \cite{foret}, indicating that the excess-modes have a wide distribution over the wave vectors $q$.  For a quantitative discussion it would be of great interest to derive the spatial correlations induced by the soft mode equation (\ref{3}).  Furthermore,  other interesting effects could in principle affect the transport. In particular, our derivation of the anomalous modes does not  preclude the presence of underlying acoustic modes, that  could  hybridize with the anomalous modes, and enhance the transport. 

At lower temperature, the properties of glasses are still a challenge to theory.  The specific heat has a nearly linear dependence with temperature, and the thermal conductivity varies quadratically. This is in general interpreted  by the presence of two-levels systems: atoms or groups of atoms can switch between two configurations by tunnel effect. This model is phenomenological and there is no consensus on what these two level systems may be. It has been often argued that the excess-modes of the boson peak are good candidates to form two-levels systems, see e.g. \cite{buchenau, shlomon}. As these modes are soft, non-harmonic terms are important, which could lead to the canonical form of 2-levels systems: two wells separated by a potential barrier. If it is so, our interpretation of the boson peak suggests that the 2-level systems are rather extended, plausibly on the length $l^*$, and that the displacement of each particle could be much smaller than a particle size. It should be possible in principle to test this possibility. As we discussed earlier, the pressure has two opposite effects on the anomalous modes: on the one hand it increases the coordination, and on the other hand the applied stress term lowers their frequency. We expect that in some systems, the destabilizing effect of pressure dominates. This can even lead to an elastic instability where anomalous modes become unstable, and where the configuration of the system changes. Such elastic instability might occur for example in silica glass at high pressure \cite{lacks}.  If the density of  anomalous modes is increased at very low temperature by tuning the pressure, the density of two-levels systems should be enhanced too. The latter could be directly checked by specific heat measurements.

\subsection{The glass transition}

We discuss the following possibility:  in fragile glass former a fast slow down of the dynamics occurs close to the temperature at which the system manages to form an extended rigid network.  We first study the role of temperature on the anomalous modes in the glass phase. Then we propose a microscopic distinction between strong  and fragile glasses \footnote[26]{The relaxation time of  strong glass as an Arrhenius dependence with temperature, and grows faster in a fragile glass.}, and we discuss a possible structural relaxation process in the fragile case. 

\subsubsection{Role of temperature}

Following the results of Chapter \ref{c8}, we can write for the onset of the anomalous modes at zero temperature $\omega_{AM}^2= A_1 B \delta z^2 -{\cal H} (p)+\zeta$, where $B$ is the bulk modulus, $\delta z$ is the effective excess-coordination number, $\zeta$ quantifies the effect of the weak interactions on the anomalous modes, and $A_1$ is a numerical constant. ${\cal H}(p)$ is the correction induced by initial stress term. For soft spheres near the jamming threshold where the distances between particles in contact   are similar  we found ${\cal H}\approx A_1 p\sim \langle f \rangle$, where $\langle f  \rangle$ is the average contact force. In a system where the distance $r$ between interacting particles can vary,  following Eq.(\ref{000}) we have  ${\cal H}\sim \langle f/r \rangle$.  In the present qualitative discussion we shall neglect these corrections  and consider ${\cal H}\approx A_1 p$. 

If the system is heated at constant pressure, we may extend this equation and write:
\be 
\label{rtt}
\omega_{AM}^2(T)= A_1 B(T) \delta z(T)^2 - A_2 p+\zeta(T)
\ee
In most glasses, when the temperature increases at constant pressure, the volume grows, and  $B$ decreases. As the typical  inter-particle distances grows, we expect both the effective coordination and the effect of the weak interactions to diminish. Thus all the positive terms in Eq.(\ref{rtt}) decrease. On the other hand, the pressure is constant. Hence $\omega_{AM}^2(T)$ decreases, as  is indeed observed in most glasses. Eventually $\omega_{AM}^2(T)$ reaches zero frequency, as it has been observed numerically \cite{kob} and empirically \cite{tao}.   We shall denote the temperature at which rigidity is lost $T_r$. Note that in  few materials, such as silica, the volume {\it decreases} and the bulk modulus grows with temperature. In this case Eq.(\ref{rtt}) predicts that the boson peak shifts to higher frequency, as observed empirically, see e.g. \cite{nakayama}. 

Note that when $T>T_r$, there are continuously unstable modes. Interestingly, such a temperature also exists in mean field spin models \cite{laloux},  which were proposed as possible scenarios of the glass transition \cite{woly}. The dynamics of these models at higher temperature is exactly described by mode coupling equations \cite{kurchan}. According to this analogy $T_r=T_{MCT}$. 

\subsubsection{Fragile and strong glasses}

In order to discuss the distinction between strong  and fragile glasses, we introduce a second characteristic temperature $T_a$.  $T_a$ corresponds to the typical energy activation of the structural relaxation processes that do not depend on the rigidity of the structure.  In particular we think about  local rearrangements, such as the displacements of coordination defects in strongly covalent networks \cite{fort}. The glass transition must occur around the smallest of the two temperature  $T_a$ and $T_r$, since by definition the structure relaxes easily at higher temperature.   Hence if $T_a\gg T_r$ the glass transition occurs in the vicinity of $T_r$.  As the curvatures of  the energy landscape  dramatically evolves with temperature near $T_r$, it is natural to expect  the corresponding dynamics to be super-activated. This supports the following scenario: glasses with $T_a\gg T_r$ are fragile. On the other hand,  if $T_a\ll T_r$, the glass transition takes place in the vicinity of $T_a$. As the local structure does not display any important changes  near the glass transition, we expect such glasses to be strong. A similar discussion  in terms of energy landscape is presented in \cite{cava}.

This scenario implies that  strong glasses are whether (i) system with a large coordination, where anomalous modes frequencies are high. This is coherent with the  empirical fact that  strongly connected covalent networks are strong. (ii) anomalous systems  which have a strong boson peak, but where an increase of temperature does not lower much the boson peak frequency, or even stabilizes the system like in silica. In these two cases we expect the rigidity of the covalent network  not to do  play any role at the glass transition. This is supported by simulations that indicate that the structural relaxation is purely local in silica glass \cite{fort}, and that the covalent network exists until 8000 K at our pressure \cite{si}, which is much larger than the glass transition temperature. Note the elastic anomaly of silica disappears at high pressure, which suggests that this glass might become fragile when the pressure increases \cite{ba}.

According to the present point of view, fragile glasses must display anomalous modes that approach zero frequency around the glass transition. A possible test could be done by considering particles with gaussian potentials. It was shown theoretically  and numerically \cite{parisi2, parisi3, mezard}  that such systems at infinite temperature display excess-modes.  These modes become stable above a finite density. Above this density this scenario predicts that the glass is strong, as the stable excess-modes at infinite temperature will stay stable as the temperature is lowered.  At smaller density we expect a fragile behavior to occur, as it is the case for soft spheres \cite{bernu}. 

\subsubsection{Heterogeneous relaxation in fragile glass}

According to the present scenario, in a fragile glass at $T_r$ the anomalous modes are characterized by a finite length scale $l^*$. According to Eq.(\ref{rtt}) $l^*\sim \delta z^{-1}$ does not diverge since the initial stress term, and the pressure, have a finite value. Thus we also expect  the shear modulus at $T_r$ to have a finite value, as it is the case for the marginally rigid system of \cite{J}.  As soon as $T\geq T_r$, the system displays unstable normal modes. Near $T_r$, only the modes with characteristic length $l^*$ are unstable. The anomalous modes confined on subsystems smaller than $l^*$ have higher, non-zero frequencies. Thus the collapse of unstable modes at $T_r$ involve rearrangements on length scale of the order of $l^*$, but not smaller. As the temperature increases, modes with smaller characteristic lengths become unstable, and rearrangements can occur on shorter length scales.  Hence this model predicts the presence of a growing length scale $l(T)$ that converges toward $l^*$ when the temperature decreases toward $T_r$.  Growing dynamical length scales were observed numerically as the temperature decreases, see e.g.\cite{lud} and ref. therein.  The curve $l(T)$ could be measured  by pinning the particles at the boundary of subsystems of size $l$, and by considering the dependence of the structural relaxation time with temperature  for given $l$. Note that a  similar test was already proposed in \cite{gu} to test dynamical length scales.

When the temperature decreases below $T_r$, we expect  activated events to relax the glass structure. This is plausibely enhanced by  the neighboring elastic instability, and by the large amount of nearly unstable modes.  Finding the relaxation process of such weak structures is  a necessary next step. Such models may lead to the prediction for the super-activated dependence of the relaxation time when the temperatures decreases and the elastic instability goes away. Once again,  rigidity is not a local criterion, but demands the existence of a relation between coordination number and pressure on length scale $l^*$. Hence it is not excluded that the fluctuations of quantities such as pressure and coordination on such distances may trigger relaxations.

\subsection{Granular matter}

There are open questions on the way force propagates in amorphous systems such as granular matter. At large distances, a granular pack should behave as a continuous elastic medium, at least in a small linear regime. As we discussed in Chapter \ref{c2}, the situation is different in anisotropic isostatic systems, whose elastic behaviors can be described by hyperbolic equations.  Nevertheless, there is as yet no description of force propagation in {\it isotropic} isostatic systems. More importantly, we do not know how force propagation evolves toward a normal continuous elastic behavior when the coordination number increases, or when friction is present.  It was proposed in \cite{Tom1} that in anisotropic system, normal elasticity is recovered above $l^*$, the distance at which soft modes disappear. Our derivation of the anomalous modes suggests that the transition toward continuous elasticity could occur at distances shorter than $l^*$. In particular for a force {\it dipole}, or a local strain, as we discussed in Chapter \ref{c8} it is plausible that  the characteristic transverse length  $l_t\sim \delta z^{-\frac{1}{2}}$ characterizes the cross-over  isostaticity/elasticity.  Note that  this does not preclude that the response to a force mono-pole has a different cross-over. It would be of much interest to investigate these subtle properties. It could be done in simulations of elastic spheres near the jamming threshold. Experimentally such tests would require to find efficient ways to modulate the coordination and the distance from isostaticity  \cite{evelyne}. 

The length scales we are talking about obviously depend on the system considered, but we expect them to be typically of the order of  a few tens of particle sizes. Thus they may not affect for example the building of sand castles, as a continuum description is expected to be valid for macroscopic objects. Nevertheless the properties of an assembly of grains at such  length scales might play a crucial role in the rheology of these systems, both in the solid and in the dense liquid phases. Similar length scales were  observed in the spatial correlations of dense granular flows on a slope, and  are probably related to the surprising dependence of the thickness $h$ of the flow with the angle of the slope$\theta$  \cite{pouliquen}. Bending a layer of sand toward its avalanche angle is equivalent to imposing a shear. Thus understanding how shear affects the anomalous modes might shed light on this problem. The presence of a fixed boundary at a distance $h$ of the free surface certainly increases the anomalous modes frequency. Making the most simple assumptions that (i) the shear decreases linearly the characteristic energy of the anomalous modes and (ii) the presence of of a fixed boundary is equivalent to an increase of coordination of order $1/h$ gives $\omega_{AM}^2\sim A_1 B (\delta z + 1/h) -A_3 \sin \theta$. Imposing $\omega_{AM}=0$ yields a dependence of $h$ of the form $h(\theta)\sim \frac{1}{\theta-\theta_0}$, which is not inconsistent with the empirical data.

Another interesting property of granular matter is compaction: if a container of sand is tapped from below, the density of the system increases. The process displays many glassy features: the dynamics becomes very slow with time, aging and memory effect are observed \cite{compaction}. When a granular pack is tapped weakly enough to avoid fluidization, there are  two main causes for the irreversible events that lead to compaction. One the one hand,  contacts on the Coulomb cone can slide. On the other hand, the destabilizing effect of the pressure wave can cause the buckling  of anomalous modes.  This structural buckling increases the coordination and the packing fraction, as sketched in Fig.(\ref{p}). Hence the structure becomes more and more stable, and the compaction dynamics slows down. As the coordination rises the length scale $l^*$ decreases. This  may be possible to test this presiction using  X-ray microtomography.

\chapter{ Acknowledgments } 

I am happy to thank Jean-Philippe Bouchaud, Carolina Brito, Sidney Nagel, Leonardo Silbert and Thomas Witten who contributed to this work.

\end{document}